\documentclass[10pt, twocolumn, amssymb, aps, prx, reprint, floatfix, superscriptaddress]{revtex4-2}
\usepackage[a4paper, total={7in, 10in}]{geometry}
\usepackage{graphicx} 
\graphicspath{{Figures/}}
\usepackage{amsmath}
\usepackage{physics}
\usepackage{dsfont}
\usepackage{dcolumn}
\usepackage{xcolor}
\usepackage{bbold}
\usepackage{appendix}
\usepackage{units}

\usepackage{pgfplots}
\pgfplotsset{compat=1.18} 

\usepackage{hyperref}
\hypersetup{colorlinks=true,linkcolor=blue}

\newcommand{\smartincludegraphics}[1]{%
    \settowidth{\dimen0}{\includegraphics{#1}} 
    \ifdim\dimen0>\linewidth
        \includegraphics[width=\linewidth]{#1} 
    \else
        \includegraphics{#1} 
    \fi
}

\begin{document}


\title{Unifying adiabatic state-transfer protocols with \texorpdfstring{$(\alpha, \beta)$}{(a, b)}-hypergeometries}

\author{Christian Ventura Meinersen}
\thanks{These authors contributed equally to this work.}
\affiliation{QuTech and Kavli Institute of Nanoscience, Delft University of Technology, PO Box 5046, 2600 GA Delft, The Netherlands
}

\author{David Fernandez-Fernandez}
\thanks{These authors contributed equally to this work.}
\affiliation{Instituto de Ciencia de Materiales de Madrid ICMM-CSIC, 28049 Madrid, Spain}

\author{Gloria Platero}
\affiliation{Instituto de Ciencia de Materiales de Madrid ICMM-CSIC, 28049 Madrid, Spain}

\author{Maximilian Rimbach-Russ}
\affiliation{QuTech and Kavli Institute of Nanoscience, Delft University of Technology, PO Box 5046, 2600 GA Delft, The Netherlands
}

\date{\today}

\begin{abstract}
    Adiabatic optimal control schemes are essential for advancing the practical implementation of quantum technologies. However, the vast array of possible adiabatic protocols, combined with their dependence on the particular quantum system and function-specific parameter ranges, complicates the task of discerning their respective strengths and limitations in arbitrary operations. In this work, we provide a unifying framework, called \textit{$(\alpha,\beta)$-hypergeometries}, that allows for flexible, noise-resistant, and easy-to-use implementation of enforced adiabatic dynamics for any multi-level quantum system. Moreover, this framework provides a comprehensive mapping of all adiabatic protocols through a universal cost function and offers an exact analytical characterization of the adiabatic dynamics. In particular, we derive precise expressions for infidelity resonances and establish performance guarantees in the adiabatic limit for any choice of $(\alpha,\beta)$. We also discuss in detail the experimental feasibility of the resulting pulse shapes through analytical and numerical methods. Finally, we test our method for the optimal control of coherent information transfer through spin shuttling in silicon quantum dots with small valley splittings.
\end{abstract}

\maketitle

\section{Introduction}
Efficient engineering of quantum states requires optimal control of the parameters describing the physical system~\cite{glaserTrainingSchrodingerCat2015}. In practically all systems of relevance, one encounters anticrossings in the energy levels, where quantum states hybridize due to controlled or uncontrolled interactions~\cite{ivakhnenkoNonadiabaticLandauZener2023}. The ability to precisely navigate through these anticrossings and prepare precise states is a universal challenge across various quantum platforms, including, superconducting resonators~\cite{izmalkovObservationMacroscopicLandauZener2004}, photonic platforms~\cite{alexanderManufacturablePlatformPhotonic2024}, graphene-based nanostructures~\cite{higuchiLightfielddrivenCurrentsGraphene2017}, topological superconductors~\cite{aghaeeInterferometricSingleshotParity2025}, semiconductor quantum dots~\cite{pettaCoherentBeamSplitter2010,ribeiroCoherentAdiabaticSpin2013, banFastLongrangeCharge2018}, quantum batteries~\cite{binderQuantacellPowerfulCharging2015}, adiabatic quantum computing platforms for quantum optimization~\cite{blekosReviewQuantumApproximate2024, abbasChallengesOpportunitiesQuantum2024a} and more. 

Control of quantum information is the cornerstone in the development and implementation of practical quantum technologies~\cite{glaserTrainingSchrodingerCat2015} like quantum computing and quantum communication. To surpass the challenge of fault-tolerant quantum operations, optimal quantum control strategies are unavoidable, i.e., by providing optimal pulse shapes of the relevant control parameters~\cite{werschnikQuantumOptimalControl2007,glaserTrainingSchrodingerCat2015,poggialiOptimalControlOneQubit2018,theisCounteractingSystemsDiabaticities2018,kochQuantumOptimalControl2022, greentreeCoherentElectronicTransfer2004}. These strategies range from using Pontryagin's Maximum Principle~\cite{khanejaTimeOptimalControl2001, carliniTimeOptimalQuantumEvolution2006,pontryaginMathematicalTheoryOptimal2017,boltyanskiGeometricMethodsOptimization1999,carliniTimeoptimalUnitaryOperations2007,boscainIntroductionPontryaginMaximum2021,koikeQuantumBrachistochrone2022, rezakhaniQuantumAdiabaticBrachistochrone2009,walelignDynamicallyCorrectedGates2024, chernovaOptimizingStateTransfer2024, stepanenkoTimeoptimalTransferQuantum2025}, various numerical techniques~\cite{khanejaOptimalControlCoupled2005,goodwinAcceleratedNewtonRaphsonGRAPE2023, canevaChoppedRandombasisQuantum2011, fauquenotEOGRAPEEODRLPEOpen2024, katiraee-farUnifiedEvolutionaryOptimization2025}, to optimizing adiabatic dynamics via shortcuts-to-adiabaticity methods~\cite{bergmannCoherentPopulationTransfer1998,motzoiSimplePulsesElimination2009,berryTransitionlessQuantumDriving2009,degrandiAdiabaticPerturbationTheory2010,ivakhnenkoNonadiabaticLandauZener2023,vuthaSimpleApproachLandauZener2010,chernovaOptimizingStateTransfer2024, stepanenkoTimeoptimalTransferQuantum2025,motzoiOptimalControlMethods2011,chenLewisRiesenfeldInvariantsTransitionless2011,ribeiroSystematicMagnusBasedApproach2017,selsMinimizingIrreversibleLosses2017,theisCounteractingSystemsDiabaticities2018,banFastLongrangeCharge2018,takahashiHamiltonianEngineeringAdiabatic2019,banSpinEntangledState2019,guery-odelinShortcutsAdiabaticityConcepts2019,setiawanAnalyticDesignAccelerated2021,zhuangNoiseresistantLandauZenerSweeps2022,takahashiDynamicalInvariantFormalism2022,glasbrennerLandauZenerFormula2023,rimbach-russSimpleFrameworkSystematic2023,dengisAcceleratedCreationNOON2025,romeroOptimizingEdgestateTransfer2024, liuAcceleratedAdiabaticPassage2024,xuImprovingCoherentPopulation2019,fehseGeneralizedFastQuasiadiabatic2023,limaSuperadiabaticLandauZenerTransitions2024,richermeExperimentalPerformanceQuantum2013, rolandQuantumSearchLocal2002,martinez-garaotFastQuasiadiabaticDynamics2015,chenSpeedingQuantumAdiabatic2022} to reduce coherent errors arising from non-adiabatic transitions while operating at short pulse times even in multi-level systems~\cite{fernandez-fernandezQuantumControlHole2022, meinersenQuantumGeometricProtocols2024}.
\begin{figure}[t!]
    \centering
    \smartincludegraphics{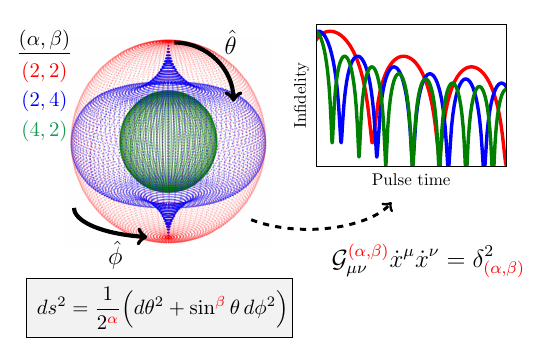}
    \caption{Hypergeometric protocols using the $(\alpha,\beta)$-hypergeometric tensor in Eq.~\eqref{eqn: generalized qgt} for a qubit Hamiltonian described in Eq.~\eqref{eqn: qubit Hamiltonian}. The $(\alpha,\beta)$-hypergeometries $\mathcal{G}_{\mu\nu}^{(\alpha,\beta)}$ unify and capture the dynamics of adiabatic protocols, including the FAQUAD (green), and the geometric fast-QUAD (red). Each protocol will result in a different adiabatic behavior, which is illustrated pictorially on the right side as the infidelity as a function of the pulse time. To get the hyper-Bloch spheres we used the embedding coordinates in Appendix~\ref{app: embedding hypergeo}.}
    \label{fig: hypergeo protocol}
\end{figure}
Given the plethora of possible adiabatic optimal control strategies, one finds that each method has its own advantages and disadvantages. Therefore, the optimal approach depends on the given physical system, the required input, computational complexity, parameter range, and the feasibility of implementing the respective optimal control pulses in a given physical system. These facts hinder the understanding of the underlying requirements for high-fidelity adiabatic optimal control. An appealing method uses the quantum geometric tensor~\cite{kolodrubetzGeometryNonadiabaticResponse2017, bukovGeometricSpeedLimit2019, liskaHiddenSymmetriesBianchi2021} for optimal adiabatic dynamics~\cite{meinersenQuantumGeometricProtocols2024}. Despite the similariy of quasi-adiabatic~\cite{fernandez-fernandezQuantumControlHole2022} and geometric methods~\cite{meinersenQuantumGeometricProtocols2024}, they constitute drastic changes in optimal pulse times and fidelities. 

Using a simple modification of the geometric picture, by introducing two parameters $(\alpha,\beta)$, we can directly map out many of the above-mentioned adiabatic protocols. We call this unifying framework \textit{$(\alpha,\beta)$-hypergeometries}, which, upon further analysis, leads us to conclude that the set of pulses yields a usable basis mapping out all possible adiabatic protocols and pulses beyond the adiabatic limit. Our $(\alpha,\beta)$-hypergeometries exhibit many advantages not seen in many of the previous optimal control strategies. Firstly, our protocol only depends on the system Hamiltonian, hence, it can be easily adjusted and used for any quantum system. Remarkably, the $(\alpha,\beta)$-hypergeometries and the corresponding equations for the control parameters only scale linearly with the number of control parameters that one wants to control simultaneously. Secondly, by changing the values $(\alpha,\beta)$, the framework is flexible toward experimental constraints by adjusting the targeted adiabaticity, pulse smoothness, and the adiabatic behavior of the infidelity. The framework allows us to provide a list of analytic equations that capture the adiabaticity, the infidelity resonance frequencies, and the upper bound of the infidelity in the adiabatic limit for any $(\alpha,\beta)$, thereby fully characterizing the adiabatic dynamics without the need for extensive time-evolution simulations. Lastly, our $(\alpha,\beta)$-hypergeometries are resistant to first-order parameter fluctuations and also allow us to find the operation sweet spot that simultaneously bypasses pure dephasing errors and residual errors arising from diabatic transitions at ultra-fast speeds.

The article is structured as follows. Starting from the general definition of the $(\alpha,\beta)$-hypergeometries in Section~\ref{sec: alphabet hypergeo intro}, we follow by providing a geometric intuition using a simple qubit model including the relation between quantum dynamics and geodesic evolution, finalizing in a conjectured minimal time bound for the $(\alpha,\beta)$-hypergeometric protocols. Subsequently, in Section~\ref{sec: single qubit optimal control}, we focus on the Landau-Zener model and provide analytic expressions to understand the adiabatic dynamics, which are then verified by in-depth numerical simulations and related to the experimental applicability. Remarkably, the pulses generated from the Landau-Zener model not only capture many known adiabatic protocols, but also provide a basis to generate arbitrary pulse shapes for beyond quasi-adiabatic control in multi-level systems (see Section~\ref{sec: N-level generalization}). After discussing and providing explicit connections between our pulses and experimental features in Section~\ref{sec: experimental feasability}, we apply our framework in Section~\ref{sec: application shuttling} to study a multi-level system capturing spin shuttling~\cite{desmetHighfidelitySinglespinShuttling2024, vanriggelen-doelmanCoherentSpinQubit2024, fernandez-fernandezFlyingSpinQubits2024, banFastRobustSpin2012}.

\section{\texorpdfstring{$(\alpha, \beta)$}{(a, b)}-hypergeometries}\label{sec: alphabet hypergeo intro}
Given the recent interpretability of adiabatic quantum optimal control problems as geometric problems~\cite{meinersenQuantumGeometricProtocols2024, nielsenOptimalControlGeometry2006, kolodrubetzGeometryNonadiabaticResponse2017, liskaHiddenSymmetriesBianchi2021}, where the geodesic equations of some suitable metric~\cite{neupertMeasuringQuantumGeometry2013, kolodrubetzGeometryNonadiabaticResponse2017, kolodrubetzClassifyingMeasuringGeometry2013, liskaHiddenSymmetriesBianchi2021, lambertClassicalQuantumInformation2023} are used to optimally solve for the adiabatic protocol, we generalize the quantum geometric tensor as follows
\begin{align}
    \label{eqn: generalized qgt}
    \mathcal{Q}^{(\alpha, \beta)}_{\mu \nu}:=  \sum_{n\neq m} \frac{\mel{\psi_m}{\partial_\mu \hat{H}}{\psi_n}^{\beta/2}\mel{\psi_n}{\partial_\nu \hat{H}}{\psi_m}^{\beta/2}}{(E_n-E_m)^{\alpha}}.
\end{align}
We call this object the \textit{$(\alpha,\beta)$-hypergeometric tensor} with respect to the initial state $\ket{\psi_m}$. Note that we assume a non-degenerate spectrum. Analogously, we have the \textit{quantum hypermetric tensor} $\mathcal{G}^{(\alpha, \beta)}_{\mu \nu}=\Re \mathcal{Q}^{(\alpha, \beta)}_{\mu \nu}$ and the associated hyper-Berry curvature $\mathcal{B}^{(\alpha, \beta)}_{\mu \nu}=-2\Im \mathcal{Q}^{(\alpha, \beta)}_{\mu \nu}$. The hypergeometric tensor exhibits the same symmetry properties, like $\hat{H}\mapsto\hat{H}+\omega(x) \hat{\mathbb{1}}$, as the quantum geometric tensor. However, we note that the conformal symmetry of~\cite{meinersenQuantumGeometricProtocols2024}, where $\hat{H}\mapsto\Omega(x)\hat{H}$ with $\Omega(x)$ a parameter-dependent scalar function, only holds when $\alpha=\beta$ as $\mathcal{Q}_{\mu \nu}^{(\alpha,\beta)}\mapsto \Omega^{\beta-\alpha}\mathcal{Q}_{\mu \nu}^{(\alpha,\beta)}$ under the conformal transformation. Remarkably, this definition captures a list of known adiabatic protocols for a given combination of values $(\alpha,\beta)$ as seen in Table~\ref{tab: adiabatic protocols and hypergeo}.
\begin{table}[t!]
    \centering
    \begin{tabular}{|c|c|c|}\hline
         $(\alpha, \beta)$  & Protocol & Reference \\ \hline \hline 
         $(0, 0)$ & Linear & \cite{degrandiAdiabaticPerturbationTheory2010,ivakhnenkoNonadiabaticLandauZener2023, vuthaSimpleApproachLandauZener2010, glasbrennerLandauZenerFormula2023} \\ \hline
         $(2, 2)$ & Geometric fast-QUAD & \cite{meinersenQuantumGeometricProtocols2024} \\ \hline
         $(4, 2)$ & FAQUAD & \cite{martinez-garaotFastQuasiadiabaticDynamics2015,petiziolQuantumControlEffective2024, fernandez-fernandezQuantumControlHole2022, fehseGeneralizedFastQuasiadiabatic2023, chenSpeedingQuantumAdiabatic2022} \\ \hline
         $(2\,\alpha_\text{LA}, 0)$ & Local adiabaticity & \cite{fernandez-fernandezQuantumControlHole2022, guery-odelinShortcutsAdiabaticityConcepts2019, richermeExperimentalPerformanceQuantum2013, rolandQuantumSearchLocal2002} \\ \hline
    \end{tabular}
    \caption{Comparison between adiabatic protocols and their respective hypergeometric subspaces in terms of ($\alpha,\beta$).}
    \label{tab: adiabatic protocols and hypergeo}
\end{table}
\subsection{Hypergeometric Bloch sphere and Berry curvature}
In the following subsections, we aim to provide the explicit correspondence between observables of the hypergeometries and adiabatic dynamics through the example of a single-qubit system. It is instructive to study a general single-qubit Hamiltonian in spherical coordinates
\begin{align}
    \label{eqn: qubit Hamiltonian}
    \hat{H}_\text{qubit}(\theta,\phi)=n(\theta,\phi)\cdot \vec{\sigma}=\begin{pmatrix}
        \cos \theta & e^{-i\phi}\sin \theta \\
        e^{i\phi}\sin \theta  & -\cos \theta
    \end{pmatrix},
\end{align}
where $n(\theta,\phi)$ is the unit vector on the 2-sphere and $\vec{\sigma}=(\sigma_x,\sigma_y,\sigma_z)$ is the Pauli vector. The angles $x^\mu=\{\theta,\phi\}$ represent the parameters of the Hamiltonian. The eigenstates (with eigenvalues $E_{0,1}=\mp 1$) of the qubit Hamiltonian in Eq.\eqref{eqn: qubit Hamiltonian} are
\begin{align}
    \ket{\psi_0} & = \begin{pmatrix}
        \sin \theta/2 \\ 
        -e^{i\phi} \cos \theta/2
    \end{pmatrix} & \ket{\psi_1} & = \begin{pmatrix}
        \cos \theta/2 \\ 
        e^{i\phi} \sin \theta/2
    \end{pmatrix}.
\end{align}
To compute the quantum hypergeometric tensor $\mathcal{Q}_{\mu \nu}^{(\alpha,\beta)}$, we need the matrix overlap elements. Without explicit dependence on $(\alpha, \beta)$ we find
\begin{align}
    \mel{\psi_0}{\partial_\theta\hat{H}_\text{qubit}}{\psi_1}&=-1\\
    \mel{\psi_0}{\partial_\phi\hat{H}_\text{qubit}}{\psi_1}&=-i\sin \theta\\
    E_1-E_0=2.
\end{align}
Therefore, we find that the quantum hypergeometric tensor (in matrix representation in parameter space $\mu,\nu$) for a single-qubit system is given by
\begin{align}
   [\mathcal{Q}_{\mu \nu}^{(\alpha,\beta)}(\theta,\phi)]= \frac{1}{2^\alpha}\begin{pmatrix}
       1 & (-i)^{\beta/2} \sin^{\beta/2} \theta\\
       i^{\beta/2} \sin^{\beta/2} \theta & \sin^\beta \theta
   \end{pmatrix}.
\end{align}
Hence, for the single-qubit quantum hypermetric tensor, we find the hyper-Bloch sphere
\begin{align}
    \mathcal{G}_{\mu \nu}^{(\alpha,\beta)}dx^\mu dx^\nu =\frac{1}{2^\alpha}\Big(d\theta^2+\sin^\beta \theta \,d\phi^2\Big).
\end{align}
For the geometric case $(\alpha,\beta)=(2,2)$, we find the line element for the 2-sphere representing the standard Bloch sphere. Here, we find that the value of $\alpha$ is an overall scaling factor of the Bloch sphere and that $\beta$ determines the shape close to the poles, as can be seen in Fig.~\ref{fig: hypergeo protocol}. The associated hyper-Berry curvature, which captures topological effects, is given by
\begin{align}
    \mathcal{B}_{\theta \phi}^{(\alpha,\beta)}=-\mathcal{B}_{\phi \theta}^{(\alpha,\beta)}=-2\,\frac{(-i)^{\beta/2} \sin^{\beta/2} \theta}{2^\alpha}.
\end{align}
\subsection{Hyper-geodesics and robustness} 
In this work, we are interested in utilizing the hypermetric tensor $\mathcal{G}_{\mu \nu}^{(\alpha,\beta)}$ for optimal control strategies. Given the relation between the state fidelity and the standard quantum metric tensor $g_{\mu\nu}$~\cite{liskaHiddenSymmetriesBianchi2021, kolodrubetzGeometryNonadiabaticResponse2017} we propose a similar structure for an action capturing the infidelity
\begin{align}
    \label{eq: hypergeo fidelity}
    1-\mathcal{F}\approx \int d\tau\; \sqrt{\mathcal{G}_{\mu \nu}^{(\alpha,\beta)}\dv{x^\mu}{\tau}\dv{x^\nu}{\tau}}=\mathcal{L}^{(\alpha,\beta)}[x^\mu(\tau)],
\end{align}
where $\tau=t/t_\text{f}$ is the affine parameter of the curve in the hypergeometry, $t_\text{f}$ is the pulse time, and $\mathcal{L}^{(\alpha,\beta)}[x^\mu]$ is the length of a curve traced out by the parameters $x^\mu(\tau)$. Hence, the minimum infidelity (maximum fidelity) is given by the Euler-Lagrange equations, i.e., the geodesics on $\mathcal{G}_{\mu \nu}^{(\alpha,\beta)}$. In Fig.~\ref{fig: hypergeo protocol}, we illustrate our proposed hypergeometric protocols. Starting from the hyper-Bloch spheres for different combinations of $(\alpha,\beta)$, we find different pulse shapes with their corresponding infidelity. As the geodesic equations constitute a minimum of the action, the solutions to those equations are also stable against first-order fluctuations (see Appendix~\ref{app: hypergeo fluctuations}). For adiabatic evolution, the geodesics will minimize the energy fluctuations and hence will move an eigenstate along the trajectory defined by the tangent vector $dx^\mu/d\tau$, minimizing the excitation of higher energy states. The conservation of energy can be expressed by Beltrami's identity or the Killing equation~\cite{liskaHiddenSymmetriesBianchi2021, meinersenQuantumGeometricProtocols2024} and may be extended to
\begin{align}
    \label{eq: hypergeo protocol}
    \mathcal{G}_{\mu \nu}^{(\alpha,\beta)}\dv{x^\mu}{\tau}\dv{x^\nu}{\tau}=(\delta^{(\alpha,\beta)})^2,
\end{align}
where we call $\delta^{(\alpha,\beta)}$ the hyper-adiabaticities. The hyper-adiabaticities capture how adiabatic a given protocol is, where adiabaticity is satisfied for $\delta^{(\alpha,\beta)}\ll1$. 
\subsection{Quantum hypergeometric minimal time bound}
The $(\alpha,\beta)$-hypergeometries allow us to study adiabatic dynamics from a geometric perspective. By comparing equations~\eqref{eq: hypergeo fidelity} with Eq.~\eqref{eq: hypergeo protocol}, we see that the hyper-adiabaticities capture the length of the curve in parameter space. Hence, the shortest paths are also the ones that are most adiabatic. Note that the dynamics of the geodesic are not dependent on overall scaling, as the conformal factor $\Omega^{\beta-\alpha}$ cancels in the hypergeometric protocol. If we constrain ourselves to a single parameter subspace of the single-qubit system, we can compute the path lengths over the entire hyper-Bloch sphere analytically
\begin{align}
    \mathcal{L}^{(\alpha,\beta)}_\theta&=\int_0^\pi d\theta\, \sqrt{\mathcal{G}_{\theta\theta}^{(\alpha,\beta)}}=\frac{\pi}{2^{\alpha/2}}\\
    \mathcal{L}^{(\alpha,\beta)}_\phi&=\int_0^{2\pi} d\phi\, \sqrt{\mathcal{G}_{\phi\phi}^{(\alpha,\beta)}}=\frac{\pi}{2^{\alpha/2-1}}\sin^{\beta/2}\theta.
\end{align}
The total volume of the hyper-Bloch sphere can also be expressed analytically
\begin{align}
    \text{Vol}^{(\alpha,\beta)}&=\int_0^{2\pi}d\phi\int_0^{\pi}d\theta \sqrt{\text{det } \mathcal{G}_{\mu\nu}^{(\alpha,\beta)}}\\
    &=\frac{\pi}{2^{\alpha-1}}\left[\sqrt{\pi}\frac{\Gamma(\frac{2+\beta}{4})}{\Gamma(1+\frac{\beta}{4})}\right] \,\text{ for } \beta>-2 ,
\end{align}
where $\Gamma(z)$ is the Gamma function. Note that for the qubit case, we find that for $\alpha,\beta\gg 1$ the volume decreases as $\text{Vol}^{(\alpha,\beta)}\propto 2^{-\alpha}\beta^{-1/2}$, meaning that on average the distance between any two points will be decreased. Importantly, with these geometric quantities, we are now able to find the generalized quantum minimum time bound similarly as defined in Refs.~\cite{bukovGeometricSpeedLimit2019, sunDistinctBoundQuantum2019} as $T_\text{qsl}\geq \mathcal{L}^{(2,2)}[x^\mu]/\delta^{(2,2)}$, where this bound is saturated by geodesic protocols. Henceforth, we aim to compare the lengths of the hypergeometries with the standard geometry of Hilbert space to see for which values of $(\alpha,\beta)$, we find a deviation of the length and, hence, the minimum time bound. The intuition is that smaller geodesic paths require a smaller operation time. However, comparing two lengths on manifolds of different geometry requires proper renormalization. Otherwise, for instance, in the case of the hyper-Bloch sphere, the protocol of $\alpha\to\infty$ would generate the minimal time bound as $\lim_{\alpha\to \infty}\mathcal{L}_{\theta,\phi}^{(\alpha,\beta)}=0$. Therefore, we normalize the lengths by the total volume of the hypergeometries
\begin{align}
    \frac{T_\text{qsl}^{(\alpha,\beta)}}{T_\text{qsl}}&=\frac{\mathcal{L}^{(\alpha,\beta)}[x^\mu]/\delta^{(\alpha,\beta)}}{\mathcal{L}^{(2,2)}[x^\mu]/\delta^{(2,2)}}\frac{\text{Vol}^{(2,2)}}{\text{Vol}^{(\alpha,\beta)}}\\&=\frac{\text{Vol}^{(2,2)}}{\text{Vol}^{(\alpha,\beta)}}=\frac{2^{\alpha-1}}{\sqrt{\pi}}\frac{\Gamma(1+\frac{\beta}{4})}{\Gamma(\frac{2+\beta}{4})},
\end{align}
as given by the hypergeometric protocol in Eq.~\eqref{eq: hypergeo protocol}.  For values of $\alpha,\beta\geq 2$ we find that $T_\text{qsl}^{(\alpha,\beta)}\geq T_\text{qsl}$, hence the standard geometric bound provides the minimal bound. When $\alpha,\beta< 2$, we find that the hypergeometric protocol obtains a lower bound, i.e., $T_\text{qsl}^{(\alpha,\beta)}<T_\text{qsl}$. Importantly, there is a constraint for valid combinations of $(\alpha,\beta)$. In order to have a well-defined distance measure, the hypermetric tensor $\mathcal{G}$ must be symmetric and non-degenerate, i.e., $\text{det }\mathcal{G}\neq 0$. However, this requirement is not satisfied for values $\beta<0$, where for the qubit system $\text{det }\mathcal{G}\propto \sin^\beta \theta = 0$ for $\theta=\pi/2$. For the adiabatic transfer problem, this corresponds to a middle point. Henceforth, only values $\beta\geq 0$ have a well-defined metric for the single-qubit case. A well-defined smaller minimal time bound $T_\text{qsl}^{(\alpha,\beta)}<T_\text{qsl}$ is then guaranteed for $\alpha,\beta<2$ with $\beta\geq 0$. In Appendix~\ref{app: geometric invariants}, we compute some other geometric properties that can be related to topological invariants~\cite{kolodrubetzClassifyingMeasuringGeometry2013, sharipovHilbertSpaceGeometry2024}. Note that the explicit analytic expressions and restrictions above only hold for the single-qubit case. 
\section{Two-level optimal control}
\label{sec: single qubit optimal control}
In experimental settings, the ability to control multiple parameters simultaneously requires a lot of fine-tuning. Therefore, we aim to study in-depth the optimal control of a single parameter using our $(\alpha,\beta)$-hypergeometries. In particular, we completely characterize the adiabatic behavior of the fidelities for any $(\alpha,\beta)$ combination, including the exact resonance frequencies and the upper bound on the infidelity. In addition, it will allow us to provide a clear comparison to known adiabatic protocols as laid out in Table~\ref{tab: adiabatic protocols and hypergeo}.
\subsection{Single parameter hyper-geodesics}
For a single control parameter $\lambda$, the quantum hypermetric tensor takes the form
\begin{align}
    \label{eqn: generalized qgt single param}
    \mathcal{G}_{\lambda\lambda}^{(\alpha, \beta)}= \Re \sum_{n\neq m} \frac{|\mel{\psi_m}{\partial_\lambda \hat{H}}{\psi_n}|^{\beta}}{(E_n-E_m)^{\alpha}},
\end{align}
where $ \{E_n,\ket{\psi_n}\}_{n}$ is the set of eigenvalues and eigenvectors of the control Hamiltonian $\hat{H}(\lambda)$ that we want to study, with $\ket{\psi_m}$ the initial state. In many physical systems, it is instructive to study the Landau-Zener problem as a proxy~\cite{limaSuperadiabaticLandauZenerTransitions2024, ivakhnenkoNonadiabaticLandauZener2023}. The Landau-Zener Hamiltonian reads
\begin{align}
    \label{eq: Landau-Zener Hamiltonian}
    \hat{H}(t)=z(t)\sigma_z+x\,\sigma_x=\begin{pmatrix}
        z(t) & x\\
        x& -z(t)
    \end{pmatrix},
\end{align}
where $x$ is the coupling and $z(t)$ the control parameter. For this model, we can compute the hypermetric tensor analytically, resulting in
\begin{align}
    \mathcal{G}_{zz}^{(\alpha, \beta)}=\frac{1}{2^\alpha}\frac{x^\beta}{(x^2+z(t)^2)^{(\alpha + \beta)/2}}.
\end{align}
One can confirm that for $(\alpha,\beta)=(2,2)$ one achieves the geometric fast-QUAD~\cite{meinersenQuantumGeometricProtocols2024} and for $(\alpha,\beta)=(4,2)$ one retrieves the solution for the FAQUAD~\cite{fehseGeneralizedFastQuasiadiabatic2023, chenSpeedingQuantumAdiabatic2022, fernandez-fernandezQuantumControlHole2022, martinez-garaotFastQuasiadiabaticDynamics2015}. To generate optimal pulse shapes, we need to compute the hyper-adiabaticities to solve for the hyper-geodesics using Eq.~\eqref{eq: hypergeo protocol} by separation of variables
\begin{align}
    \int_{t_0}^tdt\;\delta^{(\alpha,\beta)}=\int_{z(t_0)}^{z(t)}dz\; \sqrt{\mathcal{G}_{zz}^{(\alpha, \beta)}}. 
\end{align}
We can find the exact hyper-adiabaticities for the Landau-Zener model
\begin{align}
\begin{split}
    \delta^{(\alpha,\beta)} \, (t-t_0) = \,_2F_1\left(1, \frac{3-n_+}{2}, \frac{3}{2}; -\frac{z(t)^2}{x^2}\right) \\
    \times \; z(t) \sqrt{\mathcal{G}_{zz}^{(\alpha, \beta)}} \left(1 + \frac{z(t)^2}{x^2}\right),
\end{split}
\end{align}
where $n_\pm=(\alpha\pm\beta)/2$, $t_0$ captures the boundary conditions of the pulse, and $_2F_1(a,b,c;z)$ is the hypergeometric function, defined as
\begin{align}
    _2F_1(a,b,c;z)=\frac{\Gamma(c)}{\Gamma(a)\Gamma(b)}\sum_{s=0}^\infty \frac{\Gamma(a+s)\Gamma(b+s)}{\Gamma(c+s)}\frac{z^s}{s!},
\end{align}
which motivates the name $(\alpha,\beta)$-hypergeometries. 
\subsection{Hypergeometric resonances and fidelity analysis}
\begin{figure*}[tb!]
    \centering
    \smartincludegraphics{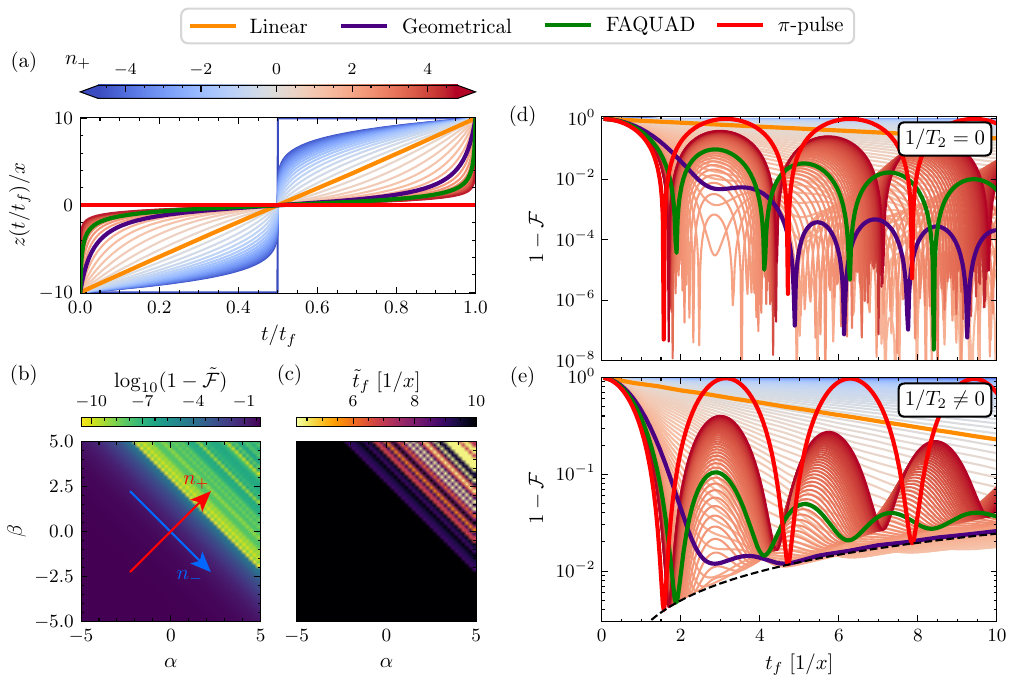}
    \caption{(a) Hypergeometric pulses shapes as a function of $t/t_\text{f}$ for different values of $n_+=(\alpha+\beta)/2$.
    Different protocols given by $n_+$ are color-coded, where the linear ($n_+=0$), the geometric fast-QUAD ($n_+=2$), and FAQUAD ($n_+=3$) are highlighted with an orange, purple, and green line, respectively.
     For completion, we find the $\pi$-pulse (red line) as the limit of $n_+\to\infty$. 
    The boundary conditions of the pulse are given by $z(t=t_\text{f})/x=-z(t=0) / x = 10$.
    (b) Minimal infidelity ($1 - \tilde{\mathcal{F}})$ as a function of $(\alpha,\beta)$.
    The red and blue arrows denote the two perpendicular directions defined as $n_\pm=(\alpha \pm \beta) / 2$.
    (c) Optimal pulse time $\tilde{t}_\text{f}$ needed to reach the minimum infidelity, where the possible pulse times investigated are $t_\text{f}\in [0,10]$ in units of $1 / x$.
    Infidelity as a function of the total pulse time $t_\text{f}$ is plotted without noise (d) and with dephasing (e). 
    The dephasing is included via the Lindblad master equation $\dot{\rho}=-i[H,\rho]+\mathcal{D}[\rho]$ using the jump operator $\mathcal{D}[\rho]=(1/T_2)(\text{diag}(\rho)-\rho)$ with $T_2=100/x$. The dashed black line in (e) corresponds to the theoretical bound of a $\pi$-pulse, which reads $\mathcal{F}_{\pi-\text{pulse}}\leq(1+\exp(-t/2T_2))/2$.}
    \label{fig:two_level_system}
\end{figure*}

The $(\alpha,\beta)$-hypergeometries allow for the generation of a set of pulses labeled by $(\alpha,\beta)$. Here, we will study the properties of these pulses, and we will fully characterize the key features of the infidelity in the adiabatic limit. Note that given the hyper-geodesic protocol in Eq.~\eqref{eq: hypergeo protocol} we find that the pulse shapes for the Landau-Zener problem only depend on the value $n_+=(\alpha+\beta)/2$, as the factor of $\sqrt{\mathcal{G}}$ cancels in the hypergeometric protocol in Eq.~\eqref{eq: hypergeo protocol}. Hence, any two pulses with the same value of $n_+$, but with different combinations of $(\alpha,\beta)$ will result in an identical pulse and infidelity. In Fig.~\ref{fig:two_level_system}, we can sweep the values of $(\alpha,\beta)$ for some trial pulse times $t_\text{f}\in [0,10]$ in units of $x$ (Fig.~\ref{fig:two_level_system}~(a)) and evaluate the minimum infidelity $1-\tilde{\mathcal{F}}\equiv1-\mathcal{F}(\tilde{t}_\text{f})$ (Fig.~\ref{fig:two_level_system}~(b)), with optimal time $\tilde{t}_\text{f}$ (Fig.~\ref{fig:two_level_system}~(c)). 
We can observe a clear dependence only on the value of $n_+$, as the diagonal lines of the constant $n_+$ have the same minimum infidelity. Furthermore, it is clear that we find the pulse shapes with value $n_+>0$ to outperform the ones with $n_+<0$, as they are more adiabatic and hence reduce the errors coming from diabatic transitions. In addition, using the pulses in Fig.~\ref{fig:two_level_system}~(a) we can illustrate their corresponding infidelity in Fig.~\ref{fig:two_level_system}~(d-e),  as a function of the pulse time $t_\text{f}$ without (Fig.~\ref{fig:two_level_system}~(d)) and with dephasing noise (Fig.~\ref{fig:two_level_system}~(e)). We highlight certain known pulses, including the linear $(n_+=0)$ in orange, the geometric fast-QUAD $(n_+=2)$ in purple, the FAQUAD $(n_+=3)$ in green, and the $\pi$-pulse $(n_+\to \infty)$ in red. The pulses with $n_+<0$ also include similar pulse shapes as the superadiabatic transitions in~\cite{limaSuperadiabaticLandauZenerTransitions2024}. Therefore, we are able to, with the use of $(\alpha,\beta)$-hypergeometries, capture many known adiabatic protocols, which are summarized in Table~\ref{tab: adiabatic protocols and hypergeo}.

There are two main features in the infidelity that we want to study: the fidelity resonances and the infidelity upper bound in the adiabatic limit. These two features allow us to fully characterize the adiabatic dynamics. To study these features, we invoke adiabatic perturbation theory for a general multi-level system
\begin{align}
\begin{split}
    \dot{c}_m=&-c_m\braket{\psi_m}{\dot{\psi}_m}\\
    &-\sum_{n\neq m}c_n\frac{\mel{\psi_m}{\dot{H}}{\psi_n}}{E_n-E_m}e^{-i\int_0^t (E_n-E_m) dt'},
    \label{eq:time_evolution}
\end{split}
\end{align}
where $c_m$ are the expansion coefficients of the wave function, $\ket{\psi_n}$ the eigenvectors with their corresponding eigenvalues $E_n$. The second term on the right consists of the non-adiabatic/diabatic corrections. To ensure adiabatic dynamics, we need to fulfill the following condition
\begin{align}
   \frac{\mel{\psi_m}{\dot{H}}{\psi_n}}{E_n-E_m} \ll (E_n-E_m)=\omega_{nm}.
\end{align}
or, in other words, we need
\begin{equation}
   \tilde{a}(\tau) = \frac{\bra{\psi_m(\tau)}\partial_\tau H(\tau)\ket{\psi_n(\tau)}}{|E_n(\tau) - E_m(\tau)|^2}\ll 1.
\end{equation}
Here, $\tilde{a}(\tau)$ is a rescaled adiabaticity $\tilde{a}(\tau) =t_\text{f}\,a(\tau t_\text{f})$~\cite{martinez-garaotFastQuasiadiabaticDynamics2015,fernandez-fernandezQuantumControlHole2022}, which is a function of the rescaled time $\tau=t/t_\text{f}$. Hence, it differs from the hyper-adiabaticities above, as those are constant throughout the time evolution. During the pulse duration, we need to minimize the second term in adiabatic perturbation theory. We are interested in the fidelity of the adiabatic transfer at the end of the protocol $t=t_\text{f}$.
In the adiabatic limit (See Appendix~\ref{app: adiabaticity}), we can approximate the coefficients as
\begin{equation}
    c_n(t_\text{f})\sim i\left[e^{it_\text{f}\Phi_{nm}}a(t_f)-a(0)\right],
\end{equation}
with $\Phi_{nm} \equiv W_{nm}/t_\text{f}=\int_0^1\omega_{nm}(\tau)d\tau$.
Based on the above equation, the fidelity is given by
\begin{align}
    \mathcal{F}&= 1-\sum_{n\neq m}|c_n(t_\text{f})|^2\\
    &\approx1-4\,a(0)^2\sum_{n\neq m}\sin\left(\frac{t_\text{f}\Phi_{nm}}{2}\right)^2,
\end{align}
where we have used that $a(0) = a(t_\text{f})$ as we assume the adiabaticity to remain constant during the time-evolution.
In the adiabatic limit, the fidelity is therefore bounded by
\begin{equation}
    \mathcal{F}\geq 1 -4\frac{\tilde{a}(0)^2}{t_\text{f}^2}(N-1),
    \label{eq:lower_bound_limit}
\end{equation}
where $N$ is the total number of states.
In the Landau-Zener problem, we can obtain the adiabaticity at $\tau=0$ as
\begin{align}
\begin{split}
    \tilde{a}(\tau=0) = -\frac{1}{2}x^{1-n_+}&z_0\sqrt{(x^2+z_0^2)^{n_+-3}}\\
   & \times \,{_2F_1}\left(\frac{1}{2}, \frac{n_+}{2},\frac{3}{2},-\frac{z_0^2}{x^2}\right),
\end{split}
\end{align}
with $z_0 = z(\tau=0)$ being the boundary condition. Furthermore, we have assumed a symmetric boundary condition such that $z(\tau=1) = -z(\tau=0)$. The adiabatic regime is reached when $a(t) \ll 1$.
In particular, we find that the bound given in Eq.~(\ref{eq:lower_bound_limit}) is a good approximation for all the cases studied here for pulse times $t_\text{f} \gtrsim \max[\tilde{a}(\tau)] / 0.01\equiv t_\mathrm{adiab.}$, as seen in Fig.~\ref{fig:resonances}.
In Appendix~\ref{app: adiabaticity} (Fig.~\ref{fig:adiabaticity}), we show the dependency of $t_\mathrm{adiab.}$ as a function of $n_+$. As expected, the minimum is obtained for the FAQUAD ($n_+=3$), as it purposefully minimizes the adiabaticity.

Furthermore, in the adiabatic limit, the fidelity has a series of resonances, which happen at pulse times
\begin{align}
    t_\text{f}^*=\frac{2\pi k}{\Phi_{nm}}=2\pi k \left(\int_0^1 d\tau\;\omega_{nm}(\tau) \right)^{-1},
    \label{eq:t_tilde}\end{align}
for $k\in \mathbb{N}$.
Here, the explicit pulse shape enters into the function $\omega_{nm}\equiv \omega_{nm}(\alpha,\beta)$.
Using these insights, we can characterize the infidelity in the adiabatic limit as seen in Fig.~\ref{fig:resonances}.
\begin{figure}[tb!]
    \centering
    \smartincludegraphics{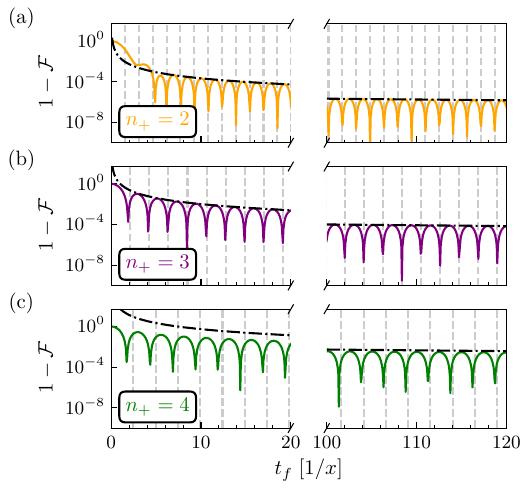}
    \caption{Infidelity for adiabatic transfer in a two-level system, using a pulse shape with $n_+=2$ (a) plotted in orange, $n_+=3$ (b) purple, and $n_+=4$ (c) green.
    The vertical gray dashed lines represent the predicted positions for the minima given by Eq.~\eqref{eq:t_tilde}, and the black dot-dashed line is the upper bound limit analytically predicted by Eq.~\eqref{eq:lower_bound_limit}.
    A broken x-axis is used to visualize both short- and long-time regimes, remarking the different (non-)adiabatic regimes.
    }
    \label{fig:resonances}
\end{figure}
We find excellent agreement for the resonant pulse times and the infidelity bound in the adiabatic limit for any $n_+$. 
Taken together, we can fully predict the adiabatic dynamics for any $(\alpha,\beta)$-hypergeometric pulse shape without the need to simulate them. 
\subsection{Quasistatic noise analysis and filter functions}
To complete our analysis, we aim to study the effects of noise on the parameters in the Hamiltonian. We firstly include the noise via an additional Hamiltonian $\delta\hat{H}=\tilde{z} \,\sigma_z+\tilde{x}\,\sigma_x,$ where both fluctuating parameters are drawn from a Gaussian distribution $\mathcal{N}(0,\delta_{x(z)})$ with zero mean and variance equals to $\delta_{x(z)}$ for the $x$- and $z$-direction, respectively. This is the so-called quasistatic approximation that well describes low-frequency noise.
We simulate quasistatic noise via Monte Carlo samples and plot the minimum infidelity in Fig.~\ref{fig:quasistatic_noise}. In Appendix~\ref{app: hypergeo fluctuations}, we study the fluctuations analytically and provide explicit constraints for the values of $(\alpha,\beta)$, such that first-order fluctuations are suppressed. We want to highlight two features of our numerical simulations. Firstly, the horizontal lines of roughly constant infidelity (as seen in Fig.~\ref{fig:quasistatic_noise}~(a)) are due to the resonances that were previously studied in Fig.~\ref{fig:resonances}. Secondly, we note that there is a common minimum infidelity for values around $n_+\approx 2$, which corresponds to the geometric fast-QUAD, which suppresses fluctuations in first order~\cite{meinersenQuantumGeometricProtocols2024}. 
\begin{figure}[tb!]
    \centering
    \smartincludegraphics{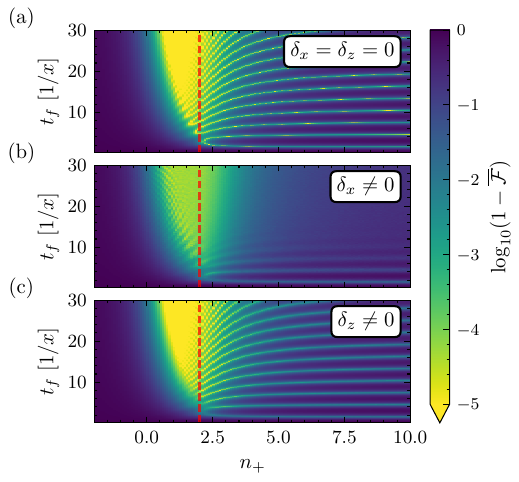}
    \caption{Average infidelity for adiabatic transfer (a) in the absence of quasi-static noise, and in the presence of quasi-static noise in the $x$-direction (b) with a noise strength $\delta_x = x / 10$, and $z$-direction (c) with $\delta_z=x / 10$.
    The vertical red dashed line denotes the geometric fast-QUAD ($n_+=2$).
    For panels (b-c), each point represents an average over 200 independent runs of the stochastic noise.
    }
    \label{fig:quasistatic_noise}
\end{figure}

Beyond the quasistatic approximation, we also compute the filter functions for each pulse. The filter function formalism~\cite{cerfontaineFilterFunctionsQuantum2021, burkardSemiconductorSpinQubits2023, greenArbitraryQuantumControl2013a, hansenAccessingFullCapabilities2023} allows us to extract the overlap of the control pulse through the filter function $F_j$ with the spectral density $S_j$ of the $\sigma_j$ of the noise. For a two-level system, the resulting fidelity, up to second order in the Magnus expansion, can be computed in the interacting frame as~\cite{cerfontaineFilterFunctionsQuantum2021}
\begin{align}
    \mathcal{F}= 1 - \frac{1}{3}\sum_{i}\int df\,S_i(f)F_i(f).
\end{align}
Here the noise Hamiltonian is $\delta \hat{H}(t)=\sum_j \delta\lambda_j(t)\sigma_j$ (with $j=x,y,z$), and the corresponding filter functions are given by
\begin{align}
    F_i(f)&=\sum_j\abs{R_{ij}(f)}^2,\\
    R_{ij}(f) &= \int_0^{t_\text{f}}dt \Tr[U_c^\dagger(t)\sigma_iU_c(t)\sigma_j]\,e^{i2\pi f t},
\end{align}
where $U_c(t)$ is the time evolution operator of the noiseless control Hamiltonian.
\begin{figure}[tb!]
    \centering
    \smartincludegraphics{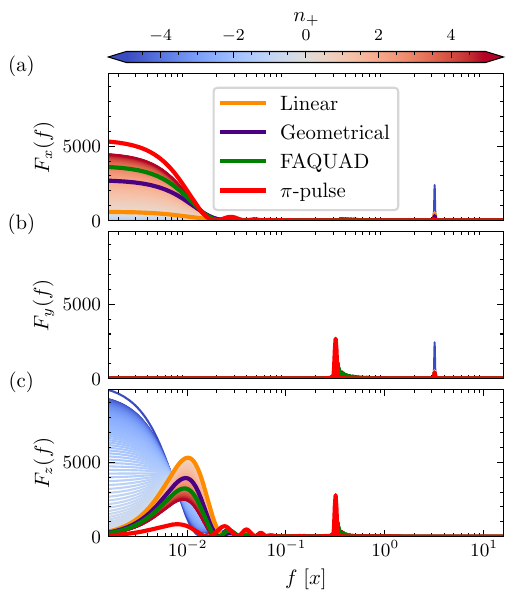}
    \caption{Filter functions $F_j$ for the Landau-Zener problem, under a stochastic noise in the $x$-, $y$-, and $z$-directions as plotted in subfigures (a, b, c), respectively.
    The total pulse time is $t_\text{f}/x = 33 \pi/2$.
    }
    \label{fig:filter_functions}
\end{figure}
\begin{figure}[htb!]
    \centering
    \smartincludegraphics{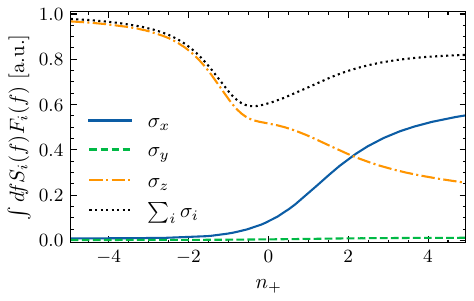}
    \caption{Noise susceptibility of the type $S_i(f)\propto1/f$, given the filter functions shown in Fig.~\ref{fig:filter_functions}, for each $i$-th component of noise symbolized by the Pauli matrix component of the noise, represented with colored lines.
    The dashed black line denotes the sum over all contributions with equal strength for all directions.
    }
    \label{fig:PSD}
\end{figure}
In Fig.~\ref{fig:filter_functions}, we plot the different filter functions $F_j$ for all Landau-Zener pulses in the same color scheme as in Fig.~\ref{fig:two_level_system}~(a).
To make a fair comparison between different protocols, we decide to use a long pulse time such that the pulses with $n_+\sim0$ reach the adiabatic regime, where the total time coincides with a resonance of the $\pi$-pulse.
Both conditions are fulfilled for a pulse time of $t_\text{f}/x = 33 \pi/2$.
The noise susceptibility for $1/f$ noise is shown in Fig.~\ref{fig:PSD}. Both figures illustrate that there is no unique pulse shape that will provide general protection against all environmental noise. Nevertheless, given certain frequency regimes, one may choose the appropriate pulse shape to provide protection in that frequency regime. 
\section{\texorpdfstring{$N$}{N}-level optimal control} \label{sec: N-level generalization}
In this section, we will see that, under certain constraints, our pulse shapes can straightforwardly be used to $N$-level systems without the need to simulate the $N$-level pulses. We will observe that the pulse shapes $\{f_n\}$ from the Landau-Zener problem, labeled by $n=n_+$, form an overcomplete basis that allows us to construct arbitrary functions in the domain where the original functions are defined. We end this section by considering multi-level uses of these pulses. In particular, we will see that when the Hamiltonian has a certain symmetry, we can directly make use of the results of the two-level system to multi-level systems of the same symmetry.

\subsection{Hyper-geodesics as overcomplete basis}
\begin{figure}[tb!]
    \centering
    \smartincludegraphics{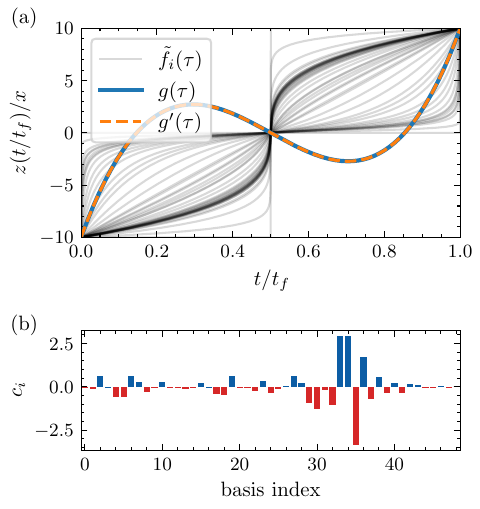}
    \caption{General pulse approximation using a linear combination of Landau-Zener pulses. (a) Gray lines denote the basis of independent functions obtained for the Landau-Zener problem.
    As a proof of concept, we use a cubic function given by $g(\tau) = -10 + 100 \tau -240 \tau^2 + 160\tau^3$ (solid blue line), which can be perfectly reproduced by a linear combination of functions in our basis (dashed orange). These types of pulse shapes have been found to be inherently resistant to low-frequency noise~\cite{barnesDynamicallyCorrectedGates2022, zhuangNoiseresistantLandauZenerSweeps2022}.
    (b) Weights for each element in the basis to reproduce $g(\tau)$ in panel (a).
    The blue and red bars represent positive and negative coefficients, respectively.
    The basis functions are sorted from negative values of $n_+$ to positive values.
    }
    \label{fig:function_approximation}
\end{figure}

\begin{figure}[tb!]
    \centering
    \smartincludegraphics{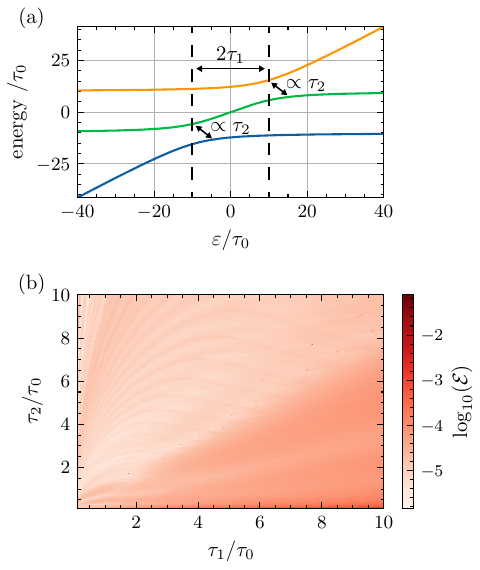}
    \caption{(a) Energy diagram of a $\Lambda$-system described by the Hamiltonian given in Eq.~\eqref{eq:lambda_system}.
    The parameter $\tau_1$ controls the position of the anticrossings, which are located at $\varepsilon= \pm \tau_1$.
    The second tunneling rate, $\tau_2$, controls the minimum gap between the first excited state (green) and the other two eigenstates (blue and orange).
    (b) Error for the reconstruction of the driving pulse for the first excited state in a $\Lambda$-system as a function of the dimensionless parameters $\tau_2/\tau_0$ and $\tau_1/\tau_0$, with $\alpha=\beta=2$, based on a linear combination of the pulses obtained for the Landau Zener problem. 
    }
    \label{fig:reproduce_three_level_system_modified}
\end{figure}

The set of pulses $\{f_i\}$ can be decomposed,  with the use of the QR decomposition and the Gram-Schmidt procedure, to construct any pulse shape $g(\tau)$ as a linear combination of the subset of orthogonal pulses $\tilde{f}_i(\tau)$
 \begin{align}
    \label{eq:linear_combination}
     g(\tau)\simeq\sum_i c_i \,\tilde{f}_i(\tau)=g'(\tau).
 \end{align}
For example, in Fig.~\ref{fig:function_approximation}, we approximate with high accuracy a third-order polynomial, which lies outside the image of the Landau-Zener pulses. 
In Appendix~\ref{app: Gramschmidt}, we provide further details on the analytical motivation and the numerical procedure. Importantly, we find that the $(\alpha,\beta)$-hypergeometries indeed unify adiabatic protocols and also describe pulses beyond the adiabatic limit by forming an overcomplete basis that allows us to construct arbitrary pulses, which fulfill the boundary conditions of the original pulses. Furthermore, we can also do a linear combination of two-level pulses for multi-level systems. For instance, take the $\Lambda$-system, which is defined by the following Hamiltonian
\begin{equation}
    \label{eq:lambda_system}
    \hat{H}_\Lambda(t) = \frac{1}{\tau_0}\begin{pmatrix}
        0 & \tau_1 & 0 \\
        \tau_1 & 0 & \tau_2 \\
        0 & \tau_2 & \varepsilon(t) \\
    \end{pmatrix},
\end{equation}
where $\tau_0$ constitutes an arbitrary overall energy scale, $\tau_1$ controls the position of the anticrossings, located at $\varepsilon= \pm \tau_1$, and $\tau_2$ controls the minimum gap between the first excited state and the other two eigenstates.
In Fig.~\ref{fig:reproduce_three_level_system_modified}~(a), we show the typical energy diagram of this system.
The pulses obtained for an adiabatic transfer can be obtained as a linear combination of the pulses obtained in the Landau-Zener problem, as shown in Fig.~\ref{fig:function_approximation}~(a).
Here, we define the total error $\mathcal{E}$ in the reconstruction of the pulses as the difference between the exact pulse shape $\varepsilon(\tau)$ obtained by solving Eq.~(\ref{eqn: generalized qgt single param}) together with the Hamiltonian of Eq.~(\ref{eq:lambda_system}), and the approximated pulse $\varepsilon'(\tau)$ using the Landau-Zener pulses as a basis
\begin{equation}
    \mathcal{E} \equiv \int_0^1 d\tau\; |\varepsilon(\tau) - \varepsilon'(\tau)|.
\end{equation}

Fig.~\ref{fig:reproduce_three_level_system_modified}~(b) shows the error as a function of the two tunneling parameters. In general, the error is small, and a good approximation can be obtained. The maximum error is obtained when the anticrossings are small, as this leads to a degenerate spectrum in the limit $\tau_2 \to 0$. For non-zero $\tau_2$, the reason for the increased error is due to the finite number of basis functions included, which are limited in the numeric approach by an error threshold in the Gram-Schmidt procedure (see Appendix~\ref{app: Gramschmidt} for details). However, the error does not heavily depend on the distance between the anticrossings, obtaining a good approximation for all the values studied here.
To improve the approximation of a general function, we can divide the basis functions into two independent ones, as
\begin{equation}
    g(\tau) \simeq \sum_n\left[c_n^{(1)}\Theta(-\tau+1/2) + c_n^{(2)}\Theta(\tau-1/2)\right]\,\tilde{f}_n(\tau),
    \label{eq:interpolation_basis}
\end{equation}
where $\Theta(\tau)$ is the Heaviside step function.
This modification, akin to the concept of interpolation, allows us to lift the previously implemented symmetry condition for the band structure and provide even lower error estimates. Given that the pulse shapes for the Landau-Zener problem serve as a useful basis, we may use this basis set in combination with machine learning techniques to find optimal expansion coefficients, using, for instance, the (dressed) CRAB algorithm~\cite{canevaChoppedRandombasisQuantum2011, rachDressingChoppedrandombasisOptimization2015}. Additionally, this basis may be used to generate pulse shapes that are derived from Lie-algebraic methods for counter-diabatic driving~\cite{guery-odelinShortcutsAdiabaticityConcepts2019} to drive specific components of the driving Hamiltonian without the need to transform into a suitable frame.
\subsection{Symmetries and \texorpdfstring{$N$}{N}-level pulse shaping}
The fact that we can generate multi-level pulses from the fundamental two-level pulses is a consequence of the symmetry of the investigated Hamiltonian. A symmetry transformation $\mathcal{S}:L(\mathcal{H})\to L(\mathcal{H})$ is an isometry on the space of linear operators on the Hilbert space $\mathcal{H}$ that preserves probabilities. If a Hamiltonian is symmetric under some specific transformation, we find $\mathcal{S}[\hat{H}]= \hat{S}\hat{H}\hat{S}^\dagger=\hat{H}$. The hypergeometric tensor is invariant if $\mathcal{S}[\mathcal{Q}_{\mu\nu}^{(\alpha,\beta)}]=\mathcal{Q}_{\mu\nu}^{(\alpha,\beta)}$, which only relies on the symmetry of the Hamiltonian itself. The simplest example is the case of $N$ decoupled two-level systems, where the Hilbert space decomposes $\mathcal{H}=\bigoplus_j \mathcal{H}_j$, such that we have
\begin{align}
    \hat{H}=\bigoplus_j \hat{H}_j.
\end{align}
In this case, we find decoupled subsectors that only obey certain boundary conditions. The relative changes purely determine the dynamics. As such, the eigenvectors and eigenvalues are also just unions of each block
\begin{align}
    \bigcup_j \; \Big\{\ket{\psi_n}\in \mathcal{H}_j, E_n \in \text{spec}(\hat{H}_j) \Big\},
\end{align}
hence, only within each $j$-th block do we find an overlap between energy eigenstates. Therefore, we find that the quantum hypergeometric tensor is a direct sum
\begin{align}
    \mathcal{Q}^{(\alpha, \beta)}_{\mu \nu}=\bigoplus_j \mathcal{Q}^{(\alpha, \beta)}_{\mu \nu; j},
\end{align}
where $\mathcal{Q}^{(\alpha, \beta)}_{\mu \nu;j}$ represents the hypergeometric tensor for the $j$-th block. However, we want to study the effects of a coupled system and see how the 2-level result can be used for the coupled $N$-level case. Here, we will study two models: a chain of particles and the anisotropic transverse-field Ising model. Both models exhibit some symmetry, which allows for the simplification of the hypergeometry. The chain of particles is described by
\begin{align}
        \hat{H}=\sum_j t_j \Big(c_j^\dagger c_{j+1} + c_{j+1}^\dagger c_j \Big),
\end{align}
where $c_j, c_j^\dagger$ are the annihilation and creation operators for the $j$-th site, respectively. The above Hamiltonian can be written in a standard matrix representation
\begin{align}
    \hat{H}=\begin{pmatrix}
        0   & t_0 &     &        & \\
        t_0 & 0   & t_1 &        & \\
            & t_1 & 0   & t_2    & \\
            &     & t_2 & 0 & \\
            &     &     &        & \ddots\\
    \end{pmatrix}\neq \bigoplus_j t_j \; \sigma_x^{(j,j+1)},
\end{align}
where $\sigma_x^{(j,j+1)}$ is the Pauli matrix that provides transitions between the $j,j+1$ levels. We see that the states are non-trivially coupled together, which can be seen by the eigenvectors
\begin{align}
    \ket{\psi_\pm^{(j)}}=\frac{1}{\sqrt{2}}\Big( \ket{j}
    \pm\ket{j+1}\Big) \;\text{ with  }\; E_\pm^{(j)}=\pm \abs{t_j}.
\end{align}
Because the Hamiltonian hosts a conserved quantity, namely the total number operator $\hat{n}=\sum_j c_j^\dagger c_j$, we can compute the hypergeometric tensor only relating to a subsector labeled by the eigenvalue of $\hat{n}$
\begin{align}
    \mathcal{Q}_{t_jt_j}^{(\alpha, \beta)}=\frac{\abs{\mel{\psi_+^{(j)}}{\partial_{t_j} \hat{H}}{\psi_-^{(j)}}}^\beta}{(2t_j)^\alpha}=\frac{1}{(2t_j)^\alpha},
\end{align}
as the matrix elements are equal to 1. We used
\begin{align}
        \partial_{t_j}\hat{H}&=\sum_j \Big(c_j^\dagger c_{j+1} + c_{j+1}^\dagger c_j \Big)\\
        &=\sum_m \Big(\ketbra{j}{j+1} + \ketbra{j+1}{j} \Big),
\end{align}
where we chose a specific basis $\ket{j}=c_j^\dagger\ket{\text{vac}}$ ,with $\ket{\text{vac}}$ is the vacuum state, and the fact that $\braket{i}{j}=\delta_{ij}$. We see that the individual hypergeometric tensors are only dependent on a 2-level subsector, described by the number $\expval{\hat{n}}$; however, they describe the optimal control of the full space. In the case of adiabatic optimal control, we find that the conserved quantity is the total energy, as the energy fluctuations are minimized~\cite{meinersenQuantumGeometricProtocols2024}. 

Another example could be the anisotropic transverse-field Ising model, whose Hamiltonian is given by
\begin{align}
    \begin{split}
    \hat{H}/J=&-\sum_j \left(\frac{1+\gamma}{2}\right)\sigma^x_{j}\sigma^x_{j+1} \\
    &+\left(\frac{1-\gamma}{2}\right)\sigma^y_{j}\sigma^y_{j+1}+h\sigma^z_j,
\end{split}
\end{align}
where $J$ is an overall energy scaling, $\gamma$ describes the extent of anisotropy, and $h$ plays the role of a (rescaled) external magnetic field. In~\cite{zanardiGroundStateOverlap2006, zanardiInformationTheoreticDifferentialGeometry2007, alvarez-jimenezQuantumInformationMetric2017, liskaHiddenSymmetriesBianchi2021}, it was found that the ground state decomposes into a set of Bloch sphere product states, which is a result of the underlying $SU(2)$ isometry of the original Hamiltonian. Henceforth, we identify that, given an underlying symmetry of the Hamiltonian, the $(\alpha,\beta)$-hypergeometries restrict to the relevant subspace, making it scalable for pulse generation in higher-dimensional systems. 
\section{Experimental feasibility}
\label{sec: experimental feasability}
The use of optimal control strategies should ultimately help the practical manipulation of quantum information~\cite{kochQuantumOptimalControl2022}. In experimental settings, this implies the use of electronic devices that possess hardware constraints such as a finite bandwidth and slew rate for the signal output. In this section, we study realistic constraints on the control electronics for pulse generation and derive analytical expressions for the bandwidth and slew rates from the hypermetric tensor $\mathcal{G}_{\mu\nu}^{(\alpha,\beta)}$. Finally, we will discuss the computational time complexity of simulating the pulse shapes for large systems.
\subsection{Impact of signal filter}
\begin{figure}[tb!]
    \centering
    \smartincludegraphics{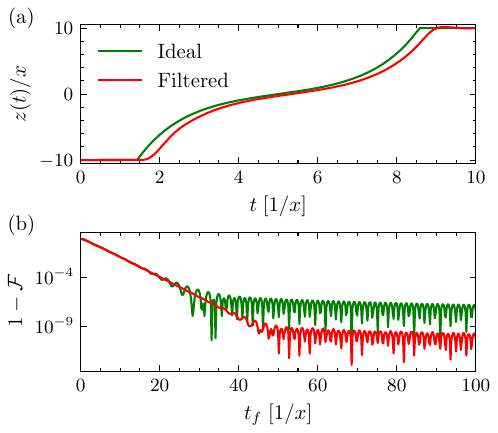}
    \caption{(a) Example of an ideal pulse (green) with $n_+=1$, and the filtered pulse (red) after a third-order Butterworth filter.
    These pulses are obtained by solving the Landau-Zener problem.
    The total pulse time is $t_\text{f}= 10 / x$, while the cutoff frequency is $f_c / x = 1$.
    (b) Infidelity of the pulse shapes shown in panel (a), sharing the same color codes.
    For this case, no dephasing has been included in the model.
    }
    \label{fig:filtered_pulse_example}
\end{figure}

\begin{figure}[tb!]
    \centering
    \smartincludegraphics{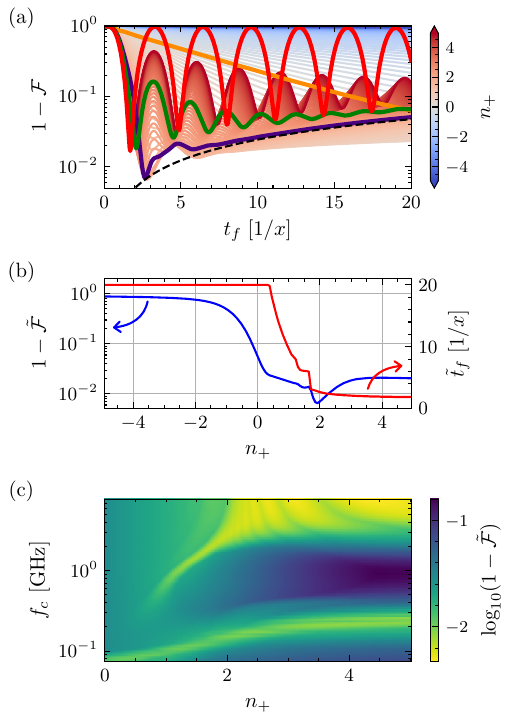}
    \caption{(a) Infidelities for the Landau-Zener problem with dephasing noise obtained with filtered pulses.
    The dashed black line corresponds to the analytical prediction for the minimum infidelity of the ideal $\pi$-pulse.
    All the pulses share the same 3rd-order Butterworth filter.
    Other parameters are the same as in Fig.~\ref{fig:two_level_system}~(e).
    (b) Minimum infidelity (left axis, blue line), and pulse time needed to reach that fidelity (right axis, red line).
    (c) Minimum infidelity for an adiabatic transfer in the Landau-Zener problem, in terms of $n_+$ and the cut-off frequency $f_c$ used in the Butterworth filter.
    Here, we have used a tunneling rate of $x = 10\;\mu$eV, and a dephasing time $T_2=65$ ns, as a representative example for the use in quantum dot qubit devices.
    }
    \label{fig:filtered_pulses_dephasing}
\end{figure}

Most control electronics filter out high-frequency components~\cite{rimbach-russSimpleFrameworkSystematic2023}. We model this behavior by adding a 3rd-order Butterworth filter. We can see how this filter removes high-frequency components by investigating a simple two-tone pulse
\begin{align}
    z(t)=A_\ell \cos(\omega_\ell t)+A_h \cos(\omega_h t).
\end{align}
Here we only have two frequency components, a low and high-frequency component with $\omega_\ell \ll \omega_h$. In more realistic settings, it will be a superposition of many frequencies. For the given pulse above, we find that the Fourier transform is a sum over Dirac delta functions
\begin{align}
    Z(\omega)&=\mathcal{F}[z(t)](\omega)\\
    &=\sum_{j=\ell,h} A_j \Big(\delta(\omega-\omega_j)+\delta(\omega+\omega_j)\Big).
\end{align}
Using the Butterworth filter $f_B(\omega)$ we find the filtered pulse
\begin{align}
    \tilde{z}(t)=\mathcal{F}^{-1}[Z(\omega)f_B(\omega)]=\sum_{j=\ell,h}\tilde{z}_j(t),
\end{align}
where the ratio of low- and high-frequency magnitudes is given by
\begin{align}
    \frac{\abs{\tilde{z}_h}}{\abs{\tilde{z}_\ell}}=\frac{\abs{A_h}\abs{f_B(\omega_h)}}{\abs{A_\ell}\abs{f_B(\omega_\ell)}}=\frac{\abs{A_h}}{\abs{A_\ell}}\cdot \sqrt{\frac{1+(\omega_\ell/\omega_c)^6}{1+(\omega_h/\omega_c)^{6}}},
\end{align}
for a 3rd-order Butterworth filter with cut-off frequency $\omega_c=2\pi f_c$.
Given that $\omega_\ell\ll \omega_h$ we find that $\abs{\tilde{z}_h}/\abs{\tilde{z}_\ell}\to 0$, and hence the high-frequency components are filtered out. As an example, we can simulate this behavior for a hypergeometric pulse in the Landau-Zener model as seen in Fig.~\ref{fig:filtered_pulse_example}~(a) for a pulse with $n_+=1$. In Fig.~\ref{fig:filtered_pulse_example}~(b), we plot the corresponding infidelity in the absence of dephasing. By reducing high-frequency contributions of the pulse, the infidelity is smoother, and lower error values are obtained. However, those values are obtained at larger total times, denoting a larger time needed to reach the adiabatic regime.
However, if we analyze the fidelities of the filtered hypergeometric pulses under finite dephasing (Fig.~\ref{fig:filtered_pulses_dephasing}), we find a similar behavior to that in Fig.~\ref{fig:two_level_system}~(e). Interestingly, we find in Fig.~\ref{fig:filtered_pulses_dephasing}~(b) that the minimum infidelity is reached for pulse shapes close to the geometric fast-QUAD with $n_+\approx 2$.
The dependence on the cutoff frequency is shown in Fig.~\ref{fig:filtered_pulses_dephasing}~(c). We can extract two trends. For $f_c\to\infty$, we find that higher $n_+$ yields better fidelity, which is exemplified in the formal limit of $n_+\to\infty$, where the $\pi$-pulse yields the highest fidelity if we allow for arbitrarily high-frequency components. In the opposite limit, when $f_c\to 0$, we expect the linear pulse to be most effective, as it contains no high-frequency components with the slew rate given directly by the slope of the linear pulse.
In Appendix~\ref{app: filter fidelity analysis}, we study the robustness of this feature and find that this is a robust feature as long as one keeps the filter cutoff frequency at a fixed value (see Fig.~\ref{fig:optimal_point_filtered_vs_nu} in Appendix~\ref{app: filter fidelity analysis}).
\subsection{Connection to slew rate and bandwidth} 
Experimental equipment relies on precise control electronics to provide the optimal pulse sequences $\lambda(t)$ derived from theory. An important subset of variables is~\cite{rimbach-russSimpleFrameworkSystematic2023}:
\begin{align}
        \text{Slew rate} &= \max_{t\in [0,t_\text{f}]} \left(\dv{V}{t}\right),\\
        \text{Bandwidth} &= f_\text{max}-f_\text{min}\Big|_{\abs{\mathcal{F}[\lambda(t)](f, t)}\geq \mathcal{F}[\lambda]_\text{max}/\sqrt{2}},
\end{align}
where $V$ is the corresponding voltage associated with the pulse, $f_\text{max, min}$ are the maximum and minimum frequencies at a 3 dB bandwidth of the pulse in the (short-time) Fourier space given by
\begin{align}
    \mathcal{F}[\lambda(t)](f, t)=\int_{-\infty}^\infty dt'\, \lambda(t')\,w(t, t')\,e^{-2\pi i f t'},
\end{align}
where $w(t, t')$ is the window function, in our case we use a Hamming window function.

\begin{figure}[tb!]
    \centering
    \smartincludegraphics{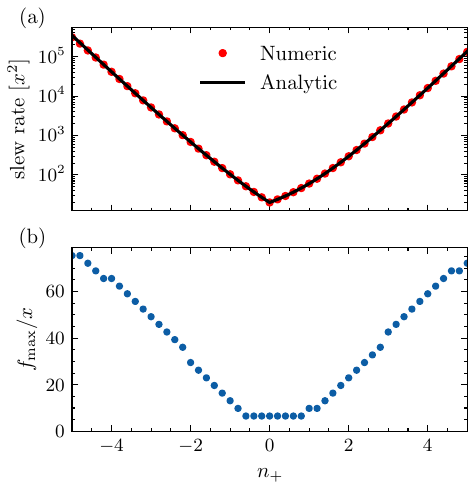}
    \caption{(a) Slew rate for the pulses obtained from the Landau-Zener problem.
    The numerical solution is plotted with red dots, while the analytical solution, given by Eq.~\eqref{eq:skew_rate_analytical}, is represented with a black line.
    (b) Maximum frequency at which the short-time Fourier transform has an intensity larger than $-20$dB.
    To compute the short-time Fourier transform we have used a Hamming window function.
    The plateau located at $n_+\sim 0$ is due to numerical constraints.
    }
    \label{fig:experimental_parameters_pulse}
\end{figure}

As such, we want to provide an explicit set of formulas to identify beforehand whether the control electronics manage to provide these pulse shapes. Given the protocol to derive the hypergeometric geodesics for a given parameter $\lambda(t)$ controlled by the voltage $V$, we find that the slew rate is given by
\begin{align}
    \text{Slew rate} &= \max_{t\in [0,t_\text{f}]} \left(\dv{V}{t}\right)\propto\max_{t\in [0,t_\text{f}]} \left(\frac{\delta^{(\alpha,\beta)}}{\sqrt{\mathcal{G}_{\lambda \lambda}^{(\alpha,\beta)}}}\right),
\end{align}
where the proportionality is given by the conversion between voltage units and energy units. For semiconductor spins, this is known as the lever arm~\cite{michielisSiliconSpinQubits2023, wangAutomatedCharacterizationDouble2023, xueQuantumLogicSpin2022, vonhorstigElectricalReadoutSpins2024}. For the symmetric two-level system, we find explicitly
\begin{align}
    \text{Slew rate} &\propto \frac{2^{\alpha/2}\delta^{(\alpha,\beta)}}{x^{\beta/2}}\max_{t\in [0,t_\text{f}]} \left((x^2+z(t)^2)^{n_+/2}\right).
    \label{eq:skew_rate_analytical}
\end{align}
A comparison between the numerical results and the analytical solution for the slew rate is given in Fig.~\ref{fig:experimental_parameters_pulse}~(a). Here, we obtain a perfect agreement between both results. 
For the bandwidth, we need to compute the short-time Fourier transform for the pulse shape. As the analytic expression of the pulse itself is the inverse of hypergeometric functions, we will make use of the derivative property of the Fourier transform
\begin{equation}
    \mathcal{F}\left[\dv{\lambda}{t}\right](f, t)=2\pi f i \,\mathcal{F}[\lambda(t)](f, t). 
\end{equation}
As such, we can find the Fourier transform by
\begin{equation}
    \mathcal{F}[\lambda(t)](f, t)=\frac{-i}{2\pi f}\,\mathcal{F}\left[\frac{\delta^{(\alpha,\beta)}}{\sqrt{\mathcal{G}_{\lambda \lambda}^{(\alpha,\beta)}}}\right](f, t).
\end{equation}
In Fig.~\ref{fig:experimental_parameters_pulse}~(b), we plot the maximum frequency for the pulse solutions, obtained numerically.
Interestingly, both the slew rate (in logarithmic scale) and the maximum frequency share a similar dependence on $n_+$.
\subsection{Time complexity of pulse simulation}
\begin{figure*}[tb!]
    \smartincludegraphics{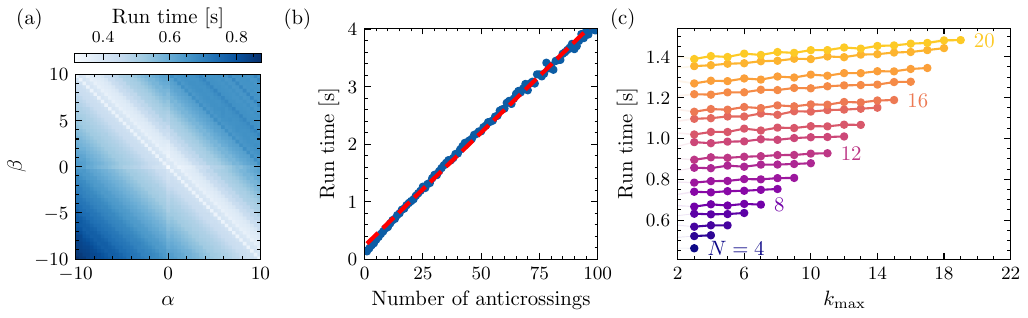}
    \caption{(a) Run time for computing the solution of the driving protocol for the Landau-Zener problem, versus the parameters $\alpha$ and $\beta$ in seconds $[\mathrm{s}]$. The run time includes the numerical diagonalization of the time-dependent Hamiltonian, the calculation of $\mathcal{G}$, and the solution of the ODE. The parameters are $z(t=0) = -z(t=t_\text{f}) = -10\,x$.
    (b) Run time for the generation of the pulse for the (periodic) Landau-Zener problem, with $\alpha=\beta=2$, as a function of the number of anticrossings present during the driving. The red dashed line denotes the fit to a linear function, obtaining a good approximation of the numerical results for $\text{Run time} = 0.23 + 0.04\,(\text{\#anticrossings})$ in seconds.
    (c) Run time to obtain the driving pulse for an adiabatic protocol following an adiabatic state in a Hamiltonian of $N$ states, with all-to-all coupling, given by Eq.~\eqref{eq:all_to_all_Hamiltonian}. Different results for different numbers of states are color-coded, with the value of $N$ written at the rightmost side of some lines for reference. The diagonalization of the obtained matrix with a total of $k_{\max}$ states, around the adiabatic state of interest. The run time includes the diagonalization of the time-dependent Hamiltonian, the calculation of $\mathcal{G}$, and the solution of the ODE. The parameters are $\Delta=5x$, $z(t=0)=-z(t=t_\text{f}) = -N(x + \Delta / 3)$, and $\alpha=\beta=2$.
    In all panels, to mitigate fluctuations in the CPU time, each point corresponds to the average of 20 runs.}
    \label{fig: run time complexity analysis}
\end{figure*}

In any practical quantum computing application, we would need to be able to generate these pulses fast on classical hardware. The simulations require two steps: the calculation of the hypergeometric tensor (see Eq.~\ref{eqn: generalized qgt single param}) and solving the first-order differential equation for the pulse (see Eq.~\ref{eq: hypergeo protocol}). Hence, we identify four possible bottlenecks in the simulation of our pulse shapes: 
\begin{enumerate}
    \item the values of $(\alpha,\beta)$,
    \item the number of anticrossings,
    \item the Hilbert space dimension, 
    \item and the number of excited states included.
\end{enumerate}
We study the run time to complete the simulation of our hypergeometric pulse shapes by using the Landau-Zener model (1), a periodic Landau-Zener model (2), and an all-to-all coupling model (3, 4), respectively, to study the above-mentioned bottlenecks (1-4). The periodic Landau-Zener and all-to-all coupling model are given by
\begin{align}
    \hat{H}_{pLZ}&=\cos z(t) \,\sigma_z + x \,\sigma_x,\\
    \hat{H}_N &= \sum_{k=1}^N \left[(-1)^kz(t)
    +k\Delta\right]+ x\sum_{k,k'}^N(c_k^\dagger c_{k'} + c_{k'}^\dagger c_k),
    \label{eq:all_to_all_Hamiltonian}
\end{align}
respectively. The parameters $x,\Delta$ control the size and the position of the anticrossings, $z(t)$ is our control field, and $c_k,c_k^\dagger$ are the annihilation and creation operators, respectively. The results of the numerical study are illustrated in Fig.~\ref{fig: run time complexity analysis}.
Important to note is that the time complexity as a function of $(\alpha,\beta)$ also only depends on $n_+$. In addition, we find that both the number of anticrossings and the number of excited states included, which serve as proxies for the dimension of the Hilbert space, scale linearly with a small slope coefficient, making it very applicable for local control of circuit-based quantum computers, where most qubit interactions are local. Furthermore, we may make use of efficient diagonalization methods like the Arnoldi package (ARPACK), the Lanczos algorithm, Locally Optimal Block Preconditions Conjugate Gradient (LOBPCG), and Filtered Eigensolver Algorithm with Spectral Transformation (FEAST)~\cite{polizziFEASTEigenvalueSolver2020}, when the relevant eigen-subspace for adiabatic transfer is sparse or separated sufficiently in energy.  Finally, note that, as the biggest constraint is solving an ordinary differential equation, we may also resort to analog hardware to compute the differential equation on the fly with, for instance, field-programmable gate arrays (FPGA's)~\cite{bartelGenerationLogicDesigns2023}. This would reduce the classical overhead and provide an even faster pulse generation.
\section{Application: Optimized shuttling with valley splitting}
\label{sec: application shuttling}
Using the $(\alpha,\beta)$-hypergeometries, we can optimize the adiabatic state-transfer protocol in any quantum system. Quantum optimal control is of special interest in emergent quantum technologies, including quantum computing.  For quantum computing platforms based on semiconductor spins~\cite{burkardSemiconductorSpinQubits2023}, which are expected to offer potential advantages such as scalability~\cite{scappucciGermaniumQuantumInformation2021, burkardSemiconductorSpinQubits2023}, long coherence times~\cite{stanoReviewPerformanceMetrics2022}, high-temperature functionality~\cite{undsethHotterEasierUnexpected2023, huangHighfidelitySpinQubit2024}, flexibility in qubit encodings and operability~\cite{burkardSemiconductorSpinQubits2023, russThreeelectronSpinQubits2017,  boscoExchangeOnlySpinOrbitQubits2024, rimbach-russSpinlessSpinQubit2024, foulkSingletonlyAlwaysonGapless2025, nguyenSinglestepHighfidelityThreequbit2025}, and fabrication processes closely aligned with those of the classical semiconductor industry~\cite{lanzaYieldVariabilityReliability2020, zwerverQubitsMadeAdvanced2022, cifuentesBoundsElectronSpin2024}. Spin shuttling, which is the coherent state transfer across a qubit array, has become an integral part of providing a clear path to scalable qubit arrays in semiconductor platforms~\cite{desmetHighfidelitySinglespinShuttling2024, vanriggelen-doelmanCoherentSpinQubit2024, siegelEarlyFaultTolerance2024a, fernandez-fernandezFlyingSpinQubits2024}, but also for other device proposals like trapped-ion quantum processors~\cite{mosesRaceTrackTrappedIonQuantum2023}. Still, there exist many bottlenecks regarding undesired excitations~\cite{ivakhnenkoNonadiabaticLandauZener2023, limaSuperadiabaticLandauZenerTransitions2024, krzywdaDecoherenceElectronSpin2024}, heating~\cite{undsethHotterEasierUnexpected2023}, and crosstalk~\cite{undsethNonlinearResponseCrosstalk2023} with their corresponding attempts to optimize shuttling fidelities~\cite{banFastRobustSpin2012,krzywdaDecoherenceElectronSpin2024, losertStrategiesEnhancingSpinShuttling2024, fehseGeneralizedFastQuasiadiabatic2023, nemethOmnidirectionalShuttlingAvoid2024, limaSuperadiabaticLandauZenerTransitions2024, thayilTheoryValleySplitting2024}. Given the flexibility of our method and the importance of reducing the above-mentioned bottlenecks, we aim to analyze the hypergeometries for the use of spin shuttling in silicon, where close-by valley states exist. Nevertheless, we stress that the framework of $(\alpha,\beta)$-hypergeometries applies to any quantum system with a non-degenerate spectrum. The Hamiltonian describing the spin shuttlings, with Hilbert space $\mathcal{H}=\mathcal{H}_\text{orbital}\otimes \mathcal{H}_\text{valley}$, is given by~\cite{nemethOmnidirectionalShuttlingAvoid2024, thayilTheoryValleySplitting2024}
\begin{align}
\begin{split}
    \hat{H}_\text{s}=&\frac{\varepsilon(t)}{2}\,\hat{\tau}_z \otimes \mathbb{1}_\text{valley}+t_c\,\hat{\tau}_x\otimes \mathbb{1}_\text{valley}\\
    &+\sum_{j=L,R}\hat{P}_j \otimes \Big(\Re \Delta_j \,\hat{\gamma}_x-\Im \Delta_j \,\hat{\gamma}_y\Big).
\end{split}
\end{align}
Here $t_c$ is the orbital tunnel coupling, $\Delta_j$ are the complex-valued inter-valley couplings, $\varepsilon(t)$ is the control parameter, the $\hat{P}_j=\ketbra{j}{j}$ are projectors onto the left or right orbitals, and $\hat{\tau}_j, \hat{\gamma}_j$ are the Pauli matrices associated with the orbital and valley degree of freedom, respectively. In real materials, the inter-valley couplings are well described by a random variable drawn from a Gaussian distribution~\cite{limaSuperadiabaticLandauZenerTransitions2024, thayilTheoryValleySplitting2024,nemethOmnidirectionalShuttlingAvoid2024}. Here, the spin degree of freedom is effectively decoupled unless we add a spin-orbit coupling term~\cite{krzywdaDecoherenceElectronSpin2024}. We can diagonalize the Hamiltonian with respect to the valley states using $U_\text{valley}=\sum_j \hat{P}_j U_j$ with $U_j=(\hat{\gamma}_0+i\hat{\gamma}_y \cos \phi_j+i\hat{\gamma}_x \sin \phi_j)/\sqrt{2}$ and $\phi_j= \arg \Delta_j$ is the valley phase. In that basis, we find the matrix representation of the Hamiltonian to be~\cite{nemethOmnidirectionalShuttlingAvoid2024}
    \begin{align}
\label{eqn: shuttling Ham}
    \hat{H}_\text{s}'&=U_\text{valley}\hat{H}_\text{s} U_\text{valley}^\dagger\\
    &=\left(\begin{smallmatrix}
        \varepsilon/2 + |\Delta_L| & 0 & t_{ee} & t_{eg} \\
        0 & \varepsilon/2-|\Delta_L| & t_{ge} & t_{gg} \\
        t_{ee}^* & t_{ge}^* & -(\varepsilon/2) + |\Delta_R| & 0 \\
        t_{eg}^* & t_{gg}^* & 0 & -(\varepsilon/2) - |\Delta_R| 
    \end{smallmatrix}\right),
\end{align}
where the tunnel coupling is now affected by the valley phase
\begin{align}
    t_{ee}&=t_{gg}^*=\frac{t_c}{2}\left(1+e^{i(\phi_L-\phi_R)}\right),\\
    t_{eg}&=-t_{ge}^*=\frac{t_c}{2}\left(e^{i\phi_L}-e^{i\phi_R}\right).
\end{align}
\begin{figure*}[tb!]
    \centering
    \smartincludegraphics{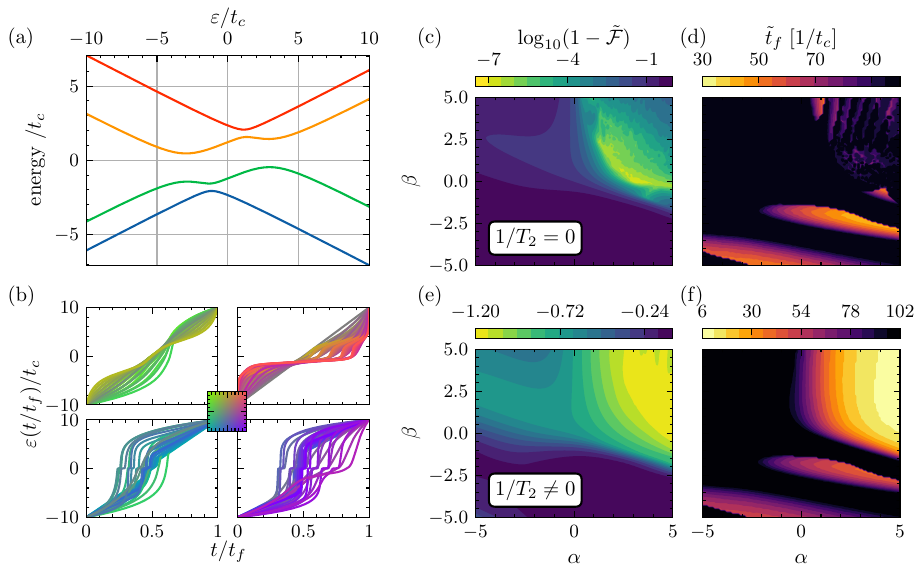}
    \caption{(a) Band structure of the shuttling Hamiltonian in Eq.~\eqref{eqn: shuttling Ham}. (b) Pulse shape solutions for the adiabatic driving of the first excited state. Here, we include all energy levels in the sum of the hypergeometric tensor. Each panel represents a different quadrant in the $(\alpha,\beta)$ space, where $\alpha,\beta \in [-5,5]$ and each quadrant is separated by 5 units, i.e. $[-5,0]\cup [0,5]$ for both $\alpha,\beta$. Different colored lines denote these quadrants of $\alpha$ and $\beta$, as given in the middle panel. We can also reconstruct these pulse shapes using our basis expansion (See Appendix~\ref{app: four level approx}). (c, d) Minimum infidelity and optimal pulse time for coherent evolution. (e, f) Non-unitary evolution including dephasing similar to Fig.~\ref{fig:two_level_system}~(e). The dephasing time is given by $T_2 = 100/t_c$. The parameters used for the energy landscape (a), pulse shapes (b), and corresponding time evolutions (c-f) are:  $\phi_L$ = 0, $\phi_R = 0.8\pi$, $\Delta_L/t_c = 1$, and $\Delta_R/t_c=2$.
    }
    \label{fig: shuttling figure}
\end{figure*}
The shuttling Hamiltonian exhibits four energy states, where the lower two encode the qubit state as seen in Fig.~\ref{fig: shuttling figure}~(a). To fully study the effectiveness of our hypergeometric framework, we will proceed to study the transfer fidelity of the excited qubit state (first excited level above ground state) as it encounters a total of three anticrossings. Similar to the two-level case, we perform a sweep over the parameters $(\alpha,\beta)$ and study the coherent and non-unitary evolution. Given that the inter-valley couplings $\Delta_j$ are drawn from normal distributions, we may attempt to study the averaged Hamiltonian. However, the hypergeometries reduce to two decoupled two-level systems (see Appendix~\ref{app: average shuttling}) and hence do not capture the intricacies of the non-averaged model. Therefore, we aim at studying a single instance of the non-averaged model and simulating the resulting pulse shapes and their time evolution. The pulse shapes generated for the Hamiltonian~\eqref{eqn: shuttling Ham} are shown in Fig.~\ref{fig: shuttling figure}~(b), where the coloring refers to different sectors in the $(\alpha,\beta)$ plane with $\alpha,\beta \in [-5,5]$. In Fig.~\ref{fig: shuttling figure}~(c, e) we plot the minimum infidelity $1-\tilde{\mathcal{F}}\equiv1-\mathcal{F}(\tilde{t}_\text{f})$ and in Fig.~\ref{fig: shuttling figure}~(d, f)  the corresponding optimal pulse time $\tilde{t}_\text{f}$ as a function of $(\alpha,\beta)$ for the coherent and non-unitary case, respectively. The non-unitary evolution is given by the Lindblad master equation with the jump operator $\mathcal{D}[\rho]=(1/T_2)(\text{diag}(\rho)-\rho)$. First, note that the best fidelities are reached for $\alpha,\beta>0$. In the coherent case, to minimize the excitations of higher energy valley states, the more adiabatic protocols will reach higher fidelities. We find that this does not necessarily imply large pulse times as seen in Fig.~\ref{fig: shuttling figure}~(d). For the non-unitary case (Fig.~\ref{fig: shuttling figure}~(e, f)), due to global dephasing affecting the entire evolution, one finds that short pulse times ($t_\text{f}\ll T_2$) are optimal to circumvent dephasing. Yet, given the advantages of flexible pulse shaping, we can still circumvent diabatic transitions while reducing dephasing effects. Given the theorem of equivalence of Lagrangians for the quantum brachistochrone~\cite{santosQuantumAdiabaticBrachistochrone2021}, we find that reducing the effects of pure dephasing noise can only be circumvented by faster dynamics (with respect to the dephasing time), making our framework the ideal, flexible method.
\section{Conclusion}
In this work, we have presented the framework of $(\alpha,\beta)$-hypergeometries, which unifies and generalizes existing adiabatic protocols such as the FAQUAD~\cite{martinez-garaotFastQuasiadiabaticDynamics2015, fehseGeneralizedFastQuasiadiabatic2023, fernandez-fernandezQuantumControlHole2022}, and geometric fast-QUAD~\cite{meinersenQuantumGeometricProtocols2024}. Using the Landau-Zener model as a pedagogical example, providing a proxy for many physical systems, we provided exact analytic expressions of important quantities and proved their validity through various numerical simulations. Beyond the Landau-Zener model, we found explicit infidelity resonance frequencies and the infidelity upper bound values in the adiabatic limit without the need for full-scale simulations. The hypergeometric protocols allowed for the generation of control pulses of the existing control Hamiltonian without the need for additional control fields as generically required in counter-diabatic driving strategies~\cite{berryTransitionlessQuantumDriving2009, selsMinimizingIrreversibleLosses2017, chenLewisRiesenfeldInvariantsTransitionless2011} or Magnus-expansion-based methods~\cite{ribeiroSystematicMagnusBasedApproach2017, figueiredoroqueEngineeringFastHighfidelity2021}. The $(\alpha,\beta)$-hypergeometries also only depend on the Hamiltonian, i.e., the generator of time translations, making it very easy to use and apply to any quantum system. The inherent complexity of the $(\alpha,\beta)$-hypergeometries is also mostly linked to the number of parameters and the difficulty in solving a first-order ordinary differential equation. By introducing the two parameters $(\alpha,\beta)$, we allow for more flexibility as we can predict, and hence adjust, the adiabaticity, pulse shape smoothness, and the behavior of the fidelity in the adiabatic limit for more robust optimal control. Being the minimum of the length action, we also find that the $(\alpha,\beta)$-hypergeometries are robust against quasistatic noise for certain combinations of $(\alpha,\beta)$~\cite{meinersenQuantumGeometricProtocols2024}. For fast dynamics and dense energy spectra, reducing diabatic transitions while reducing dephasing effects is the main challenge, which can be dealt with efficiently using the unifying framework of noise-resistant $(\alpha,\beta)$-hypergeometries. This allowed us to study the optimal control of spin shuttling in the case of valley disorder. Our protocol enabled us to coherently shuttle the first excited state through multiple anticrossings without losing coherence. Finally, we would like to mention possible avenues of the $(\alpha,\beta)$-hypergeometries beyond optimal control tasks. The standard quantum geometric tensor was shown to exhibit properties of many-body systems, ground state manifold classification, or as sensitive probes for quantum chaos, by investigation of the symmetry group~\cite{liskaHiddenSymmetriesBianchi2021}, curvature invariants~\cite{kolodrubetzClassifyingMeasuringGeometry2013, pandeyAdiabaticEigenstateDeformations2020a, sharipovHilbertSpaceGeometry2024}, and topological invariants~\cite{lambertClassicalQuantumInformation2023}. In addition, in some machine learning tasks, the quantum geometric tensor is used for the optimization of the weights of a neural quantum state~\cite{vicentiniNetKetMachineLearning2022, dawidModernApplicationsMachine2023}. Exploring how the $(\alpha,\beta)$-hypergeometries influence these domains could provide, beyond the use of quantum optimal control, deeper insights into quantum many-body physics, quantum chaos, and machine learning, paving the way for novel scientific and technological advancements. 
\section*{Acknowledgments}
G.P. and D.F.F. are supported by the Spanish Ministry of Science through the grant: PID2023-149072NB-I00 and by the CSIC Research Platform PTI-001. D.F.F. acknowledges support from FPU Program No. FPU20/04762. M.R.-R. and C.V.M. acknowledge that the EU partly supported this research through the H2024 QLSI2 project and was partly sponsored by the Army Research Office under Award Number: W911NF-23-1-0110. The views and conclusions contained in this document are those of the authors and should not be interpreted as representing the official policies, either expressed or implied, of the Army Research Office or the U.S. Government. The U.S. Government is authorized to reproduce and distribute reprints for Government purposes, notwithstanding any copyright notation herein.

We thank Stefano Bosco and Sander de Snoo, and all members of the Rimbach-Russ, Bosco, and Vandersypen group for providing valuable insights.

\section*{Data Availability}
Simulation software and data analysis scripts supporting this work are available at \url{https://doi.org/10.5281/zenodo.15173706}.

\newpage
\appendix

\section{Embedding coordinates for hyper-Bloch sphere}
\label{app: embedding hypergeo}
To plot the hyper-Bloch sphere we make the ansatz of
\begin{align}
    x(\theta,\phi)&=f(\theta)\cos \phi\\
    y(\theta,\phi)&=f(\theta)\sin \phi\\
    z(\theta)&=g(\theta),
\end{align}
where by matching the line elements
\begin{align}
    dx^2+dy^2+dz^2&=(f'^2+g'^2)d\theta^2+f^2d\phi^2\\
    &\equiv\frac{1}{2^\alpha}\Big(d\theta^2+\sin^\beta \theta \,d\phi^2\Big)
\end{align}
we find that
\begin{align}
    f(\theta)^2&=\frac{\sin^{\beta} \theta}{2^{\alpha}}\\
    g'(\theta)^2&=\frac{1}{2^\alpha}\left(1-\frac{\beta^2\sin^\beta \theta}{4 \tan^2 \theta}\right),
\end{align}
where the embedding is only well-defined if $1-\beta^2\sin^\beta \theta/4 \tan^2 \theta>0$.

\section{Hypergeometric fluctuations}
\label{app: hypergeo fluctuations}
The hypergeometric tensor can be interpreted as the generalized quantum geometric tensor, which is the second-order expansion of the state fidelity of the parameter changes and is hence robust against first-order parameter fluctuations. Given this insight, we aim to study third-order corrections to the state fidelity through first-order corrections in the hypergeometric tensor. Here, we will study how it changes when we perturb the system fully, i.e., up to first order, we have
\begin{align}
    \hat{H}'&=\hat{H}+\delta x^\sigma \,\partial_\sigma\hat{H}\\
    &\equiv \hat{H}+\delta x^\sigma \,\hat{V}_\sigma,\\
    E_n'&= E_n + \delta x^\sigma \mel{\psi_n}{\hat{V}_\sigma}{\psi_n}\\
    &=E_n+\delta x^\sigma\, (V_\sigma)_{nn},\\
    \ket{\psi_n'}&=\ket{\psi_n}+\delta x^\sigma \sum_{m\neq n}\frac{\mel{\psi_m}{\hat{V}_\sigma}{\psi_n}}{E_n-E_m}\ket{\psi_m}\\
    &\equiv\ket{\psi_n}+\delta x^\sigma \sum_{m\neq n}\frac{(V_\sigma)_{mn}}{\omega_{nm}}\ket{\psi_m},
\end{align}
where we recognize that the error gauge potential $\hat{A}$ gives the generator of translations/fluctuations of the states in parameter space
\begin{align}
    (A_\sigma)_{mn}=\mel{\psi_m}{\hat{A}_\sigma}{\psi_n}=\frac{(V_\sigma)_{mn}}{\omega_{nm}}=\frac{\mel{\psi_m}{\partial_\sigma \hat{H}}{\psi_n}}{E_n-E_m}.
\end{align}
\subsubsection*{General \texorpdfstring{$N$}{N}-level system}
The energy splitting changes as follows
\begin{align}
    E'_n-E'_m&=(E_n-E_m)+\delta x^\sigma \Big[(V_\sigma)_{nn}-(V_\sigma)_{mm}\Big]\\
    &=\omega_{nm}\Big[1 + \left(\frac{(V_\sigma)_{nn}-(V_\sigma)_{mm}}{\omega_{nm}}\right) \, \delta x^\sigma \Big]\\
    &\equiv\omega_{nm}\Big[1 + h_{nm,\sigma} \, \delta x^\sigma \Big].
\end{align}
The overlap matrix element changes, up to first order, as
\begin{widetext}
    \begin{align}
    \mel{\psi_n'}{\partial_\mu \hat{H}}{\psi_m'}&\approx \left(\bra{\psi_n}+\delta x^\sigma \sum_{k\neq n}\frac{(V_\sigma)_{kn}}{\omega_{nk}}\bra{\psi_k}\right)\left(\partial_\mu\hat{H}+\delta x^\sigma \,\partial_\mu\hat{V}_\sigma\right)\left(\ket{\psi_m}+\delta x^\sigma \sum_{\ell\neq m}\frac{(V_\sigma)_{\ell m}}{\omega_{m\ell}}\ket{\psi_\ell}\right)\\
    &\approx \mel{\psi_n}{\partial_\mu \hat{H}}{\psi_m}+\delta x^\sigma \left[\sum_{\ell\neq m}\frac{(V_\sigma)_{\ell m}(V_\mu)_{n \ell}}{\omega_{m\ell}}+\mel{\psi_n}{\partial_\mu \hat{V}_\sigma}{\psi_m}+\sum_{k\neq n}\frac{(V_\sigma)_{kn}(V_\mu)_{km}}{\omega_{nk}}\right]\\
    &\equiv\mel{\psi_n}{\partial_\mu \hat{H}}{\psi_m}\Big(1+g_{nm,\mu\sigma}\,\delta x^\sigma\Big).
\end{align}
\end{widetext}
We can now compute the perturbed hypergeometric tensor for the $m^\text{th}$ excited state
\begin{widetext}
\begin{align}
    ^{(m)}\mathcal{G'}_{\mu \nu}^{(\alpha, \beta)}&=\Re \sum_{n\neq m} \frac{\mel{\psi_m'}{\partial_\mu \hat{H}'}{\psi_n'}^{\beta/2}\mel{\psi_n'}{\partial_\nu \hat{H}'}{\psi_m'}^{\beta/2}}{(E_n'-E_m')^{\alpha}}\\
    &= \Re \sum_{n\neq m} \frac{\mel{\psi_m}{\partial_\mu \hat{H}}{\psi_n}^{\beta/2}\mel{\psi_n}{\partial_\nu \hat{H}}{\psi_m}^{\beta/2}}{(E_n-E_m)^{\alpha}}\frac{(1+g_{mn,\mu\sigma}\,\delta x^\sigma )^{\beta/2}(1+g_{nm,\nu\sigma}\,\delta x^\sigma )^{\beta/2}}{(1 + h_{nm,\sigma} \, \delta x^\sigma )^\alpha},
\end{align}
\end{widetext}
where we can expand the second fraction up to linear order in the parameter fluctuations due to quasistatic noise
\begin{widetext}
    \begin{align}
    \frac{(1+g_{mn,\mu\sigma}\,\delta x^\sigma )^{\beta/2}(1+g_{nm,\nu\sigma}\,\delta x^\sigma )^{\beta/2}}{(1 + h_{nm,\sigma} \, \delta x^\sigma )^\alpha}&\approx \Big(1+(\beta/2)g_{mn,\mu\sigma}\,\delta x^\sigma \Big)\Big(1+(\beta/2)g_{nm,\nu\sigma}\,\delta x^\sigma \Big)\Big(1 -\alpha \, h_{nm,\sigma} \, \delta x^\sigma \Big)\\
    &\approx 1+ \delta x^\sigma \Big[\beta/2\Big(g_{mn,\mu\sigma}+g_{nm,\nu\sigma}\Big)-\alpha \, h_{nm,\sigma}\Big].
\end{align}
\end{widetext}

\subsubsection*{Two level system}

For the hypergeometric tensor of the rescaled Landau-Zener system $\hat{H}/x=\tan\theta(t)\,\sigma_z+\sigma_x$, we need to find the changes in the overlap and energy splitting. For the energy splitting, we find
\begin{align}
    E'_+-E'_-&=(E_+-E_-)+\delta \theta (V_{++}-V_{--})+\mathcal{O}(\delta\theta^2)\\
    &=\omega_{+-}\Big[1 + h(\theta) \, \delta \theta \Big]+\mathcal{O}(\delta\theta^2).
\end{align}
For the overlap matrix element, we find
\begin{align}
    \mel{\psi_-'}{\partial_\theta \hat{H}'}{\psi_+'}=\mel{\psi_-}{\partial_\theta \hat{H}}{\psi_+}\Big[1+g(\theta)\,\delta \theta \Big]+\mathcal{O}(\delta\theta^2),
\end{align}
where we define
\begin{align}
\begin{split}
    g(\theta)=&\frac{1}{\mel{\psi_-}{\partial_\theta \hat{H}}{\psi_+}}\Bigg[\frac{V_{-+}}{\omega_{+-}}\mel{\psi_+}{\partial_\theta \hat{H}}{\psi_+}\\
    &+\mel{\psi_-}{\partial_\theta \hat{V}}{\psi_+}+\frac{V_{+-}}{\omega_{-+}}\mel{\psi_-}{\partial_\theta \hat{H}}{\psi_-}\Bigg].
\end{split}
\end{align}
As $1+g(\theta)\,\delta \theta > 0$, because the correction is small, then we find
\begin{align}
    \mathcal{G}'^{(\alpha, \beta)}
    &\approx \frac{|\mel{\psi_-}{\partial_\theta \hat{H}}{\psi_+}|^{\beta}}{(E_+-E_-)^{\alpha}} \Big(1-\Big[\alpha\,h(\theta)-\beta g(\theta)\Big] \delta \theta\Big)\\
    &\equiv\mathcal{G}^{(\alpha, \beta)}+\delta \mathcal{G}^{(\alpha, \beta)}.
\end{align}
The term responsible for the quasistatic noise drops out to linear order if the following constraint is fulfilled
\begin{align}
    \mathcal{C}^{(\alpha, \beta)}[\theta]=\alpha\,h(\theta)-\beta g(\theta)=0.
\end{align}
For symmetric two-level systems, we find that $h(\theta)=g(\theta)$, resulting in noise-robust pulses for all pulse shapes with $\alpha=\beta$.

\section{Single-qubit geometric and topological invariants}
\label{app: geometric invariants}
In the main text, we focused on the applications of the hypergeometric tensor to optimal control tasks. However, we may also want to investigate the applications to ground state classification or as a probe of quantum chaos. Therefore, here, we provide a list of geometric and topological invariants to aid in understanding the above-mentioned tasks. Firstly, we may compute the Chern-like number for the single-qubit as~\cite{lambertClassicalQuantumInformation2023}
\begin{align}
    c^{(\alpha,\beta)}=\frac{(-i)^{\beta/2}}{2^\alpha}\left[\sqrt{\pi}\frac{\Gamma(\frac{2+\beta}{4})}{\Gamma(1+\frac{\beta}{4})}\right],
\end{align}
which satisfies the inequality $\text{Vol}^{(\alpha,\beta)}\geq \pi \,\abs{c^{(\alpha,\beta)}}$, similar to the original quantum geometric tensor~\cite{ozawaRelationsTopologyQuantum2021, lambertClassicalQuantumInformation2023}. Note, however, that as opposed to the standard Chern number, this variant is not restricted to be integer-valued. For other topological properties, like the Euler characteristic, we may compute the Ricci scalar $R$ and Kretschmann invariant $K$, which for the single-qubit results in
\begin{align}
    R&=g^{\mu\nu} R_{\mu\nu} = g^{\mu\nu} g^{\rho\sigma} R_{\mu\rho\nu\sigma}\\
    &=\frac{2^{\alpha-2}\beta}{\sin^2 \theta}\Big(8-\beta\,[1+\cos(2\theta)]\Big)\\
    K&=R_{\mu\nu\rho\sigma}R^{\mu\nu\rho\sigma}=R^2,
\end{align}
where $R_{\mu\rho\nu\sigma}$ is the Riemann tensor and $g_{\mu\nu}\equiv\mathcal{G}_{\mu\nu}^{(\alpha,\beta)}$ for notational convenience. Interestingly, for arbitrary $(\alpha,\beta)\neq (2,2)$, we find that the Ricci scalar and Kretschmann invariant are both dependent on the angle $\theta$, which is not the case for the standard Bloch sphere $(\alpha,\beta)=(2,2)$, where both are constant. Notably, for $\beta>2$, we find that the two-dimensional Euler characteristic $\chi=2(1-g)$, which is a function of the genus $g$, results in
\begin{align}
    \chi=\int d\theta \,d\phi\, \sqrt{g} \,R = 0,
\end{align}
leading to the conclusion that a general hyper-Bloch sphere is a genus-1 surface significantly different from the genus-0 surface of the standard Bloch sphere. 
\section{Adiabaticity}
\label{app: adiabaticity}
\begin{figure}[htb!]
    \centering
    \smartincludegraphics{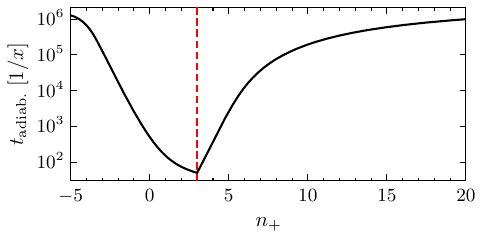}
    \caption{Pulse time threshold to reach the adiabatic regime.
    A red dashed line marks the FAQUAD protocol, at $n_+=3$, which by design, tries to minimize the time to achieve adiabaticity.
    }
    \label{fig:adiabaticity}
\end{figure}

In the main text, especially in Section~\ref{sec: single qubit optimal control}, we discussed explicit formulas that, in the adiabatic limit, provide the behavior of the resonance frequencies and the upper bound on the infidelity. To derive the expressions we note that at first order in an adiabatic expansion, the coefficients of Eq.~(\ref{eq:time_evolution}) read
\begin{equation}
\begin{split}
    c_n(t) =& -\int_0^t\braket{\psi_n(t')}{\dot{\psi}_m(t')}e^{iW_{nm}(t')}dt'\\
           =&-\int_0^ta(t')\omega_{nm}(t')e^{iW_{nm}(t')}dt'\\
           =& i\left[e^{iW_{nm}(t)}a(t)-a(0)\right]\\
           &+ i\int_0^t\left[e^{iW_{nm}(t')}-1\right]\dot{a}(t')dt',
\end{split}
\end{equation}
which is a result of the Feynman-Hellman theorem. As we assume that the adiabaticity remains constant for our protocol, we can neglect the fast-oscillating second term, yielding the result in the main text. The adiabatic limit is reached when the physical adiabaticity parameter $a(t)\ll 1$. The pulse time at which this occurs is the adiabatic time $t_\text{adiab.}$, which we plot in Fig.~\ref{fig:adiabaticity}. 
\section{Details on Gram-Schmidt procedure}
\label{app: Gramschmidt}
\begin{figure}[tb!]
    \centering
    \smartincludegraphics{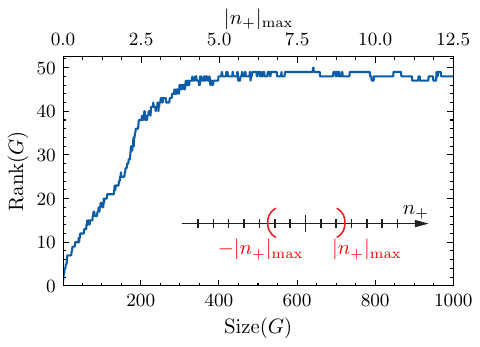}
    \caption{Rank of the Gram matrix as a function of its size (bottom axis). The number of initial elements in the basis is computed by solving the Landau-Zener problem for $n_+$ between $-|n_+|_\text{max}$ and $|n_+|_\text{max}$ (upper axis), as shown schematically in the bottom of the figure.
    }
    \label{fig:rank_vs_size_modified}
\end{figure}

To get a better intuition for the set of pulses, we can rescale the Landau-Zener Hamiltonian $\hat{H}/x=\tan\theta(t)\,\sigma_z+\sigma_x$, which leads to a simpler hypermetric tensor
\begin{align}
    \mathcal{G}_{\theta\theta}^{(\alpha, \beta)}=\frac{1}{2^\alpha}\abs{\cos \theta}^{\alpha-\beta}
\end{align}
with the corresponding hypergeometric protocol
\begin{align}
    \dv{\theta}{\tau}=\frac{2^{\alpha/2}\delta^{(\alpha,\beta)}}{\abs{\cos \theta}^{n_-}},
\end{align}
where $n_-=(\alpha-\beta)/2$ and the hyper-adiabaticity for the state transfer $\ket{0} \leftrightarrow \ket{1}$ is given by 
\begin{align}
    \delta^{(\alpha,\beta)} = \frac{1}{2^{\alpha/2}}\int_0^\pi d\theta\,\abs{\cos \theta}^{n_-} =\frac{\sqrt{\pi}}{2^{\alpha/2}} \frac{\Gamma(\frac{1+n_-}{2})}{\Gamma(1+\frac{n_-}{2})},
\end{align}
for $\Re\,n_->-1$. Opposite to the normal Landau-Zener model here we find an explicit dependence on $n_-$. We can now write an Ansatz for the basis expansion as follows. The tangents of the hyper-geodesics  $\dot{\theta}_n\propto\abs{\cos \theta}^n=\abs{\cos^n \theta}$ in the interval $\theta \in [0,\pi]$, which allows us to expand any tangent vector $v(\theta)$, if $n\in 2\mathbb{N}$, as
\begin{align}
    v(\theta)=\sum_{n\in2 \mathbb{N}} c_n \cos^n \theta.
\end{align}
 Given that the basis functions are made up of a polynomial sequence $\{1,z^2,z^4,\dots\}$ for $z^2=\cos^2 \theta$, these functions then also form a basis in the space of symmetric functions. Note, however, that they do not form an orthogonal subset, but an overcomplete set in the space of symmetric functions. Given a set of monomials $\{1,z,z^2,\dots\}$ we can use the Gram-Schmidt procedure to generate a set of orthogonal functions of the tangent vectors defined by the inner product
\begin{equation}
    \innerproduct{f_i}{f_j}=\int_a^b d\mu(z) f_i(z)f_j(z),
\end{equation}
where depending on the measure $d\mu(z)$ and the interval $z\in [a,b]$ we can generate different polynomials via
\begin{equation}
    \hat{f}_k=f_k-\sum_{j=1}^{k-1}\frac{\innerproduct{f_k}{f_j}}{\innerproduct{f_j}{f_j}}f_j.
\end{equation}
For instance, we generate Legendre polynomials from a trivial measure $d\mu(z)=dz$ on the interval $z\in[-1,1]$. The Chebyshev polynomials form a basis on the same interval but with the measure $d\mu(z)=(1-z^2)^{-1/2}$. As the interval of interest for our (tangent) hyper-geodesics is $z \in [0,1]$, we might need to adjust the procedure. One possibility would be to use the odd powers of half-range Chebyshev polynomials~\cite{orelComputationsHalfrangeChebyshev2012}, which are defined on this interval and serve as an orthogonal basis. Here we opt for numerical methods, where we define the Gram matrix $G$ in terms of the inner products of the basis functions as
\begin{equation}
    G_{i, j} = \innerproduct{f_i}{f_j} = \int_0^1 f_i(\tau)f_j(\tau) d\tau,
\end{equation}
where $f_i$ are the different pulse solutions for the Landau-Zener problem. In our case, the Gram matrix is non-invertible, so the functions are not linearly independent. Since $\operatorname{rank}(G) < \operatorname{size}(G)$, we have redundant functions (see Fig.~\ref{fig:rank_vs_size_modified}). To eliminate unnecessary functions, we use a QR decomposition such that $G = Q \cdot R$, with $Q$ orthogonal and $R$ upper triangular matrices. By choosing indices where the absolute value of the diagonal of $R$ is above a given threshold ($10^{-10}$ in our case), we can reconstruct a basis of linearly independent functions. These functions are shown in Fig.~\ref{fig:function_approximation}~(a) with gray lines. This subset of functions is denoted by $\tilde{f}_i$, and the Gram matrix as $\tilde{G}$. A given function can be expanded as
\begin{equation}
    g(s) \simeq \sum_i c_i \tilde{f}_i(\tau).
\end{equation}
In the case of a non-orthogonal basis, the coefficients are computed by solving the linear system
\begin{equation}
    \tilde{G} \vec{c} = \vec{b},
\end{equation}
where $\vec{c}$ is the vector of coefficients, and $b_i=\langle g|\tilde{f}_i\rangle$.

\section{Filter-fidelity analysis}
\label{app: filter fidelity analysis}
\begin{figure}[tb!]
    \centering
    \smartincludegraphics{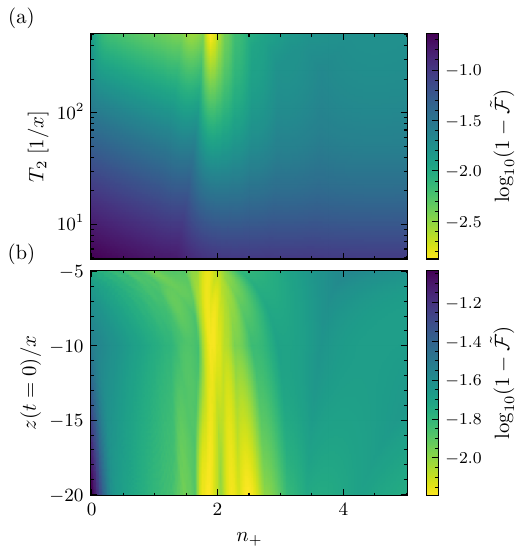}
    \caption{(Right) Minimum infidelity in terms of the pure dephasing time (a), and the boundary condition for the driving pulse (b), with $z(t_f) = -z(t=0)$.
    In (a) we have set $z(t=0) / x = 10$, while in (b) $T_2 = 100/x$.
    In both cases, the cut-off frequency is $f_c / x = 1$.}
    \label{fig:optimal_point_filtered_vs_nu}
\end{figure}

In the main text, we analyze the infidelity as a response to adding a filter to the ideal pulses, obtaining a non-trivial dependence of the optimal $n_+$ with the cut-off frequency. Here, we provide a more in-depth look at that feature. In Fig.~\ref{fig:optimal_point_filtered_vs_nu}, we plot the infidelity as a function of $n_+$ (x-axis) and the noise strength in Fig.~\ref{fig:optimal_point_filtered_vs_nu}~(a), and the boundary condition for the driving parameter $z(t=0)$ in Fig.~\ref{fig:optimal_point_filtered_vs_nu}~(b). We find that for a cutoff frequency set to $f_c/x=1$, the ideal pulse is the one for which $n_+\approx 2$, independently of the dephasing strength and the boundary condition.
\section{Averaged shuttling Hamiltonian}
\label{app: average shuttling}
The averaged Hamiltonian takes the form
\begin{align}
    \expval{\hat{H}_\text{s}'}&=\prod_{j=L,R}\,\int d^2\Delta_j \, p(\Delta_j) \, \hat{H}_\text{s}'(\Delta_j)\\
    &= \prod_{j=L,R}\,\int_0^{2\pi} d\phi_j \int_0^\infty  d|\Delta_j|\,|\Delta_j| \frac{e^{-|\Delta_j|^2/2\sigma^2}}{2\pi \sigma^2}\, \hat{H}_\text{s}'(\Delta_j),
\end{align}
where we assume the variance of the probability distributions $p(\Re\Delta_j), p(\Im\Delta_j)$ to be the same. Then we can change them into polar coordinates. The averaged Hamiltonian requires the computation of the two following integrals
\begin{align}
    \expval{|\Delta_j|}&=\frac{1}{\sigma^2}\int_0^\infty d|\Delta_j|\, |\Delta_j|^2\,  e^{-\frac{|\Delta_j|^2}{2\sigma^2}}=\sqrt{\frac{\pi \sigma^2}{2}}, \\
    \expval{e^{i\phi_j}}&=\frac{1}{2\pi}\int_0^{2\pi} d\phi_j\, e^{i\phi_j}  = 0.
\end{align}
This yields an effective average Hamiltonian
\begin{align}
\begin{split}
    \expval{\hat{H}_\text{s}'}=&\frac{\varepsilon}{2}\,\hat{\tau}_z \otimes \mathbb{1}_\text{valley}+\frac{t_c}{2}\,\hat{\tau}_x\otimes \mathbb{1}_\text{valley}\\
    &+ \sqrt{\frac{\pi \sigma^2}{2}} \,\mathbb{1}_\text{orbital} \otimes \gamma_z.
\end{split}
\end{align}
The Hamiltonian can be split into two subsectors and hence, the eigenvalues of this matrix can be written as
\begin{align}
    E_\pm &= \pm \frac{1}{2}\sqrt{\varepsilon^2+t_c^2}\\
    E_{\pm,\sigma} &= E_\pm + \sqrt{2\pi \sigma^2},
\end{align}
which are the energies of two $2\times2$ systems. At zero detuning, there are different levels that overlap, namely at values $\sigma^2=0$ and $\sigma^2=t_c^2/2\pi$. The eigenvectors are independent of $\sigma$ and are the same in the different subsectors. By rescaling the Hamiltonian by $t_c/2$ with $\tilde{\varepsilon}(t)=\varepsilon/t_c$ and $\tilde{\sigma}=\sigma/t_c$ one finds that due to the splitting into decoupled two-level subsystems that the hypergeometric tensor with respect to the ground takes the same form as for the simple two-level system
\begin{align}
    \mathcal{G}^{(\alpha,\beta)}_{\tilde{\varepsilon}\tilde{\varepsilon}}=\frac{1}{2^\alpha}\Big(1+\tilde{\varepsilon}(t)^2\Big)^{n_+}.
\end{align}
\section{Four-level pulse approximation}
\label{app: four level approx}
\begin{figure}[htb!]
    \centering
    \smartincludegraphics{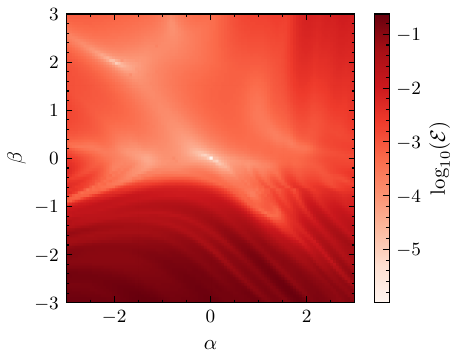}
    \caption{Approximation of the four-level pulses for the shuttling Hamiltonian in Eq.~\eqref{eqn: shuttling Ham} using the Landau-Zener pulses as a basis. The same parameters are used as in Fig.~\ref{fig: shuttling figure}.}
    \label{fig: four level pulse approx}
\end{figure}

Similar to Section~\ref{sec: N-level generalization}, where we studied the overcomplete basis of pulses to reconstruct the pulse shapes from the $\Lambda$-system, we verify this reconstruction method for the pulse shapes of the shuttling Hamiltonian in Eq.~\eqref{eqn: shuttling Ham}. We find in Fig.~\ref{fig: four level pulse approx} that the pulses which have $\beta>0$ allow for the best approximation yielding the lowest values of the total error $\mathcal{E}$. Importantly, these pulses also provide the highest fidelity protocols with and without noise.

\bibliography{references}

\begin{thebibliography}{117}%
\makeatletter
\providecommand \@ifxundefined [1]{%
 \@ifx{#1\undefined}
}%
\providecommand \@ifnum [1]{%
 \ifnum #1\expandafter \@firstoftwo
 \else \expandafter \@secondoftwo
 \fi
}%
\providecommand \@ifx [1]{%
 \ifx #1\expandafter \@firstoftwo
 \else \expandafter \@secondoftwo
 \fi
}%
\providecommand \natexlab [1]{#1}%
\providecommand \enquote  [1]{``#1''}%
\providecommand \bibnamefont  [1]{#1}%
\providecommand \bibfnamefont [1]{#1}%
\providecommand \citenamefont [1]{#1}%
\providecommand \href@noop [0]{\@secondoftwo}%
\providecommand \href [0]{\begingroup \@sanitize@url \@href}%
\providecommand \@href[1]{\@@startlink{#1}\@@href}%
\providecommand \@@href[1]{\endgroup#1\@@endlink}%
\providecommand \@sanitize@url [0]{\catcode `\\12\catcode `\$12\catcode `\&12\catcode `\#12\catcode `\^12\catcode `\_12\catcode `\%12\relax}%
\providecommand \@@startlink[1]{}%
\providecommand \@@endlink[0]{}%
\providecommand \url  [0]{\begingroup\@sanitize@url \@url }%
\providecommand \@url [1]{\endgroup\@href {#1}{\urlprefix }}%
\providecommand \urlprefix  [0]{URL }%
\providecommand \Eprint [0]{\href }%
\providecommand \doibase [0]{https://doi.org/}%
\providecommand \selectlanguage [0]{\@gobble}%
\providecommand \bibinfo  [0]{\@secondoftwo}%
\providecommand \bibfield  [0]{\@secondoftwo}%
\providecommand \translation [1]{[#1]}%
\providecommand \BibitemOpen [0]{}%
\providecommand \bibitemStop [0]{}%
\providecommand \bibitemNoStop [0]{.\EOS\space}%
\providecommand \EOS [0]{\spacefactor3000\relax}%
\providecommand \BibitemShut  [1]{\csname bibitem#1\endcsname}%
\let\auto@bib@innerbib\@empty
\bibitem [{\citenamefont {Glaser}\ \emph {et~al.}(2015)\citenamefont {Glaser}, \citenamefont {Boscain}, \citenamefont {Calarco}, \citenamefont {Koch}, \citenamefont {K{\"o}ckenberger}, \citenamefont {Kosloff}, \citenamefont {Kuprov}, \citenamefont {Luy}, \citenamefont {Schirmer}, \citenamefont {{Schulte-Herbr{\"u}ggen}}, \citenamefont {Sugny},\ and\ \citenamefont {Wilhelm}}]{glaserTrainingSchrodingerCat2015}%
  \BibitemOpen
  \bibfield  {author} {\bibinfo {author} {\bibfnamefont {S.~J.}\ \bibnamefont {Glaser}}, \bibinfo {author} {\bibfnamefont {U.}~\bibnamefont {Boscain}}, \bibinfo {author} {\bibfnamefont {T.}~\bibnamefont {Calarco}}, \bibinfo {author} {\bibfnamefont {C.~P.}\ \bibnamefont {Koch}}, \bibinfo {author} {\bibfnamefont {W.}~\bibnamefont {K{\"o}ckenberger}}, \bibinfo {author} {\bibfnamefont {R.}~\bibnamefont {Kosloff}}, \bibinfo {author} {\bibfnamefont {I.}~\bibnamefont {Kuprov}}, \bibinfo {author} {\bibfnamefont {B.}~\bibnamefont {Luy}}, \bibinfo {author} {\bibfnamefont {S.}~\bibnamefont {Schirmer}}, \bibinfo {author} {\bibfnamefont {T.}~\bibnamefont {{Schulte-Herbr{\"u}ggen}}}, \bibinfo {author} {\bibfnamefont {D.}~\bibnamefont {Sugny}},\ and\ \bibinfo {author} {\bibfnamefont {F.~K.}\ \bibnamefont {Wilhelm}},\ }\bibfield  {title} {\bibinfo {title} {Training {{Schr{\"o}dinger}}'s cat: Quantum optimal control},\ }\href {https://doi.org/10.1140/epjd/e2015-60464-1} {\bibfield  {journal} {\bibinfo  {journal} {The European
  Physical Journal D}\ }\textbf {\bibinfo {volume} {69}},\ \bibinfo {pages} {279} (\bibinfo {year} {2015})}\BibitemShut {NoStop}%
\bibitem [{\citenamefont {Ivakhnenko}\ \emph {et~al.}(2023)\citenamefont {Ivakhnenko}, \citenamefont {Shevchenko},\ and\ \citenamefont {Nori}}]{ivakhnenkoNonadiabaticLandauZener2023}%
  \BibitemOpen
  \bibfield  {author} {\bibinfo {author} {\bibfnamefont {O.~V.}\ \bibnamefont {Ivakhnenko}}, \bibinfo {author} {\bibfnamefont {S.~N.}\ \bibnamefont {Shevchenko}},\ and\ \bibinfo {author} {\bibfnamefont {F.}~\bibnamefont {Nori}},\ }\bibfield  {title} {\bibinfo {title} {Nonadiabatic {{Landau}}--{{Zener}}--{{St{\"u}ckelberg}}--{{Majorana}} transitions, dynamics, and interference},\ }\href {https://doi.org/10.1016/j.physrep.2022.10.002} {\bibfield  {journal} {\bibinfo  {journal} {Physics Reports}\ }\bibinfo {series} {Nonadiabatic {{Landau-Zener-St{\"u}ckelberg-Majorana}} Transitions, Dynamics, and Interference},\ \textbf {\bibinfo {volume} {995}},\ \bibinfo {pages} {1} (\bibinfo {year} {2023})}\BibitemShut {NoStop}%
\bibitem [{\citenamefont {Izmalkov}\ \emph {et~al.}(2004)\citenamefont {Izmalkov}, \citenamefont {Grajcar}, \citenamefont {Il'ichev}, \citenamefont {Oukhanski}, \citenamefont {Wagner}, \citenamefont {Meyer}, \citenamefont {Krech}, \citenamefont {Amin}, \citenamefont {van~den Brink},\ and\ \citenamefont {Zagoskin}}]{izmalkovObservationMacroscopicLandauZener2004}%
  \BibitemOpen
  \bibfield  {author} {\bibinfo {author} {\bibfnamefont {A.}~\bibnamefont {Izmalkov}}, \bibinfo {author} {\bibfnamefont {M.}~\bibnamefont {Grajcar}}, \bibinfo {author} {\bibfnamefont {E.}~\bibnamefont {Il'ichev}}, \bibinfo {author} {\bibfnamefont {N.}~\bibnamefont {Oukhanski}}, \bibinfo {author} {\bibfnamefont {T.}~\bibnamefont {Wagner}}, \bibinfo {author} {\bibfnamefont {H.-G.}\ \bibnamefont {Meyer}}, \bibinfo {author} {\bibfnamefont {W.}~\bibnamefont {Krech}}, \bibinfo {author} {\bibfnamefont {M.~H.~S.}\ \bibnamefont {Amin}}, \bibinfo {author} {\bibfnamefont {A.~M.}\ \bibnamefont {van~den Brink}},\ and\ \bibinfo {author} {\bibfnamefont {A.~M.}\ \bibnamefont {Zagoskin}},\ }\bibfield  {title} {\bibinfo {title} {Observation of macroscopic {{Landau-Zener}} transitions in a superconducting device},\ }\href {https://doi.org/10.1209/epl/i2003-10200-6} {\bibfield  {journal} {\bibinfo  {journal} {Europhysics Letters}\ }\textbf {\bibinfo {volume} {65}},\ \bibinfo {pages} {844} (\bibinfo {year} {2004})}\BibitemShut
  {NoStop}%
\bibitem [{\citenamefont {Alexander}\ \emph {et~al.}(2024)\citenamefont {Alexander}, \citenamefont {Bahgat}, \citenamefont {Benyamini}, \citenamefont {Black}, \citenamefont {Bonneau}, \citenamefont {Burgos}, \citenamefont {Burridge}, \citenamefont {Campbell}, \citenamefont {Catalano}, \citenamefont {Ceballos}, \citenamefont {Chang}, \citenamefont {Chung}, \citenamefont {Danesh}, \citenamefont {Dauer}, \citenamefont {Davis}, \citenamefont {Dudley}, \citenamefont {{Er-Xuan}}, \citenamefont {Fargas}, \citenamefont {Farsi}, \citenamefont {Fenrich}, \citenamefont {Frazer}, \citenamefont {Fukami}, \citenamefont {Ganesan}, \citenamefont {Gibson}, \citenamefont {{Gimeno-Segovia}}, \citenamefont {Goeldi}, \citenamefont {Goley}, \citenamefont {Haislmaier}, \citenamefont {Halimi}, \citenamefont {Hansen}, \citenamefont {Hardy}, \citenamefont {Horng}, \citenamefont {House}, \citenamefont {Hu}, \citenamefont {Jadidi}, \citenamefont {Johansson}, \citenamefont {Jones}, \citenamefont {Kamineni}, \citenamefont {Kelez},
  \citenamefont {Koustuban}, \citenamefont {Kovall}, \citenamefont {Krogen}, \citenamefont {Kumar}, \citenamefont {Liang}, \citenamefont {LiCausi}, \citenamefont {Llewellyn}, \citenamefont {Lokovic}, \citenamefont {Lovelady}, \citenamefont {Manfrinato}, \citenamefont {Melnichuk}, \citenamefont {Souza}, \citenamefont {Mendoza}, \citenamefont {Moores}, \citenamefont {Mukherjee}, \citenamefont {Munns}, \citenamefont {Musalem}, \citenamefont {Najafi}, \citenamefont {O'Brien}, \citenamefont {Ortmann}, \citenamefont {Pai}, \citenamefont {Park}, \citenamefont {Peng}, \citenamefont {Penthorn}, \citenamefont {Peterson}, \citenamefont {Poush}, \citenamefont {Pryde}, \citenamefont {Ramprasad}, \citenamefont {Ray}, \citenamefont {Rodriguez}, \citenamefont {Roxworthy}, \citenamefont {Rudolph}, \citenamefont {Saunders}, \citenamefont {Shadbolt}, \citenamefont {Shah}, \citenamefont {Shin}, \citenamefont {Smith}, \citenamefont {Sohn}, \citenamefont {Sohn}, \citenamefont {Son}, \citenamefont {Sparrow}, \citenamefont
  {Staffaroni}, \citenamefont {Stavrakas}, \citenamefont {Sukumaran}, \citenamefont {Tamborini}, \citenamefont {Thompson}, \citenamefont {Tran}, \citenamefont {Triplet}, \citenamefont {Tung}, \citenamefont {Vert}, \citenamefont {Vidrighin}, \citenamefont {Vorobeichik}, \citenamefont {Weigel}, \citenamefont {Wingert}, \citenamefont {Wooding},\ and\ \citenamefont {Zhou}}]{alexanderManufacturablePlatformPhotonic2024}%
  \BibitemOpen
  \bibfield  {author} {\bibinfo {author} {\bibfnamefont {K.}~\bibnamefont {Alexander}}, \bibinfo {author} {\bibfnamefont {A.}~\bibnamefont {Bahgat}}, \bibinfo {author} {\bibfnamefont {A.}~\bibnamefont {Benyamini}}, \bibinfo {author} {\bibfnamefont {D.}~\bibnamefont {Black}}, \bibinfo {author} {\bibfnamefont {D.}~\bibnamefont {Bonneau}}, \bibinfo {author} {\bibfnamefont {S.}~\bibnamefont {Burgos}}, \bibinfo {author} {\bibfnamefont {B.}~\bibnamefont {Burridge}}, \bibinfo {author} {\bibfnamefont {G.}~\bibnamefont {Campbell}}, \bibinfo {author} {\bibfnamefont {G.}~\bibnamefont {Catalano}}, \bibinfo {author} {\bibfnamefont {A.}~\bibnamefont {Ceballos}}, \bibinfo {author} {\bibfnamefont {C.-M.}\ \bibnamefont {Chang}}, \bibinfo {author} {\bibfnamefont {C.~J.}\ \bibnamefont {Chung}}, \bibinfo {author} {\bibfnamefont {F.}~\bibnamefont {Danesh}}, \bibinfo {author} {\bibfnamefont {T.}~\bibnamefont {Dauer}}, \bibinfo {author} {\bibfnamefont {M.}~\bibnamefont {Davis}}, \bibinfo {author} {\bibfnamefont {E.}~\bibnamefont
  {Dudley}}, \bibinfo {author} {\bibfnamefont {P.}~\bibnamefont {{Er-Xuan}}}, \bibinfo {author} {\bibfnamefont {J.}~\bibnamefont {Fargas}}, \bibinfo {author} {\bibfnamefont {A.}~\bibnamefont {Farsi}}, \bibinfo {author} {\bibfnamefont {C.}~\bibnamefont {Fenrich}}, \bibinfo {author} {\bibfnamefont {J.}~\bibnamefont {Frazer}}, \bibinfo {author} {\bibfnamefont {M.}~\bibnamefont {Fukami}}, \bibinfo {author} {\bibfnamefont {Y.}~\bibnamefont {Ganesan}}, \bibinfo {author} {\bibfnamefont {G.}~\bibnamefont {Gibson}}, \bibinfo {author} {\bibfnamefont {M.}~\bibnamefont {{Gimeno-Segovia}}}, \bibinfo {author} {\bibfnamefont {S.}~\bibnamefont {Goeldi}}, \bibinfo {author} {\bibfnamefont {P.}~\bibnamefont {Goley}}, \bibinfo {author} {\bibfnamefont {R.}~\bibnamefont {Haislmaier}}, \bibinfo {author} {\bibfnamefont {S.}~\bibnamefont {Halimi}}, \bibinfo {author} {\bibfnamefont {P.}~\bibnamefont {Hansen}}, \bibinfo {author} {\bibfnamefont {S.}~\bibnamefont {Hardy}}, \bibinfo {author} {\bibfnamefont {J.}~\bibnamefont {Horng}},
  \bibinfo {author} {\bibfnamefont {M.}~\bibnamefont {House}}, \bibinfo {author} {\bibfnamefont {H.}~\bibnamefont {Hu}}, \bibinfo {author} {\bibfnamefont {M.}~\bibnamefont {Jadidi}}, \bibinfo {author} {\bibfnamefont {H.}~\bibnamefont {Johansson}}, \bibinfo {author} {\bibfnamefont {T.}~\bibnamefont {Jones}}, \bibinfo {author} {\bibfnamefont {V.}~\bibnamefont {Kamineni}}, \bibinfo {author} {\bibfnamefont {N.}~\bibnamefont {Kelez}}, \bibinfo {author} {\bibfnamefont {R.}~\bibnamefont {Koustuban}}, \bibinfo {author} {\bibfnamefont {G.}~\bibnamefont {Kovall}}, \bibinfo {author} {\bibfnamefont {P.}~\bibnamefont {Krogen}}, \bibinfo {author} {\bibfnamefont {N.}~\bibnamefont {Kumar}}, \bibinfo {author} {\bibfnamefont {Y.}~\bibnamefont {Liang}}, \bibinfo {author} {\bibfnamefont {N.}~\bibnamefont {LiCausi}}, \bibinfo {author} {\bibfnamefont {D.}~\bibnamefont {Llewellyn}}, \bibinfo {author} {\bibfnamefont {K.}~\bibnamefont {Lokovic}}, \bibinfo {author} {\bibfnamefont {M.}~\bibnamefont {Lovelady}}, \bibinfo {author}
  {\bibfnamefont {V.}~\bibnamefont {Manfrinato}}, \bibinfo {author} {\bibfnamefont {A.}~\bibnamefont {Melnichuk}}, \bibinfo {author} {\bibfnamefont {M.}~\bibnamefont {Souza}}, \bibinfo {author} {\bibfnamefont {G.}~\bibnamefont {Mendoza}}, \bibinfo {author} {\bibfnamefont {B.}~\bibnamefont {Moores}}, \bibinfo {author} {\bibfnamefont {S.}~\bibnamefont {Mukherjee}}, \bibinfo {author} {\bibfnamefont {J.}~\bibnamefont {Munns}}, \bibinfo {author} {\bibfnamefont {F.-X.}\ \bibnamefont {Musalem}}, \bibinfo {author} {\bibfnamefont {F.}~\bibnamefont {Najafi}}, \bibinfo {author} {\bibfnamefont {J.~L.}\ \bibnamefont {O'Brien}}, \bibinfo {author} {\bibfnamefont {J.~E.}\ \bibnamefont {Ortmann}}, \bibinfo {author} {\bibfnamefont {S.}~\bibnamefont {Pai}}, \bibinfo {author} {\bibfnamefont {B.}~\bibnamefont {Park}}, \bibinfo {author} {\bibfnamefont {H.-T.}\ \bibnamefont {Peng}}, \bibinfo {author} {\bibfnamefont {N.}~\bibnamefont {Penthorn}}, \bibinfo {author} {\bibfnamefont {B.}~\bibnamefont {Peterson}}, \bibinfo {author}
  {\bibfnamefont {M.}~\bibnamefont {Poush}}, \bibinfo {author} {\bibfnamefont {G.~J.}\ \bibnamefont {Pryde}}, \bibinfo {author} {\bibfnamefont {T.}~\bibnamefont {Ramprasad}}, \bibinfo {author} {\bibfnamefont {G.}~\bibnamefont {Ray}}, \bibinfo {author} {\bibfnamefont {A.}~\bibnamefont {Rodriguez}}, \bibinfo {author} {\bibfnamefont {B.}~\bibnamefont {Roxworthy}}, \bibinfo {author} {\bibfnamefont {T.}~\bibnamefont {Rudolph}}, \bibinfo {author} {\bibfnamefont {D.~J.}\ \bibnamefont {Saunders}}, \bibinfo {author} {\bibfnamefont {P.}~\bibnamefont {Shadbolt}}, \bibinfo {author} {\bibfnamefont {D.}~\bibnamefont {Shah}}, \bibinfo {author} {\bibfnamefont {H.}~\bibnamefont {Shin}}, \bibinfo {author} {\bibfnamefont {J.}~\bibnamefont {Smith}}, \bibinfo {author} {\bibfnamefont {B.}~\bibnamefont {Sohn}}, \bibinfo {author} {\bibfnamefont {Y.-I.}\ \bibnamefont {Sohn}}, \bibinfo {author} {\bibfnamefont {G.}~\bibnamefont {Son}}, \bibinfo {author} {\bibfnamefont {C.}~\bibnamefont {Sparrow}}, \bibinfo {author} {\bibfnamefont
  {M.}~\bibnamefont {Staffaroni}}, \bibinfo {author} {\bibfnamefont {C.}~\bibnamefont {Stavrakas}}, \bibinfo {author} {\bibfnamefont {V.}~\bibnamefont {Sukumaran}}, \bibinfo {author} {\bibfnamefont {D.}~\bibnamefont {Tamborini}}, \bibinfo {author} {\bibfnamefont {M.~G.}\ \bibnamefont {Thompson}}, \bibinfo {author} {\bibfnamefont {K.}~\bibnamefont {Tran}}, \bibinfo {author} {\bibfnamefont {M.}~\bibnamefont {Triplet}}, \bibinfo {author} {\bibfnamefont {M.}~\bibnamefont {Tung}}, \bibinfo {author} {\bibfnamefont {A.}~\bibnamefont {Vert}}, \bibinfo {author} {\bibfnamefont {M.~D.}\ \bibnamefont {Vidrighin}}, \bibinfo {author} {\bibfnamefont {I.}~\bibnamefont {Vorobeichik}}, \bibinfo {author} {\bibfnamefont {P.}~\bibnamefont {Weigel}}, \bibinfo {author} {\bibfnamefont {M.}~\bibnamefont {Wingert}}, \bibinfo {author} {\bibfnamefont {J.}~\bibnamefont {Wooding}},\ and\ \bibinfo {author} {\bibfnamefont {X.}~\bibnamefont {Zhou}},\ }\href {https://doi.org/10.48550/arXiv.2404.17570} {\bibinfo {title} {A manufacturable
  platform for photonic quantum computing}} (\bibinfo {year} {2024}),\ \Eprint {https://arxiv.org/abs/2404.17570} {2404.17570} \BibitemShut {NoStop}%
\bibitem [{\citenamefont {Higuchi}\ \emph {et~al.}(2017)\citenamefont {Higuchi}, \citenamefont {Heide}, \citenamefont {Ullmann}, \citenamefont {Weber},\ and\ \citenamefont {Hommelhoff}}]{higuchiLightfielddrivenCurrentsGraphene2017}%
  \BibitemOpen
  \bibfield  {author} {\bibinfo {author} {\bibfnamefont {T.}~\bibnamefont {Higuchi}}, \bibinfo {author} {\bibfnamefont {C.}~\bibnamefont {Heide}}, \bibinfo {author} {\bibfnamefont {K.}~\bibnamefont {Ullmann}}, \bibinfo {author} {\bibfnamefont {H.~B.}\ \bibnamefont {Weber}},\ and\ \bibinfo {author} {\bibfnamefont {P.}~\bibnamefont {Hommelhoff}},\ }\bibfield  {title} {\bibinfo {title} {Light-field-driven currents in graphene},\ }\href {https://doi.org/10.1038/nature23900} {\bibfield  {journal} {\bibinfo  {journal} {Nature}\ }\textbf {\bibinfo {volume} {550}},\ \bibinfo {pages} {224} (\bibinfo {year} {2017})}\BibitemShut {NoStop}%
\bibitem [{\citenamefont {Aghaee}\ \emph {et~al.}(2025)\citenamefont {Aghaee}, \citenamefont {Alcaraz~Ramirez}, \citenamefont {Alam}, \citenamefont {Ali}, \citenamefont {Andrzejczuk}, \citenamefont {Antipov}, \citenamefont {Astafev}, \citenamefont {Barzegar}, \citenamefont {Bauer}, \citenamefont {Becker}, \citenamefont {Bhaskar}, \citenamefont {Bocharov}, \citenamefont {Boddapati}, \citenamefont {Bohn}, \citenamefont {Bommer}, \citenamefont {Bourdet}, \citenamefont {Bousquet}, \citenamefont {Boutin}, \citenamefont {Casparis}, \citenamefont {Chapman}, \citenamefont {Chatoor}, \citenamefont {Christensen}, \citenamefont {Chua}, \citenamefont {Codd}, \citenamefont {Cole}, \citenamefont {Cooper}, \citenamefont {Corsetti}, \citenamefont {Cui}, \citenamefont {Dalpasso}, \citenamefont {Dehollain}, \citenamefont {{de Lange}}, \citenamefont {{de Moor}}, \citenamefont {Ekefj{\"a}rd}, \citenamefont {El~Dandachi}, \citenamefont {Estrada~Salda{\~n}a}, \citenamefont {Fallahi}, \citenamefont {Galletti}, \citenamefont
  {Gardner}, \citenamefont {Govender}, \citenamefont {Griggio}, \citenamefont {Grigoryan}, \citenamefont {Grijalva}, \citenamefont {Gronin}, \citenamefont {Gukelberger}, \citenamefont {Hamdast}, \citenamefont {Hamze}, \citenamefont {Hansen}, \citenamefont {Heedt}, \citenamefont {Heidarnia}, \citenamefont {Herranz~Zamorano}, \citenamefont {Ho}, \citenamefont {Holgaard}, \citenamefont {Hornibrook}, \citenamefont {Indrapiromkul}, \citenamefont {Ingerslev}, \citenamefont {Ivancevic}, \citenamefont {Jensen}, \citenamefont {Jhoja}, \citenamefont {Jones}, \citenamefont {Kalashnikov}, \citenamefont {Kallaher}, \citenamefont {Kalra}, \citenamefont {Karimi}, \citenamefont {Karzig}, \citenamefont {King}, \citenamefont {Kloster}, \citenamefont {Knapp}, \citenamefont {Kocon}, \citenamefont {Koski}, \citenamefont {Kostamo}, \citenamefont {Kumar}, \citenamefont {Laeven}, \citenamefont {Larsen}, \citenamefont {Lee}, \citenamefont {Lee}, \citenamefont {Leum}, \citenamefont {Li}, \citenamefont {Lindemann}, \citenamefont
  {Looij}, \citenamefont {Love}, \citenamefont {Lucas}, \citenamefont {Lutchyn}, \citenamefont {Madsen}, \citenamefont {Madulid}, \citenamefont {Malmros}, \citenamefont {Manfra}, \citenamefont {Mantri}, \citenamefont {Markussen}, \citenamefont {Martinez}, \citenamefont {Mattila}, \citenamefont {McNeil}, \citenamefont {Mei}, \citenamefont {Mishmash}, \citenamefont {Mohandas}, \citenamefont {Mollgaard}, \citenamefont {Morgan}, \citenamefont {Moussa}, \citenamefont {Nayak}, \citenamefont {Nielsen}, \citenamefont {Nielsen}, \citenamefont {Nielsen}, \citenamefont {Nijholt}, \citenamefont {Nystrom}, \citenamefont {O'Farrell}, \citenamefont {Ohki}, \citenamefont {Otani}, \citenamefont {Paquelet~W{\"u}tz}, \citenamefont {Pauka}, \citenamefont {Petersson}, \citenamefont {Petit}, \citenamefont {Pikulin}, \citenamefont {Prawiroatmodjo}, \citenamefont {Preiss}, \citenamefont {Puchol~Morejon}, \citenamefont {Rajpalke}, \citenamefont {Ranta}, \citenamefont {Rasmussen}, \citenamefont {Razmadze}, \citenamefont {Reentila},
  \citenamefont {Reilly}, \citenamefont {Ren}, \citenamefont {Reneris}, \citenamefont {Rouse}, \citenamefont {Sadovskyy}, \citenamefont {Sainiemi}, \citenamefont {Sanlorenzo}, \citenamefont {Schmidgall}, \citenamefont {Sfiligoj}, \citenamefont {Shah}, \citenamefont {Simoes}, \citenamefont {Singh}, \citenamefont {Sinha}, \citenamefont {Soerensen}, \citenamefont {Sohr}, \citenamefont {Stankevic}, \citenamefont {Stek}, \citenamefont {Stuppard}, \citenamefont {Suominen}, \citenamefont {Suter}, \citenamefont {Teicher}, \citenamefont {Thiyagarajah}, \citenamefont {Tholapi}, \citenamefont {Thomas}, \citenamefont {Toomey}, \citenamefont {Tracy}, \citenamefont {Turley}, \citenamefont {Upadhyay}, \citenamefont {Urban}, \citenamefont {Van~Hoogdalem}, \citenamefont {Van~Woerkom}, \citenamefont {Viazmitinov}, \citenamefont {Vogel}, \citenamefont {Watson}, \citenamefont {Webster}, \citenamefont {Weston}, \citenamefont {Winkler}, \citenamefont {Xu}, \citenamefont {Yang}, \citenamefont {Yucelen}, \citenamefont {Zeisel},
  \citenamefont {Zheng},\ and\ \citenamefont {Zilke}}]{aghaeeInterferometricSingleshotParity2025}%
  \BibitemOpen
  \bibfield  {author} {\bibinfo {author} {\bibfnamefont {M.}~\bibnamefont {Aghaee}}, \bibinfo {author} {\bibfnamefont {A.}~\bibnamefont {Alcaraz~Ramirez}}, \bibinfo {author} {\bibfnamefont {Z.}~\bibnamefont {Alam}}, \bibinfo {author} {\bibfnamefont {R.}~\bibnamefont {Ali}}, \bibinfo {author} {\bibfnamefont {M.}~\bibnamefont {Andrzejczuk}}, \bibinfo {author} {\bibfnamefont {A.}~\bibnamefont {Antipov}}, \bibinfo {author} {\bibfnamefont {M.}~\bibnamefont {Astafev}}, \bibinfo {author} {\bibfnamefont {A.}~\bibnamefont {Barzegar}}, \bibinfo {author} {\bibfnamefont {B.}~\bibnamefont {Bauer}}, \bibinfo {author} {\bibfnamefont {J.}~\bibnamefont {Becker}}, \bibinfo {author} {\bibfnamefont {U.~K.}\ \bibnamefont {Bhaskar}}, \bibinfo {author} {\bibfnamefont {A.}~\bibnamefont {Bocharov}}, \bibinfo {author} {\bibfnamefont {S.}~\bibnamefont {Boddapati}}, \bibinfo {author} {\bibfnamefont {D.}~\bibnamefont {Bohn}}, \bibinfo {author} {\bibfnamefont {J.}~\bibnamefont {Bommer}}, \bibinfo {author} {\bibfnamefont {L.}~\bibnamefont
  {Bourdet}}, \bibinfo {author} {\bibfnamefont {A.}~\bibnamefont {Bousquet}}, \bibinfo {author} {\bibfnamefont {S.}~\bibnamefont {Boutin}}, \bibinfo {author} {\bibfnamefont {L.}~\bibnamefont {Casparis}}, \bibinfo {author} {\bibfnamefont {B.~J.}\ \bibnamefont {Chapman}}, \bibinfo {author} {\bibfnamefont {S.}~\bibnamefont {Chatoor}}, \bibinfo {author} {\bibfnamefont {A.~W.}\ \bibnamefont {Christensen}}, \bibinfo {author} {\bibfnamefont {C.}~\bibnamefont {Chua}}, \bibinfo {author} {\bibfnamefont {P.}~\bibnamefont {Codd}}, \bibinfo {author} {\bibfnamefont {W.}~\bibnamefont {Cole}}, \bibinfo {author} {\bibfnamefont {P.}~\bibnamefont {Cooper}}, \bibinfo {author} {\bibfnamefont {F.}~\bibnamefont {Corsetti}}, \bibinfo {author} {\bibfnamefont {A.}~\bibnamefont {Cui}}, \bibinfo {author} {\bibfnamefont {P.}~\bibnamefont {Dalpasso}}, \bibinfo {author} {\bibfnamefont {J.~P.}\ \bibnamefont {Dehollain}}, \bibinfo {author} {\bibfnamefont {G.}~\bibnamefont {{de Lange}}}, \bibinfo {author} {\bibfnamefont {M.}~\bibnamefont {{de
  Moor}}}, \bibinfo {author} {\bibfnamefont {A.}~\bibnamefont {Ekefj{\"a}rd}}, \bibinfo {author} {\bibfnamefont {T.}~\bibnamefont {El~Dandachi}}, \bibinfo {author} {\bibfnamefont {J.~C.}\ \bibnamefont {Estrada~Salda{\~n}a}}, \bibinfo {author} {\bibfnamefont {S.}~\bibnamefont {Fallahi}}, \bibinfo {author} {\bibfnamefont {L.}~\bibnamefont {Galletti}}, \bibinfo {author} {\bibfnamefont {G.}~\bibnamefont {Gardner}}, \bibinfo {author} {\bibfnamefont {D.}~\bibnamefont {Govender}}, \bibinfo {author} {\bibfnamefont {F.}~\bibnamefont {Griggio}}, \bibinfo {author} {\bibfnamefont {R.}~\bibnamefont {Grigoryan}}, \bibinfo {author} {\bibfnamefont {S.}~\bibnamefont {Grijalva}}, \bibinfo {author} {\bibfnamefont {S.}~\bibnamefont {Gronin}}, \bibinfo {author} {\bibfnamefont {J.}~\bibnamefont {Gukelberger}}, \bibinfo {author} {\bibfnamefont {M.}~\bibnamefont {Hamdast}}, \bibinfo {author} {\bibfnamefont {F.}~\bibnamefont {Hamze}}, \bibinfo {author} {\bibfnamefont {E.~B.}\ \bibnamefont {Hansen}}, \bibinfo {author} {\bibfnamefont
  {S.}~\bibnamefont {Heedt}}, \bibinfo {author} {\bibfnamefont {Z.}~\bibnamefont {Heidarnia}}, \bibinfo {author} {\bibfnamefont {J.}~\bibnamefont {Herranz~Zamorano}}, \bibinfo {author} {\bibfnamefont {S.}~\bibnamefont {Ho}}, \bibinfo {author} {\bibfnamefont {L.}~\bibnamefont {Holgaard}}, \bibinfo {author} {\bibfnamefont {J.}~\bibnamefont {Hornibrook}}, \bibinfo {author} {\bibfnamefont {J.}~\bibnamefont {Indrapiromkul}}, \bibinfo {author} {\bibfnamefont {H.}~\bibnamefont {Ingerslev}}, \bibinfo {author} {\bibfnamefont {L.}~\bibnamefont {Ivancevic}}, \bibinfo {author} {\bibfnamefont {T.}~\bibnamefont {Jensen}}, \bibinfo {author} {\bibfnamefont {J.}~\bibnamefont {Jhoja}}, \bibinfo {author} {\bibfnamefont {J.}~\bibnamefont {Jones}}, \bibinfo {author} {\bibfnamefont {K.~V.}\ \bibnamefont {Kalashnikov}}, \bibinfo {author} {\bibfnamefont {R.}~\bibnamefont {Kallaher}}, \bibinfo {author} {\bibfnamefont {R.}~\bibnamefont {Kalra}}, \bibinfo {author} {\bibfnamefont {F.}~\bibnamefont {Karimi}}, \bibinfo {author}
  {\bibfnamefont {T.}~\bibnamefont {Karzig}}, \bibinfo {author} {\bibfnamefont {E.}~\bibnamefont {King}}, \bibinfo {author} {\bibfnamefont {M.~E.}\ \bibnamefont {Kloster}}, \bibinfo {author} {\bibfnamefont {C.}~\bibnamefont {Knapp}}, \bibinfo {author} {\bibfnamefont {D.}~\bibnamefont {Kocon}}, \bibinfo {author} {\bibfnamefont {J.~V.}\ \bibnamefont {Koski}}, \bibinfo {author} {\bibfnamefont {P.}~\bibnamefont {Kostamo}}, \bibinfo {author} {\bibfnamefont {M.}~\bibnamefont {Kumar}}, \bibinfo {author} {\bibfnamefont {T.}~\bibnamefont {Laeven}}, \bibinfo {author} {\bibfnamefont {T.}~\bibnamefont {Larsen}}, \bibinfo {author} {\bibfnamefont {J.}~\bibnamefont {Lee}}, \bibinfo {author} {\bibfnamefont {K.}~\bibnamefont {Lee}}, \bibinfo {author} {\bibfnamefont {G.}~\bibnamefont {Leum}}, \bibinfo {author} {\bibfnamefont {K.}~\bibnamefont {Li}}, \bibinfo {author} {\bibfnamefont {T.}~\bibnamefont {Lindemann}}, \bibinfo {author} {\bibfnamefont {M.}~\bibnamefont {Looij}}, \bibinfo {author} {\bibfnamefont {J.}~\bibnamefont
  {Love}}, \bibinfo {author} {\bibfnamefont {M.}~\bibnamefont {Lucas}}, \bibinfo {author} {\bibfnamefont {R.}~\bibnamefont {Lutchyn}}, \bibinfo {author} {\bibfnamefont {M.~H.}\ \bibnamefont {Madsen}}, \bibinfo {author} {\bibfnamefont {N.}~\bibnamefont {Madulid}}, \bibinfo {author} {\bibfnamefont {A.}~\bibnamefont {Malmros}}, \bibinfo {author} {\bibfnamefont {M.}~\bibnamefont {Manfra}}, \bibinfo {author} {\bibfnamefont {D.}~\bibnamefont {Mantri}}, \bibinfo {author} {\bibfnamefont {S.~B.}\ \bibnamefont {Markussen}}, \bibinfo {author} {\bibfnamefont {E.}~\bibnamefont {Martinez}}, \bibinfo {author} {\bibfnamefont {M.}~\bibnamefont {Mattila}}, \bibinfo {author} {\bibfnamefont {R.}~\bibnamefont {McNeil}}, \bibinfo {author} {\bibfnamefont {A.~B.}\ \bibnamefont {Mei}}, \bibinfo {author} {\bibfnamefont {R.~V.}\ \bibnamefont {Mishmash}}, \bibinfo {author} {\bibfnamefont {G.}~\bibnamefont {Mohandas}}, \bibinfo {author} {\bibfnamefont {C.}~\bibnamefont {Mollgaard}}, \bibinfo {author} {\bibfnamefont {T.}~\bibnamefont
  {Morgan}}, \bibinfo {author} {\bibfnamefont {G.}~\bibnamefont {Moussa}}, \bibinfo {author} {\bibfnamefont {C.}~\bibnamefont {Nayak}}, \bibinfo {author} {\bibfnamefont {J.~H.}\ \bibnamefont {Nielsen}}, \bibinfo {author} {\bibfnamefont {J.~M.}\ \bibnamefont {Nielsen}}, \bibinfo {author} {\bibfnamefont {W.~H.~P.}\ \bibnamefont {Nielsen}}, \bibinfo {author} {\bibfnamefont {B.}~\bibnamefont {Nijholt}}, \bibinfo {author} {\bibfnamefont {M.}~\bibnamefont {Nystrom}}, \bibinfo {author} {\bibfnamefont {E.}~\bibnamefont {O'Farrell}}, \bibinfo {author} {\bibfnamefont {T.}~\bibnamefont {Ohki}}, \bibinfo {author} {\bibfnamefont {K.}~\bibnamefont {Otani}}, \bibinfo {author} {\bibfnamefont {B.}~\bibnamefont {Paquelet~W{\"u}tz}}, \bibinfo {author} {\bibfnamefont {S.}~\bibnamefont {Pauka}}, \bibinfo {author} {\bibfnamefont {K.}~\bibnamefont {Petersson}}, \bibinfo {author} {\bibfnamefont {L.}~\bibnamefont {Petit}}, \bibinfo {author} {\bibfnamefont {D.}~\bibnamefont {Pikulin}}, \bibinfo {author} {\bibfnamefont
  {G.}~\bibnamefont {Prawiroatmodjo}}, \bibinfo {author} {\bibfnamefont {F.}~\bibnamefont {Preiss}}, \bibinfo {author} {\bibfnamefont {E.}~\bibnamefont {Puchol~Morejon}}, \bibinfo {author} {\bibfnamefont {M.}~\bibnamefont {Rajpalke}}, \bibinfo {author} {\bibfnamefont {C.}~\bibnamefont {Ranta}}, \bibinfo {author} {\bibfnamefont {K.}~\bibnamefont {Rasmussen}}, \bibinfo {author} {\bibfnamefont {D.}~\bibnamefont {Razmadze}}, \bibinfo {author} {\bibfnamefont {O.}~\bibnamefont {Reentila}}, \bibinfo {author} {\bibfnamefont {D.~J.}\ \bibnamefont {Reilly}}, \bibinfo {author} {\bibfnamefont {Y.}~\bibnamefont {Ren}}, \bibinfo {author} {\bibfnamefont {K.}~\bibnamefont {Reneris}}, \bibinfo {author} {\bibfnamefont {R.}~\bibnamefont {Rouse}}, \bibinfo {author} {\bibfnamefont {I.}~\bibnamefont {Sadovskyy}}, \bibinfo {author} {\bibfnamefont {L.}~\bibnamefont {Sainiemi}}, \bibinfo {author} {\bibfnamefont {I.}~\bibnamefont {Sanlorenzo}}, \bibinfo {author} {\bibfnamefont {E.}~\bibnamefont {Schmidgall}}, \bibinfo {author}
  {\bibfnamefont {C.}~\bibnamefont {Sfiligoj}}, \bibinfo {author} {\bibfnamefont {M.~B.}\ \bibnamefont {Shah}}, \bibinfo {author} {\bibfnamefont {K.}~\bibnamefont {Simoes}}, \bibinfo {author} {\bibfnamefont {S.}~\bibnamefont {Singh}}, \bibinfo {author} {\bibfnamefont {S.}~\bibnamefont {Sinha}}, \bibinfo {author} {\bibfnamefont {T.}~\bibnamefont {Soerensen}}, \bibinfo {author} {\bibfnamefont {P.}~\bibnamefont {Sohr}}, \bibinfo {author} {\bibfnamefont {T.}~\bibnamefont {Stankevic}}, \bibinfo {author} {\bibfnamefont {L.}~\bibnamefont {Stek}}, \bibinfo {author} {\bibfnamefont {E.}~\bibnamefont {Stuppard}}, \bibinfo {author} {\bibfnamefont {H.}~\bibnamefont {Suominen}}, \bibinfo {author} {\bibfnamefont {J.}~\bibnamefont {Suter}}, \bibinfo {author} {\bibfnamefont {S.}~\bibnamefont {Teicher}}, \bibinfo {author} {\bibfnamefont {N.}~\bibnamefont {Thiyagarajah}}, \bibinfo {author} {\bibfnamefont {R.}~\bibnamefont {Tholapi}}, \bibinfo {author} {\bibfnamefont {M.}~\bibnamefont {Thomas}}, \bibinfo {author} {\bibfnamefont
  {E.}~\bibnamefont {Toomey}}, \bibinfo {author} {\bibfnamefont {J.}~\bibnamefont {Tracy}}, \bibinfo {author} {\bibfnamefont {M.}~\bibnamefont {Turley}}, \bibinfo {author} {\bibfnamefont {S.}~\bibnamefont {Upadhyay}}, \bibinfo {author} {\bibfnamefont {I.}~\bibnamefont {Urban}}, \bibinfo {author} {\bibfnamefont {K.}~\bibnamefont {Van~Hoogdalem}}, \bibinfo {author} {\bibfnamefont {D.~J.}\ \bibnamefont {Van~Woerkom}}, \bibinfo {author} {\bibfnamefont {D.~V.}\ \bibnamefont {Viazmitinov}}, \bibinfo {author} {\bibfnamefont {D.}~\bibnamefont {Vogel}}, \bibinfo {author} {\bibfnamefont {J.}~\bibnamefont {Watson}}, \bibinfo {author} {\bibfnamefont {A.}~\bibnamefont {Webster}}, \bibinfo {author} {\bibfnamefont {J.}~\bibnamefont {Weston}}, \bibinfo {author} {\bibfnamefont {G.~W.}\ \bibnamefont {Winkler}}, \bibinfo {author} {\bibfnamefont {D.}~\bibnamefont {Xu}}, \bibinfo {author} {\bibfnamefont {C.~K.}\ \bibnamefont {Yang}}, \bibinfo {author} {\bibfnamefont {E.}~\bibnamefont {Yucelen}}, \bibinfo {author} {\bibfnamefont
  {R.}~\bibnamefont {Zeisel}}, \bibinfo {author} {\bibfnamefont {G.}~\bibnamefont {Zheng}},\ and\ \bibinfo {author} {\bibfnamefont {J.}~\bibnamefont {Zilke}},\ }\bibfield  {title} {\bibinfo {title} {Interferometric single-shot parity measurement in {{InAs}}--{{Al}} hybrid devices},\ }\href {https://doi.org/10.1038/s41586-024-08445-2} {\bibfield  {journal} {\bibinfo  {journal} {Nature}\ }\textbf {\bibinfo {volume} {638}},\ \bibinfo {pages} {651} (\bibinfo {year} {2025})}\BibitemShut {NoStop}%
\bibitem [{\citenamefont {Petta}\ \emph {et~al.}(2010)\citenamefont {Petta}, \citenamefont {Lu},\ and\ \citenamefont {Gossard}}]{pettaCoherentBeamSplitter2010}%
  \BibitemOpen
  \bibfield  {author} {\bibinfo {author} {\bibfnamefont {J.~R.}\ \bibnamefont {Petta}}, \bibinfo {author} {\bibfnamefont {H.}~\bibnamefont {Lu}},\ and\ \bibinfo {author} {\bibfnamefont {A.~C.}\ \bibnamefont {Gossard}},\ }\bibfield  {title} {\bibinfo {title} {A {{Coherent Beam Splitter}} for {{Electronic Spin States}}},\ }\href {https://doi.org/10.1126/science.1183628} {\bibfield  {journal} {\bibinfo  {journal} {Science}\ }\textbf {\bibinfo {volume} {327}},\ \bibinfo {pages} {669} (\bibinfo {year} {2010})}\BibitemShut {NoStop}%
\bibitem [{\citenamefont {Ribeiro}\ \emph {et~al.}(2013)\citenamefont {Ribeiro}, \citenamefont {Burkard}, \citenamefont {Petta}, \citenamefont {Lu},\ and\ \citenamefont {Gossard}}]{ribeiroCoherentAdiabaticSpin2013}%
  \BibitemOpen
  \bibfield  {author} {\bibinfo {author} {\bibfnamefont {H.}~\bibnamefont {Ribeiro}}, \bibinfo {author} {\bibfnamefont {G.}~\bibnamefont {Burkard}}, \bibinfo {author} {\bibfnamefont {J.~R.}\ \bibnamefont {Petta}}, \bibinfo {author} {\bibfnamefont {H.}~\bibnamefont {Lu}},\ and\ \bibinfo {author} {\bibfnamefont {A.~C.}\ \bibnamefont {Gossard}},\ }\bibfield  {title} {\bibinfo {title} {Coherent {{Adiabatic Spin Control}} in the {{Presence}} of {{Charge Noise Using Tailored Pulses}}},\ }\href {https://doi.org/10.1103/PhysRevLett.110.086804} {\bibfield  {journal} {\bibinfo  {journal} {Physical Review Letters}\ }\textbf {\bibinfo {volume} {110}},\ \bibinfo {pages} {086804} (\bibinfo {year} {2013})}\BibitemShut {NoStop}%
\bibitem [{\citenamefont {Ban}\ \emph {et~al.}(2018)\citenamefont {Ban}, \citenamefont {Chen},\ and\ \citenamefont {Platero}}]{banFastLongrangeCharge2018}%
  \BibitemOpen
  \bibfield  {author} {\bibinfo {author} {\bibfnamefont {Y.}~\bibnamefont {Ban}}, \bibinfo {author} {\bibfnamefont {X.}~\bibnamefont {Chen}},\ and\ \bibinfo {author} {\bibfnamefont {G.}~\bibnamefont {Platero}},\ }\bibfield  {title} {\bibinfo {title} {Fast long-range charge transfer in quantum dot arrays},\ }\href {https://doi.org/10.1088/1361-6528/aae0ce} {\bibfield  {journal} {\bibinfo  {journal} {Nanotechnology}\ }\textbf {\bibinfo {volume} {29}},\ \bibinfo {pages} {505201} (\bibinfo {year} {2018})}\BibitemShut {NoStop}%
\bibitem [{\citenamefont {Binder}\ \emph {et~al.}(2015)\citenamefont {Binder}, \citenamefont {Vinjanampathy}, \citenamefont {Modi},\ and\ \citenamefont {Goold}}]{binderQuantacellPowerfulCharging2015}%
  \BibitemOpen
  \bibfield  {author} {\bibinfo {author} {\bibfnamefont {F.~C.}\ \bibnamefont {Binder}}, \bibinfo {author} {\bibfnamefont {S.}~\bibnamefont {Vinjanampathy}}, \bibinfo {author} {\bibfnamefont {K.}~\bibnamefont {Modi}},\ and\ \bibinfo {author} {\bibfnamefont {J.}~\bibnamefont {Goold}},\ }\bibfield  {title} {\bibinfo {title} {Quantacell: Powerful charging of quantum batteries},\ }\href {https://doi.org/10.1088/1367-2630/17/7/075015} {\bibfield  {journal} {\bibinfo  {journal} {New Journal of Physics}\ }\textbf {\bibinfo {volume} {17}},\ \bibinfo {pages} {075015} (\bibinfo {year} {2015})}\BibitemShut {NoStop}%
\bibitem [{\citenamefont {Blekos}\ \emph {et~al.}(2024)\citenamefont {Blekos}, \citenamefont {Brand}, \citenamefont {Ceschini}, \citenamefont {Chou}, \citenamefont {Li}, \citenamefont {Pandya},\ and\ \citenamefont {Summer}}]{blekosReviewQuantumApproximate2024}%
  \BibitemOpen
  \bibfield  {author} {\bibinfo {author} {\bibfnamefont {K.}~\bibnamefont {Blekos}}, \bibinfo {author} {\bibfnamefont {D.}~\bibnamefont {Brand}}, \bibinfo {author} {\bibfnamefont {A.}~\bibnamefont {Ceschini}}, \bibinfo {author} {\bibfnamefont {C.-H.}\ \bibnamefont {Chou}}, \bibinfo {author} {\bibfnamefont {R.-H.}\ \bibnamefont {Li}}, \bibinfo {author} {\bibfnamefont {K.}~\bibnamefont {Pandya}},\ and\ \bibinfo {author} {\bibfnamefont {A.}~\bibnamefont {Summer}},\ }\bibfield  {title} {\bibinfo {title} {A review on {{Quantum Approximate Optimization Algorithm}} and its variants},\ }\href {https://doi.org/10.1016/j.physrep.2024.03.002} {\bibfield  {journal} {\bibinfo  {journal} {Physics Reports}\ }\bibinfo {series} {A Review on {{Quantum Approximate Optimization Algorithm}} and Its Variants},\ \textbf {\bibinfo {volume} {1068}},\ \bibinfo {pages} {1} (\bibinfo {year} {2024})}\BibitemShut {NoStop}%
\bibitem [{\citenamefont {Abbas}\ \emph {et~al.}(2024)\citenamefont {Abbas}, \citenamefont {Ambainis}, \citenamefont {Augustino}, \citenamefont {B{\"a}rtschi}, \citenamefont {Buhrman}, \citenamefont {Coffrin}, \citenamefont {Cortiana}, \citenamefont {Dunjko}, \citenamefont {Egger}, \citenamefont {Elmegreen}, \citenamefont {Franco}, \citenamefont {Fratini}, \citenamefont {Fuller}, \citenamefont {Gacon}, \citenamefont {Gonciulea}, \citenamefont {Gribling}, \citenamefont {Gupta}, \citenamefont {Hadfield}, \citenamefont {Heese}, \citenamefont {Kircher}, \citenamefont {Kleinert}, \citenamefont {Koch}, \citenamefont {Korpas}, \citenamefont {Lenk}, \citenamefont {Marecek}, \citenamefont {Markov}, \citenamefont {Mazzola}, \citenamefont {Mensa}, \citenamefont {Mohseni}, \citenamefont {Nannicini}, \citenamefont {O'Meara}, \citenamefont {Tapia}, \citenamefont {Pokutta}, \citenamefont {Proissl}, \citenamefont {Rebentrost}, \citenamefont {Sahin}, \citenamefont {Symons}, \citenamefont {Tornow}, \citenamefont {Valls},
  \citenamefont {Woerner}, \citenamefont {{Wolf-Bauwens}}, \citenamefont {Yard}, \citenamefont {Yarkoni}, \citenamefont {Zechiel}, \citenamefont {Zhuk},\ and\ \citenamefont {Zoufal}}]{abbasChallengesOpportunitiesQuantum2024a}%
  \BibitemOpen
  \bibfield  {author} {\bibinfo {author} {\bibfnamefont {A.}~\bibnamefont {Abbas}}, \bibinfo {author} {\bibfnamefont {A.}~\bibnamefont {Ambainis}}, \bibinfo {author} {\bibfnamefont {B.}~\bibnamefont {Augustino}}, \bibinfo {author} {\bibfnamefont {A.}~\bibnamefont {B{\"a}rtschi}}, \bibinfo {author} {\bibfnamefont {H.}~\bibnamefont {Buhrman}}, \bibinfo {author} {\bibfnamefont {C.}~\bibnamefont {Coffrin}}, \bibinfo {author} {\bibfnamefont {G.}~\bibnamefont {Cortiana}}, \bibinfo {author} {\bibfnamefont {V.}~\bibnamefont {Dunjko}}, \bibinfo {author} {\bibfnamefont {D.~J.}\ \bibnamefont {Egger}}, \bibinfo {author} {\bibfnamefont {B.~G.}\ \bibnamefont {Elmegreen}}, \bibinfo {author} {\bibfnamefont {N.}~\bibnamefont {Franco}}, \bibinfo {author} {\bibfnamefont {F.}~\bibnamefont {Fratini}}, \bibinfo {author} {\bibfnamefont {B.}~\bibnamefont {Fuller}}, \bibinfo {author} {\bibfnamefont {J.}~\bibnamefont {Gacon}}, \bibinfo {author} {\bibfnamefont {C.}~\bibnamefont {Gonciulea}}, \bibinfo {author} {\bibfnamefont
  {S.}~\bibnamefont {Gribling}}, \bibinfo {author} {\bibfnamefont {S.}~\bibnamefont {Gupta}}, \bibinfo {author} {\bibfnamefont {S.}~\bibnamefont {Hadfield}}, \bibinfo {author} {\bibfnamefont {R.}~\bibnamefont {Heese}}, \bibinfo {author} {\bibfnamefont {G.}~\bibnamefont {Kircher}}, \bibinfo {author} {\bibfnamefont {T.}~\bibnamefont {Kleinert}}, \bibinfo {author} {\bibfnamefont {T.}~\bibnamefont {Koch}}, \bibinfo {author} {\bibfnamefont {G.}~\bibnamefont {Korpas}}, \bibinfo {author} {\bibfnamefont {S.}~\bibnamefont {Lenk}}, \bibinfo {author} {\bibfnamefont {J.}~\bibnamefont {Marecek}}, \bibinfo {author} {\bibfnamefont {V.}~\bibnamefont {Markov}}, \bibinfo {author} {\bibfnamefont {G.}~\bibnamefont {Mazzola}}, \bibinfo {author} {\bibfnamefont {S.}~\bibnamefont {Mensa}}, \bibinfo {author} {\bibfnamefont {N.}~\bibnamefont {Mohseni}}, \bibinfo {author} {\bibfnamefont {G.}~\bibnamefont {Nannicini}}, \bibinfo {author} {\bibfnamefont {C.}~\bibnamefont {O'Meara}}, \bibinfo {author} {\bibfnamefont {E.~P.}\ \bibnamefont
  {Tapia}}, \bibinfo {author} {\bibfnamefont {S.}~\bibnamefont {Pokutta}}, \bibinfo {author} {\bibfnamefont {M.}~\bibnamefont {Proissl}}, \bibinfo {author} {\bibfnamefont {P.}~\bibnamefont {Rebentrost}}, \bibinfo {author} {\bibfnamefont {E.}~\bibnamefont {Sahin}}, \bibinfo {author} {\bibfnamefont {B.~C.~B.}\ \bibnamefont {Symons}}, \bibinfo {author} {\bibfnamefont {S.}~\bibnamefont {Tornow}}, \bibinfo {author} {\bibfnamefont {V.}~\bibnamefont {Valls}}, \bibinfo {author} {\bibfnamefont {S.}~\bibnamefont {Woerner}}, \bibinfo {author} {\bibfnamefont {M.~L.}\ \bibnamefont {{Wolf-Bauwens}}}, \bibinfo {author} {\bibfnamefont {J.}~\bibnamefont {Yard}}, \bibinfo {author} {\bibfnamefont {S.}~\bibnamefont {Yarkoni}}, \bibinfo {author} {\bibfnamefont {D.}~\bibnamefont {Zechiel}}, \bibinfo {author} {\bibfnamefont {S.}~\bibnamefont {Zhuk}},\ and\ \bibinfo {author} {\bibfnamefont {C.}~\bibnamefont {Zoufal}},\ }\bibfield  {title} {\bibinfo {title} {Challenges and opportunities in quantum optimization},\ }\href
  {https://doi.org/10.1038/s42254-024-00770-9} {\bibfield  {journal} {\bibinfo  {journal} {Nature Reviews Physics}\ }\textbf {\bibinfo {volume} {6}},\ \bibinfo {pages} {718} (\bibinfo {year} {2024})}\BibitemShut {NoStop}%
\bibitem [{\citenamefont {Werschnik}\ and\ \citenamefont {Gross}(2007)}]{werschnikQuantumOptimalControl2007}%
  \BibitemOpen
  \bibfield  {author} {\bibinfo {author} {\bibfnamefont {J.}~\bibnamefont {Werschnik}}\ and\ \bibinfo {author} {\bibfnamefont {E.~K.~U.}\ \bibnamefont {Gross}},\ }\href@noop {} {\bibinfo {title} {Quantum {{Optimal Control Theory}}}},\ \bibinfo {howpublished} {https://arxiv.org/abs/0707.1883v1} (\bibinfo {year} {2007})\BibitemShut {NoStop}%
\bibitem [{\citenamefont {Poggiali}\ \emph {et~al.}(2018)\citenamefont {Poggiali}, \citenamefont {Cappellaro},\ and\ \citenamefont {Fabbri}}]{poggialiOptimalControlOneQubit2018}%
  \BibitemOpen
  \bibfield  {author} {\bibinfo {author} {\bibfnamefont {F.}~\bibnamefont {Poggiali}}, \bibinfo {author} {\bibfnamefont {P.}~\bibnamefont {Cappellaro}},\ and\ \bibinfo {author} {\bibfnamefont {N.}~\bibnamefont {Fabbri}},\ }\bibfield  {title} {\bibinfo {title} {Optimal {{Control}} for {{One-Qubit Quantum Sensing}}},\ }\href {https://doi.org/10.1103/PhysRevX.8.021059} {\bibfield  {journal} {\bibinfo  {journal} {Physical Review X}\ }\textbf {\bibinfo {volume} {8}},\ \bibinfo {pages} {021059} (\bibinfo {year} {2018})}\BibitemShut {NoStop}%
\bibitem [{\citenamefont {Theis}\ \emph {et~al.}(2018)\citenamefont {Theis}, \citenamefont {Motzoi}, \citenamefont {Machnes},\ and\ \citenamefont {Wilhelm}}]{theisCounteractingSystemsDiabaticities2018}%
  \BibitemOpen
  \bibfield  {author} {\bibinfo {author} {\bibfnamefont {L.~S.}\ \bibnamefont {Theis}}, \bibinfo {author} {\bibfnamefont {F.}~\bibnamefont {Motzoi}}, \bibinfo {author} {\bibfnamefont {S.}~\bibnamefont {Machnes}},\ and\ \bibinfo {author} {\bibfnamefont {F.~K.}\ \bibnamefont {Wilhelm}},\ }\bibfield  {title} {\bibinfo {title} {Counteracting systems of diabaticities using {{DRAG}} controls: {{The}} status after 10 years},\ }\href {https://doi.org/10.1209/0295-5075/123/60001} {\bibfield  {journal} {\bibinfo  {journal} {EPL (Europhysics Letters)}\ }\textbf {\bibinfo {volume} {123}},\ \bibinfo {pages} {60001} (\bibinfo {year} {2018})}\BibitemShut {NoStop}%
\bibitem [{\citenamefont {Koch}\ \emph {et~al.}(2022)\citenamefont {Koch}, \citenamefont {Boscain}, \citenamefont {Calarco}, \citenamefont {Dirr}, \citenamefont {Filipp}, \citenamefont {Glaser}, \citenamefont {Kosloff}, \citenamefont {Montangero}, \citenamefont {{Schulte-Herbr{\"u}ggen}}, \citenamefont {Sugny},\ and\ \citenamefont {Wilhelm}}]{kochQuantumOptimalControl2022}%
  \BibitemOpen
  \bibfield  {author} {\bibinfo {author} {\bibfnamefont {C.~P.}\ \bibnamefont {Koch}}, \bibinfo {author} {\bibfnamefont {U.}~\bibnamefont {Boscain}}, \bibinfo {author} {\bibfnamefont {T.}~\bibnamefont {Calarco}}, \bibinfo {author} {\bibfnamefont {G.}~\bibnamefont {Dirr}}, \bibinfo {author} {\bibfnamefont {S.}~\bibnamefont {Filipp}}, \bibinfo {author} {\bibfnamefont {S.~J.}\ \bibnamefont {Glaser}}, \bibinfo {author} {\bibfnamefont {R.}~\bibnamefont {Kosloff}}, \bibinfo {author} {\bibfnamefont {S.}~\bibnamefont {Montangero}}, \bibinfo {author} {\bibfnamefont {T.}~\bibnamefont {{Schulte-Herbr{\"u}ggen}}}, \bibinfo {author} {\bibfnamefont {D.}~\bibnamefont {Sugny}},\ and\ \bibinfo {author} {\bibfnamefont {F.~K.}\ \bibnamefont {Wilhelm}},\ }\bibfield  {title} {\bibinfo {title} {Quantum optimal control in quantum technologies. {{Strategic}} report on current status, visions and goals for research in {{Europe}}},\ }\href {https://doi.org/10.1140/epjqt/s40507-022-00138-x} {\bibfield  {journal} {\bibinfo  {journal}
  {EPJ Quantum Technology}\ }\textbf {\bibinfo {volume} {9}},\ \bibinfo {pages} {19} (\bibinfo {year} {2022})}\BibitemShut {NoStop}%
\bibitem [{\citenamefont {Greentree}\ \emph {et~al.}(2004)\citenamefont {Greentree}, \citenamefont {Cole}, \citenamefont {Hamilton},\ and\ \citenamefont {Hollenberg}}]{greentreeCoherentElectronicTransfer2004}%
  \BibitemOpen
  \bibfield  {author} {\bibinfo {author} {\bibfnamefont {A.~D.}\ \bibnamefont {Greentree}}, \bibinfo {author} {\bibfnamefont {J.~H.}\ \bibnamefont {Cole}}, \bibinfo {author} {\bibfnamefont {A.~R.}\ \bibnamefont {Hamilton}},\ and\ \bibinfo {author} {\bibfnamefont {L.~C.~L.}\ \bibnamefont {Hollenberg}},\ }\bibfield  {title} {\bibinfo {title} {Coherent electronic transfer in quantum dot systems using adiabatic passage},\ }\href {https://doi.org/10.1103/PhysRevB.70.235317} {\bibfield  {journal} {\bibinfo  {journal} {Physical Review B}\ }\textbf {\bibinfo {volume} {70}},\ \bibinfo {pages} {235317} (\bibinfo {year} {2004})}\BibitemShut {NoStop}%
\bibitem [{\citenamefont {Khaneja}\ \emph {et~al.}(2001)\citenamefont {Khaneja}, \citenamefont {Brockett},\ and\ \citenamefont {Glaser}}]{khanejaTimeOptimalControl2001}%
  \BibitemOpen
  \bibfield  {author} {\bibinfo {author} {\bibfnamefont {N.}~\bibnamefont {Khaneja}}, \bibinfo {author} {\bibfnamefont {R.}~\bibnamefont {Brockett}},\ and\ \bibinfo {author} {\bibfnamefont {S.~J.}\ \bibnamefont {Glaser}},\ }\bibfield  {title} {\bibinfo {title} {Time optimal control in spin systems},\ }\href {https://doi.org/10.1103/PhysRevA.63.032308} {\bibfield  {journal} {\bibinfo  {journal} {Physical Review A}\ }\textbf {\bibinfo {volume} {63}},\ \bibinfo {pages} {032308} (\bibinfo {year} {2001})}\BibitemShut {NoStop}%
\bibitem [{\citenamefont {Carlini}\ \emph {et~al.}(2006)\citenamefont {Carlini}, \citenamefont {Hosoya}, \citenamefont {Koike},\ and\ \citenamefont {Okudaira}}]{carliniTimeOptimalQuantumEvolution2006}%
  \BibitemOpen
  \bibfield  {author} {\bibinfo {author} {\bibfnamefont {A.}~\bibnamefont {Carlini}}, \bibinfo {author} {\bibfnamefont {A.}~\bibnamefont {Hosoya}}, \bibinfo {author} {\bibfnamefont {T.}~\bibnamefont {Koike}},\ and\ \bibinfo {author} {\bibfnamefont {Y.}~\bibnamefont {Okudaira}},\ }\bibfield  {title} {\bibinfo {title} {Time-{{Optimal Quantum Evolution}}},\ }\href {https://doi.org/10.1103/PhysRevLett.96.060503} {\bibfield  {journal} {\bibinfo  {journal} {Physical Review Letters}\ }\textbf {\bibinfo {volume} {96}},\ \bibinfo {pages} {060503} (\bibinfo {year} {2006})}\BibitemShut {NoStop}%
\bibitem [{\citenamefont {Pontryagin}(2017)}]{pontryaginMathematicalTheoryOptimal2017}%
  \BibitemOpen
  \bibfield  {author} {\bibinfo {author} {\bibfnamefont {L.~S.}\ \bibnamefont {Pontryagin}},\ }\href {https://doi.org/10.1201/9780203749319} {\emph {\bibinfo {title} {Mathematical {{Theory}} of {{Optimal Processes}}}}}\ (\bibinfo  {publisher} {Routledge},\ \bibinfo {address} {London},\ \bibinfo {year} {2017})\BibitemShut {NoStop}%
\bibitem [{\citenamefont {Boltyanski}\ \emph {et~al.}(1999)\citenamefont {Boltyanski}, \citenamefont {Martini},\ and\ \citenamefont {Soltan}}]{boltyanskiGeometricMethodsOptimization1999}%
  \BibitemOpen
  \bibfield  {author} {\bibinfo {author} {\bibfnamefont {V.}~\bibnamefont {Boltyanski}}, \bibinfo {author} {\bibfnamefont {H.}~\bibnamefont {Martini}},\ and\ \bibinfo {author} {\bibfnamefont {V.}~\bibnamefont {Soltan}},\ }\href {https://doi.org/10.1007/978-1-4615-5319-9} {\emph {\bibinfo {title} {Geometric {{Methods}} and {{Optimization Problems}}}}},\ edited by\ \bibinfo {editor} {\bibfnamefont {D.-Z.}\ \bibnamefont {Du}}\ and\ \bibinfo {editor} {\bibfnamefont {P.~M.}\ \bibnamefont {Pardalos}},\ \bibinfo {series} {Combinatorial {{Optimization}}}, Vol.~\bibinfo {volume} {4}\ (\bibinfo  {publisher} {Springer US},\ \bibinfo {address} {Boston, MA},\ \bibinfo {year} {1999})\BibitemShut {NoStop}%
\bibitem [{\citenamefont {Carlini}\ \emph {et~al.}(2007)\citenamefont {Carlini}, \citenamefont {Hosoya}, \citenamefont {Koike},\ and\ \citenamefont {Okudaira}}]{carliniTimeoptimalUnitaryOperations2007}%
  \BibitemOpen
  \bibfield  {author} {\bibinfo {author} {\bibfnamefont {A.}~\bibnamefont {Carlini}}, \bibinfo {author} {\bibfnamefont {A.}~\bibnamefont {Hosoya}}, \bibinfo {author} {\bibfnamefont {T.}~\bibnamefont {Koike}},\ and\ \bibinfo {author} {\bibfnamefont {Y.}~\bibnamefont {Okudaira}},\ }\bibfield  {title} {\bibinfo {title} {Time-optimal unitary operations},\ }\href {https://doi.org/10.1103/PhysRevA.75.042308} {\bibfield  {journal} {\bibinfo  {journal} {Physical Review A}\ }\textbf {\bibinfo {volume} {75}},\ \bibinfo {pages} {042308} (\bibinfo {year} {2007})}\BibitemShut {NoStop}%
\bibitem [{\citenamefont {Boscain}\ \emph {et~al.}(2021)\citenamefont {Boscain}, \citenamefont {Sigalotti},\ and\ \citenamefont {Sugny}}]{boscainIntroductionPontryaginMaximum2021}%
  \BibitemOpen
  \bibfield  {author} {\bibinfo {author} {\bibfnamefont {U.}~\bibnamefont {Boscain}}, \bibinfo {author} {\bibfnamefont {M.}~\bibnamefont {Sigalotti}},\ and\ \bibinfo {author} {\bibfnamefont {D.}~\bibnamefont {Sugny}},\ }\bibfield  {title} {\bibinfo {title} {Introduction to the {{Pontryagin Maximum Principle}} for {{Quantum Optimal Control}}},\ }\href {https://doi.org/10.1103/PRXQuantum.2.030203} {\bibfield  {journal} {\bibinfo  {journal} {PRX Quantum}\ }\textbf {\bibinfo {volume} {2}},\ \bibinfo {pages} {030203} (\bibinfo {year} {2021})}\BibitemShut {NoStop}%
\bibitem [{\citenamefont {Koike}(2022)}]{koikeQuantumBrachistochrone2022}%
  \BibitemOpen
  \bibfield  {author} {\bibinfo {author} {\bibfnamefont {T.}~\bibnamefont {Koike}},\ }\bibfield  {title} {\bibinfo {title} {Quantum brachistochrone},\ }\href {https://doi.org/10.1098/rsta.2021.0273} {\bibfield  {journal} {\bibinfo  {journal} {Philosophical Transactions of the Royal Society A: Mathematical, Physical and Engineering Sciences}\ }\textbf {\bibinfo {volume} {380}},\ \bibinfo {pages} {20210273} (\bibinfo {year} {2022})}\BibitemShut {NoStop}%
\bibitem [{\citenamefont {Rezakhani}\ \emph {et~al.}(2009)\citenamefont {Rezakhani}, \citenamefont {Kuo}, \citenamefont {Hamma}, \citenamefont {Lidar},\ and\ \citenamefont {Zanardi}}]{rezakhaniQuantumAdiabaticBrachistochrone2009}%
  \BibitemOpen
  \bibfield  {author} {\bibinfo {author} {\bibfnamefont {A.~T.}\ \bibnamefont {Rezakhani}}, \bibinfo {author} {\bibfnamefont {W.-J.}\ \bibnamefont {Kuo}}, \bibinfo {author} {\bibfnamefont {A.}~\bibnamefont {Hamma}}, \bibinfo {author} {\bibfnamefont {D.~A.}\ \bibnamefont {Lidar}},\ and\ \bibinfo {author} {\bibfnamefont {P.}~\bibnamefont {Zanardi}},\ }\bibfield  {title} {\bibinfo {title} {Quantum {{Adiabatic Brachistochrone}}},\ }\href {https://doi.org/10.1103/PhysRevLett.103.080502} {\bibfield  {journal} {\bibinfo  {journal} {Physical Review Letters}\ }\textbf {\bibinfo {volume} {103}},\ \bibinfo {pages} {080502} (\bibinfo {year} {2009})}\BibitemShut {NoStop}%
\bibitem [{\citenamefont {Walelign}(2024)}]{walelignDynamicallyCorrectedGates2024}%
  \BibitemOpen
  \bibfield  {author} {\bibinfo {author} {\bibfnamefont {H.~Y.}\ \bibnamefont {Walelign}},\ }\bibfield  {title} {\bibinfo {title} {Dynamically corrected gates in silicon singlet-triplet spin qubits},\ }\bibfield  {journal} {\bibinfo  {journal} {Physical Review Applied}\ }\textbf {\bibinfo {volume} {22}},\ \href {https://doi.org/10.1103/PhysRevApplied.22.064029} {10.1103/PhysRevApplied.22.064029} (\bibinfo {year} {2024})\BibitemShut {NoStop}%
\bibitem [{\citenamefont {Chernova}\ \emph {et~al.}(2024)\citenamefont {Chernova}, \citenamefont {Stepanenko},\ and\ \citenamefont {Gorlach}}]{chernovaOptimizingStateTransfer2024}%
  \BibitemOpen
  \bibfield  {author} {\bibinfo {author} {\bibfnamefont {K.~S.}\ \bibnamefont {Chernova}}, \bibinfo {author} {\bibfnamefont {A.~A.}\ \bibnamefont {Stepanenko}},\ and\ \bibinfo {author} {\bibfnamefont {M.~A.}\ \bibnamefont {Gorlach}},\ }\href@noop {} {\bibinfo {title} {Optimizing state transfer in a three-qubit array via quantum brachistochrone method}} (\bibinfo {year} {2024}),\ \Eprint {https://arxiv.org/abs/2411.08644} {2411.08644} \BibitemShut {NoStop}%
\bibitem [{\citenamefont {Stepanenko}\ \emph {et~al.}(2025)\citenamefont {Stepanenko}, \citenamefont {Chernova},\ and\ \citenamefont {Gorlach}}]{stepanenkoTimeoptimalTransferQuantum2025}%
  \BibitemOpen
  \bibfield  {author} {\bibinfo {author} {\bibfnamefont {A.~A.}\ \bibnamefont {Stepanenko}}, \bibinfo {author} {\bibfnamefont {K.~S.}\ \bibnamefont {Chernova}},\ and\ \bibinfo {author} {\bibfnamefont {M.~A.}\ \bibnamefont {Gorlach}},\ }\href {https://doi.org/10.48550/arXiv.2501.11933} {\bibinfo {title} {Time-optimal transfer of the quantum state in long qubit arrays}} (\bibinfo {year} {2025}),\ \Eprint {https://arxiv.org/abs/2501.11933} {2501.11933} \BibitemShut {NoStop}%
\bibitem [{\citenamefont {Khaneja}\ \emph {et~al.}(2005)\citenamefont {Khaneja}, \citenamefont {Reiss}, \citenamefont {Kehlet}, \citenamefont {{Schulte-Herbr{\"u}ggen}},\ and\ \citenamefont {Glaser}}]{khanejaOptimalControlCoupled2005}%
  \BibitemOpen
  \bibfield  {author} {\bibinfo {author} {\bibfnamefont {N.}~\bibnamefont {Khaneja}}, \bibinfo {author} {\bibfnamefont {T.}~\bibnamefont {Reiss}}, \bibinfo {author} {\bibfnamefont {C.}~\bibnamefont {Kehlet}}, \bibinfo {author} {\bibfnamefont {T.}~\bibnamefont {{Schulte-Herbr{\"u}ggen}}},\ and\ \bibinfo {author} {\bibfnamefont {S.~J.}\ \bibnamefont {Glaser}},\ }\bibfield  {title} {\bibinfo {title} {Optimal control of coupled spin dynamics: Design of {{NMR}} pulse sequences by gradient ascent algorithms},\ }\href {https://doi.org/10.1016/j.jmr.2004.11.004} {\bibfield  {journal} {\bibinfo  {journal} {Journal of Magnetic Resonance}\ }\textbf {\bibinfo {volume} {172}},\ \bibinfo {pages} {296} (\bibinfo {year} {2005})}\BibitemShut {NoStop}%
\bibitem [{\citenamefont {Goodwin}\ and\ \citenamefont {Vinding}(2023)}]{goodwinAcceleratedNewtonRaphsonGRAPE2023}%
  \BibitemOpen
  \bibfield  {author} {\bibinfo {author} {\bibfnamefont {D.~L.}\ \bibnamefont {Goodwin}}\ and\ \bibinfo {author} {\bibfnamefont {M.~S.}\ \bibnamefont {Vinding}},\ }\bibfield  {title} {\bibinfo {title} {Accelerated {{Newton-Raphson GRAPE}} methods for optimal control},\ }\href {https://doi.org/10.1103/PhysRevResearch.5.L012042} {\bibfield  {journal} {\bibinfo  {journal} {Physical Review Research}\ }\textbf {\bibinfo {volume} {5}},\ \bibinfo {pages} {L012042} (\bibinfo {year} {2023})}\BibitemShut {NoStop}%
\bibitem [{\citenamefont {Caneva}\ \emph {et~al.}(2011)\citenamefont {Caneva}, \citenamefont {Calarco},\ and\ \citenamefont {Montangero}}]{canevaChoppedRandombasisQuantum2011}%
  \BibitemOpen
  \bibfield  {author} {\bibinfo {author} {\bibfnamefont {T.}~\bibnamefont {Caneva}}, \bibinfo {author} {\bibfnamefont {T.}~\bibnamefont {Calarco}},\ and\ \bibinfo {author} {\bibfnamefont {S.}~\bibnamefont {Montangero}},\ }\bibfield  {title} {\bibinfo {title} {Chopped random-basis quantum optimization},\ }\href {https://doi.org/10.1103/PhysRevA.84.022326} {\bibfield  {journal} {\bibinfo  {journal} {Physical Review A}\ }\textbf {\bibinfo {volume} {84}},\ \bibinfo {pages} {022326} (\bibinfo {year} {2011})}\BibitemShut {NoStop}%
\bibitem [{\citenamefont {Fauquenot}\ \emph {et~al.}(2024)\citenamefont {Fauquenot}, \citenamefont {Sarkar},\ and\ \citenamefont {Feld}}]{fauquenotEOGRAPEEODRLPEOpen2024}%
  \BibitemOpen
  \bibfield  {author} {\bibinfo {author} {\bibfnamefont {S.}~\bibnamefont {Fauquenot}}, \bibinfo {author} {\bibfnamefont {A.}~\bibnamefont {Sarkar}},\ and\ \bibinfo {author} {\bibfnamefont {S.}~\bibnamefont {Feld}},\ }\href {https://doi.org/10.48550/arXiv.2411.06556} {\bibinfo {title} {{{EO-GRAPE}} and {{EO-DRLPE}}: {{Open}} and {{Closed Loop Approaches}} for {{Energy Efficient Quantum Optimal Control}}}} (\bibinfo {year} {2024}),\ \Eprint {https://arxiv.org/abs/2411.06556} {2411.06556} \BibitemShut {NoStop}%
\bibitem [{\citenamefont {{Katiraee-Far}}\ \emph {et~al.}(2025)\citenamefont {{Katiraee-Far}}, \citenamefont {Matsumoto}, \citenamefont {Undseth}, \citenamefont {Smet}, \citenamefont {Gualtieri}, \citenamefont {Meinersen}, \citenamefont {de~Fuentes}, \citenamefont {Capannelli}, \citenamefont {{Rimbach-Russ}}, \citenamefont {Scappucci}, \citenamefont {Vandersypen},\ and\ \citenamefont {Greplova}}]{katiraee-farUnifiedEvolutionaryOptimization2025}%
  \BibitemOpen
  \bibfield  {author} {\bibinfo {author} {\bibfnamefont {S.~R.}\ \bibnamefont {{Katiraee-Far}}}, \bibinfo {author} {\bibfnamefont {Y.}~\bibnamefont {Matsumoto}}, \bibinfo {author} {\bibfnamefont {B.}~\bibnamefont {Undseth}}, \bibinfo {author} {\bibfnamefont {M.~D.}\ \bibnamefont {Smet}}, \bibinfo {author} {\bibfnamefont {V.}~\bibnamefont {Gualtieri}}, \bibinfo {author} {\bibfnamefont {C.~V.}\ \bibnamefont {Meinersen}}, \bibinfo {author} {\bibfnamefont {I.~F.}\ \bibnamefont {de~Fuentes}}, \bibinfo {author} {\bibfnamefont {K.}~\bibnamefont {Capannelli}}, \bibinfo {author} {\bibfnamefont {M.}~\bibnamefont {{Rimbach-Russ}}}, \bibinfo {author} {\bibfnamefont {G.}~\bibnamefont {Scappucci}}, \bibinfo {author} {\bibfnamefont {L.~M.~K.}\ \bibnamefont {Vandersypen}},\ and\ \bibinfo {author} {\bibfnamefont {E.}~\bibnamefont {Greplova}},\ }\href {https://doi.org/10.48550/arXiv.2503.12256} {\bibinfo {title} {Unified evolutionary optimization for high-fidelity spin qubit operations}} (\bibinfo {year} {2025}),\ \Eprint
  {https://arxiv.org/abs/2503.12256} {2503.12256} \BibitemShut {NoStop}%
\bibitem [{\citenamefont {Bergmann}\ \emph {et~al.}(1998)\citenamefont {Bergmann}, \citenamefont {Theuer},\ and\ \citenamefont {Shore}}]{bergmannCoherentPopulationTransfer1998}%
  \BibitemOpen
  \bibfield  {author} {\bibinfo {author} {\bibfnamefont {K.}~\bibnamefont {Bergmann}}, \bibinfo {author} {\bibfnamefont {H.}~\bibnamefont {Theuer}},\ and\ \bibinfo {author} {\bibfnamefont {B.~W.}\ \bibnamefont {Shore}},\ }\bibfield  {title} {\bibinfo {title} {Coherent population transfer among quantum states of atoms and molecules},\ }\href {https://doi.org/10.1103/RevModPhys.70.1003} {\bibfield  {journal} {\bibinfo  {journal} {Reviews of Modern Physics}\ }\textbf {\bibinfo {volume} {70}},\ \bibinfo {pages} {1003} (\bibinfo {year} {1998})}\BibitemShut {NoStop}%
\bibitem [{\citenamefont {Motzoi}\ \emph {et~al.}(2009)\citenamefont {Motzoi}, \citenamefont {Gambetta}, \citenamefont {Rebentrost},\ and\ \citenamefont {Wilhelm}}]{motzoiSimplePulsesElimination2009}%
  \BibitemOpen
  \bibfield  {author} {\bibinfo {author} {\bibfnamefont {F.}~\bibnamefont {Motzoi}}, \bibinfo {author} {\bibfnamefont {J.~M.}\ \bibnamefont {Gambetta}}, \bibinfo {author} {\bibfnamefont {P.}~\bibnamefont {Rebentrost}},\ and\ \bibinfo {author} {\bibfnamefont {F.~K.}\ \bibnamefont {Wilhelm}},\ }\bibfield  {title} {\bibinfo {title} {Simple {{Pulses}} for {{Elimination}} of {{Leakage}} in {{Weakly Nonlinear Qubits}}},\ }\href {https://doi.org/10.1103/PhysRevLett.103.110501} {\bibfield  {journal} {\bibinfo  {journal} {Physical Review Letters}\ }\textbf {\bibinfo {volume} {103}},\ \bibinfo {pages} {110501} (\bibinfo {year} {2009})}\BibitemShut {NoStop}%
\bibitem [{\citenamefont {Berry}(2009)}]{berryTransitionlessQuantumDriving2009}%
  \BibitemOpen
  \bibfield  {author} {\bibinfo {author} {\bibfnamefont {M.~V.}\ \bibnamefont {Berry}},\ }\bibfield  {title} {\bibinfo {title} {Transitionless quantum driving},\ }\href {https://doi.org/10.1088/1751-8113/42/36/365303} {\bibfield  {journal} {\bibinfo  {journal} {Journal of Physics A: Mathematical and Theoretical}\ }\textbf {\bibinfo {volume} {42}},\ \bibinfo {pages} {365303} (\bibinfo {year} {2009})}\BibitemShut {NoStop}%
\bibitem [{\citenamefont {De~Grandi}\ and\ \citenamefont {Polkovnikov}(2010)}]{degrandiAdiabaticPerturbationTheory2010}%
  \BibitemOpen
  \bibfield  {author} {\bibinfo {author} {\bibfnamefont {C.}~\bibnamefont {De~Grandi}}\ and\ \bibinfo {author} {\bibfnamefont {A.}~\bibnamefont {Polkovnikov}},\ }\bibfield  {title} {\bibinfo {title} {Adiabatic perturbation theory: From {{Landau-Zener}} problem to quenching through a quantum critical point}\ }(\bibinfo  {publisher} {Springer, Berlin, Heidelberg},\ \bibinfo {year} {2010})\ pp.\ \bibinfo {pages} {75--114},\ \Eprint {https://arxiv.org/abs/0910.2236} {0910.2236} \BibitemShut {NoStop}%
\bibitem [{\citenamefont {Vutha}(2010)}]{vuthaSimpleApproachLandauZener2010}%
  \BibitemOpen
  \bibfield  {author} {\bibinfo {author} {\bibfnamefont {A.~C.}\ \bibnamefont {Vutha}},\ }\bibfield  {title} {\bibinfo {title} {A simple approach to the {{Landau-Zener}} formula},\ }\href {https://doi.org/10.1088/0143-0807/31/2/016} {\bibfield  {journal} {\bibinfo  {journal} {European Journal of Physics}\ }\textbf {\bibinfo {volume} {31}},\ \bibinfo {pages} {389} (\bibinfo {year} {2010})},\ \Eprint {https://arxiv.org/abs/1001.3322} {1001.3322} \BibitemShut {NoStop}%
\bibitem [{\citenamefont {Motzoi}\ \emph {et~al.}(2011)\citenamefont {Motzoi}, \citenamefont {Gambetta}, \citenamefont {Merkel},\ and\ \citenamefont {Wilhelm}}]{motzoiOptimalControlMethods2011}%
  \BibitemOpen
  \bibfield  {author} {\bibinfo {author} {\bibfnamefont {F.}~\bibnamefont {Motzoi}}, \bibinfo {author} {\bibfnamefont {J.~M.}\ \bibnamefont {Gambetta}}, \bibinfo {author} {\bibfnamefont {S.~T.}\ \bibnamefont {Merkel}},\ and\ \bibinfo {author} {\bibfnamefont {F.~K.}\ \bibnamefont {Wilhelm}},\ }\bibfield  {title} {\bibinfo {title} {Optimal control methods for rapidly time-varying {{Hamiltonians}}},\ }\href {https://doi.org/10.1103/PhysRevA.84.022307} {\bibfield  {journal} {\bibinfo  {journal} {Physical Review A}\ }\textbf {\bibinfo {volume} {84}},\ \bibinfo {pages} {022307} (\bibinfo {year} {2011})}\BibitemShut {NoStop}%
\bibitem [{\citenamefont {Chen}\ \emph {et~al.}(2011)\citenamefont {Chen}, \citenamefont {Torrontegui},\ and\ \citenamefont {Muga}}]{chenLewisRiesenfeldInvariantsTransitionless2011}%
  \BibitemOpen
  \bibfield  {author} {\bibinfo {author} {\bibfnamefont {X.}~\bibnamefont {Chen}}, \bibinfo {author} {\bibfnamefont {E.}~\bibnamefont {Torrontegui}},\ and\ \bibinfo {author} {\bibfnamefont {J.~G.}\ \bibnamefont {Muga}},\ }\bibfield  {title} {\bibinfo {title} {Lewis-{{Riesenfeld}} invariants and transitionless quantum driving},\ }\href {https://doi.org/10.1103/PhysRevA.83.062116} {\bibfield  {journal} {\bibinfo  {journal} {Physical Review A}\ }\textbf {\bibinfo {volume} {83}},\ \bibinfo {pages} {062116} (\bibinfo {year} {2011})}\BibitemShut {NoStop}%
\bibitem [{\citenamefont {Ribeiro}\ \emph {et~al.}(2017)\citenamefont {Ribeiro}, \citenamefont {Baksic},\ and\ \citenamefont {Clerk}}]{ribeiroSystematicMagnusBasedApproach2017}%
  \BibitemOpen
  \bibfield  {author} {\bibinfo {author} {\bibfnamefont {H.}~\bibnamefont {Ribeiro}}, \bibinfo {author} {\bibfnamefont {A.}~\bibnamefont {Baksic}},\ and\ \bibinfo {author} {\bibfnamefont {A.~A.}\ \bibnamefont {Clerk}},\ }\bibfield  {title} {\bibinfo {title} {Systematic {{Magnus-Based Approach}} for {{Suppressing Leakage}} and {{Nonadiabatic Errors}} in {{Quantum Dynamics}}},\ }\href {https://doi.org/10.1103/PhysRevX.7.011021} {\bibfield  {journal} {\bibinfo  {journal} {Physical Review X}\ }\textbf {\bibinfo {volume} {7}},\ \bibinfo {pages} {011021} (\bibinfo {year} {2017})}\BibitemShut {NoStop}%
\bibitem [{\citenamefont {Sels}\ and\ \citenamefont {Polkovnikov}(2017)}]{selsMinimizingIrreversibleLosses2017}%
  \BibitemOpen
  \bibfield  {author} {\bibinfo {author} {\bibfnamefont {D.}~\bibnamefont {Sels}}\ and\ \bibinfo {author} {\bibfnamefont {A.}~\bibnamefont {Polkovnikov}},\ }\bibfield  {title} {\bibinfo {title} {Minimizing irreversible losses in quantum systems by local counterdiabatic driving},\ }\href {https://doi.org/10.1073/pnas.1619826114} {\bibfield  {journal} {\bibinfo  {journal} {Proceedings of the National Academy of Sciences}\ }\textbf {\bibinfo {volume} {114}},\ \bibinfo {pages} {E3909} (\bibinfo {year} {2017})}\BibitemShut {NoStop}%
\bibitem [{\citenamefont {Takahashi}(2019)}]{takahashiHamiltonianEngineeringAdiabatic2019}%
  \BibitemOpen
  \bibfield  {author} {\bibinfo {author} {\bibfnamefont {K.}~\bibnamefont {Takahashi}},\ }\bibfield  {title} {\bibinfo {title} {Hamiltonian {{Engineering}} for {{Adiabatic Quantum Computation}}: {{Lessons}} from {{Shortcuts}} to {{Adiabaticity}}},\ }\href {https://doi.org/10.7566/JPSJ.88.061002} {\bibfield  {journal} {\bibinfo  {journal} {Journal of the Physical Society of Japan}\ }\textbf {\bibinfo {volume} {88}},\ \bibinfo {pages} {061002} (\bibinfo {year} {2019})}\BibitemShut {NoStop}%
\bibitem [{\citenamefont {Ban}\ \emph {et~al.}(2019)\citenamefont {Ban}, \citenamefont {Chen}, \citenamefont {Kohler},\ and\ \citenamefont {Platero}}]{banSpinEntangledState2019}%
  \BibitemOpen
  \bibfield  {author} {\bibinfo {author} {\bibfnamefont {Y.}~\bibnamefont {Ban}}, \bibinfo {author} {\bibfnamefont {X.}~\bibnamefont {Chen}}, \bibinfo {author} {\bibfnamefont {S.}~\bibnamefont {Kohler}},\ and\ \bibinfo {author} {\bibfnamefont {G.}~\bibnamefont {Platero}},\ }\bibfield  {title} {\bibinfo {title} {Spin {{Entangled State Transfer}} in {{Quantum Dot Arrays}}: {{Coherent Adiabatic}} and {{Speed-Up Protocols}}},\ }\href {https://doi.org/10.1002/qute.201900048} {\bibfield  {journal} {\bibinfo  {journal} {Advanced Quantum Technologies}\ }\textbf {\bibinfo {volume} {2}},\ \bibinfo {pages} {1900048} (\bibinfo {year} {2019})}\BibitemShut {NoStop}%
\bibitem [{\citenamefont {{Gu{\'e}ry-Odelin}}\ \emph {et~al.}(2019)\citenamefont {{Gu{\'e}ry-Odelin}}, \citenamefont {Ruschhaupt}, \citenamefont {Kiely}, \citenamefont {Torrontegui}, \citenamefont {{Mart{\'i}nez-Garaot}},\ and\ \citenamefont {Muga}}]{guery-odelinShortcutsAdiabaticityConcepts2019}%
  \BibitemOpen
  \bibfield  {author} {\bibinfo {author} {\bibfnamefont {D.}~\bibnamefont {{Gu{\'e}ry-Odelin}}}, \bibinfo {author} {\bibfnamefont {A.}~\bibnamefont {Ruschhaupt}}, \bibinfo {author} {\bibfnamefont {A.}~\bibnamefont {Kiely}}, \bibinfo {author} {\bibfnamefont {E.}~\bibnamefont {Torrontegui}}, \bibinfo {author} {\bibfnamefont {S.}~\bibnamefont {{Mart{\'i}nez-Garaot}}},\ and\ \bibinfo {author} {\bibfnamefont {J.~G.}\ \bibnamefont {Muga}},\ }\bibfield  {title} {\bibinfo {title} {Shortcuts to adiabaticity: {{Concepts}}, methods, and applications},\ }\href {https://doi.org/10.1103/RevModPhys.91.045001} {\bibfield  {journal} {\bibinfo  {journal} {Reviews of Modern Physics}\ }\textbf {\bibinfo {volume} {91}},\ \bibinfo {pages} {045001} (\bibinfo {year} {2019})}\BibitemShut {NoStop}%
\bibitem [{\citenamefont {Setiawan}\ \emph {et~al.}(2021)\citenamefont {Setiawan}, \citenamefont {Groszkowski}, \citenamefont {Ribeiro},\ and\ \citenamefont {Clerk}}]{setiawanAnalyticDesignAccelerated2021}%
  \BibitemOpen
  \bibfield  {author} {\bibinfo {author} {\bibfnamefont {F.}~\bibnamefont {Setiawan}}, \bibinfo {author} {\bibfnamefont {P.}~\bibnamefont {Groszkowski}}, \bibinfo {author} {\bibfnamefont {H.}~\bibnamefont {Ribeiro}},\ and\ \bibinfo {author} {\bibfnamefont {A.~A.}\ \bibnamefont {Clerk}},\ }\bibfield  {title} {\bibinfo {title} {Analytic {{Design}} of {{Accelerated Adiabatic Gates}} in {{Realistic Qubits}}: {{General Theory}} and {{Applications}} to {{Superconducting Circuits}}},\ }\href {https://doi.org/10.1103/PRXQuantum.2.030306} {\bibfield  {journal} {\bibinfo  {journal} {PRX Quantum}\ }\textbf {\bibinfo {volume} {2}},\ \bibinfo {pages} {030306} (\bibinfo {year} {2021})}\BibitemShut {NoStop}%
\bibitem [{\citenamefont {Zhuang}\ \emph {et~al.}(2022)\citenamefont {Zhuang}, \citenamefont {Zeng}, \citenamefont {Economou},\ and\ \citenamefont {Barnes}}]{zhuangNoiseresistantLandauZenerSweeps2022}%
  \BibitemOpen
  \bibfield  {author} {\bibinfo {author} {\bibfnamefont {F.}~\bibnamefont {Zhuang}}, \bibinfo {author} {\bibfnamefont {J.}~\bibnamefont {Zeng}}, \bibinfo {author} {\bibfnamefont {S.~E.}\ \bibnamefont {Economou}},\ and\ \bibinfo {author} {\bibfnamefont {E.}~\bibnamefont {Barnes}},\ }\bibfield  {title} {\bibinfo {title} {Noise-resistant {{Landau-Zener}} sweeps from geometrical curves},\ }\href {https://doi.org/10.22331/q-2022-02-02-639} {\bibfield  {journal} {\bibinfo  {journal} {Quantum}\ }\textbf {\bibinfo {volume} {6}},\ \bibinfo {pages} {639} (\bibinfo {year} {2022})}\BibitemShut {NoStop}%
\bibitem [{\citenamefont {Takahashi}(2022)}]{takahashiDynamicalInvariantFormalism2022}%
  \BibitemOpen
  \bibfield  {author} {\bibinfo {author} {\bibfnamefont {K.}~\bibnamefont {Takahashi}},\ }\bibfield  {title} {\bibinfo {title} {Dynamical invariant formalism of shortcuts to adiabaticity},\ }\href {https://doi.org/10.1098/rsta.2022.0301} {\bibfield  {journal} {\bibinfo  {journal} {Philosophical Transactions of the Royal Society A: Mathematical, Physical and Engineering Sciences}\ }\textbf {\bibinfo {volume} {380}},\ \bibinfo {pages} {20220301} (\bibinfo {year} {2022})}\BibitemShut {NoStop}%
\bibitem [{\citenamefont {Glasbrenner}\ and\ \citenamefont {Schleich}(2023)}]{glasbrennerLandauZenerFormula2023}%
  \BibitemOpen
  \bibfield  {author} {\bibinfo {author} {\bibfnamefont {E.~P.}\ \bibnamefont {Glasbrenner}}\ and\ \bibinfo {author} {\bibfnamefont {W.~P.}\ \bibnamefont {Schleich}},\ }\bibfield  {title} {\bibinfo {title} {The {{Landau}}--{{Zener}} formula made simple},\ }\href {https://doi.org/10.1088/1361-6455/acc774} {\bibfield  {journal} {\bibinfo  {journal} {Journal of Physics B: Atomic, Molecular and Optical Physics}\ }\textbf {\bibinfo {volume} {56}},\ \bibinfo {pages} {104001} (\bibinfo {year} {2023})}\BibitemShut {NoStop}%
\bibitem [{\citenamefont {{Rimbach-Russ}}\ \emph {et~al.}(2023)\citenamefont {{Rimbach-Russ}}, \citenamefont {Philips}, \citenamefont {Xue},\ and\ \citenamefont {Vandersypen}}]{rimbach-russSimpleFrameworkSystematic2023}%
  \BibitemOpen
  \bibfield  {author} {\bibinfo {author} {\bibfnamefont {M.}~\bibnamefont {{Rimbach-Russ}}}, \bibinfo {author} {\bibfnamefont {S.~G.~J.}\ \bibnamefont {Philips}}, \bibinfo {author} {\bibfnamefont {X.}~\bibnamefont {Xue}},\ and\ \bibinfo {author} {\bibfnamefont {L.~M.~K.}\ \bibnamefont {Vandersypen}},\ }\bibfield  {title} {\bibinfo {title} {Simple framework for systematic high-fidelity gate operations},\ }\href {https://doi.org/10.1088/2058-9565/acf786} {\bibfield  {journal} {\bibinfo  {journal} {Quantum Science and Technology}\ }\textbf {\bibinfo {volume} {8}},\ \bibinfo {pages} {045025} (\bibinfo {year} {2023})}\BibitemShut {NoStop}%
\bibitem [{\citenamefont {Dengis}\ \emph {et~al.}(2025)\citenamefont {Dengis}, \citenamefont {Wimberger},\ and\ \citenamefont {Schlagheck}}]{dengisAcceleratedCreationNOON2025}%
  \BibitemOpen
  \bibfield  {author} {\bibinfo {author} {\bibfnamefont {S.}~\bibnamefont {Dengis}}, \bibinfo {author} {\bibfnamefont {S.}~\bibnamefont {Wimberger}},\ and\ \bibinfo {author} {\bibfnamefont {P.}~\bibnamefont {Schlagheck}},\ }\bibfield  {title} {\bibinfo {title} {Accelerated creation of {{NOON}} states with ultracold atoms via counterdiabatic driving},\ }\href {https://doi.org/10.1103/PhysRevA.111.L031301} {\bibfield  {journal} {\bibinfo  {journal} {Physical Review A}\ }\textbf {\bibinfo {volume} {111}},\ \bibinfo {pages} {L031301} (\bibinfo {year} {2025})}\BibitemShut {NoStop}%
\bibitem [{\citenamefont {Romero}\ \emph {et~al.}(2024)\citenamefont {Romero}, \citenamefont {Chen}, \citenamefont {Platero},\ and\ \citenamefont {Ban}}]{romeroOptimizingEdgestateTransfer2024}%
  \BibitemOpen
  \bibfield  {author} {\bibinfo {author} {\bibfnamefont {S.~V.}\ \bibnamefont {Romero}}, \bibinfo {author} {\bibfnamefont {X.}~\bibnamefont {Chen}}, \bibinfo {author} {\bibfnamefont {G.}~\bibnamefont {Platero}},\ and\ \bibinfo {author} {\bibfnamefont {Y.}~\bibnamefont {Ban}},\ }\bibfield  {title} {\bibinfo {title} {Optimizing edge-state transfer in a {{Su-Schrieffer-Heeger}} chain via hybrid analog-digital strategies},\ }\href {https://doi.org/10.1103/PhysRevApplied.21.034033} {\bibfield  {journal} {\bibinfo  {journal} {Physical Review Applied}\ }\textbf {\bibinfo {volume} {21}},\ \bibinfo {pages} {034033} (\bibinfo {year} {2024})}\BibitemShut {NoStop}%
\bibitem [{\citenamefont {Liu}\ \emph {et~al.}(2024)\citenamefont {Liu}, \citenamefont {Matsumoto}, \citenamefont {Fujita}, \citenamefont {Ludwig}, \citenamefont {Wieck},\ and\ \citenamefont {Oiwa}}]{liuAcceleratedAdiabaticPassage2024}%
  \BibitemOpen
  \bibfield  {author} {\bibinfo {author} {\bibfnamefont {X.-F.}\ \bibnamefont {Liu}}, \bibinfo {author} {\bibfnamefont {Y.}~\bibnamefont {Matsumoto}}, \bibinfo {author} {\bibfnamefont {T.}~\bibnamefont {Fujita}}, \bibinfo {author} {\bibfnamefont {A.}~\bibnamefont {Ludwig}}, \bibinfo {author} {\bibfnamefont {A.~D.}\ \bibnamefont {Wieck}},\ and\ \bibinfo {author} {\bibfnamefont {A.}~\bibnamefont {Oiwa}},\ }\bibfield  {title} {\bibinfo {title} {Accelerated adiabatic passage of a single electron spin qubit in quantum dots},\ }\href {https://doi.org/10.1103/PhysRevLett.132.027002} {\bibfield  {journal} {\bibinfo  {journal} {Physical Review Letters}\ }\textbf {\bibinfo {volume} {132}},\ \bibinfo {pages} {027002} (\bibinfo {year} {2024})}\BibitemShut {NoStop}%
\bibitem [{\citenamefont {Xu}\ \emph {et~al.}(2019)\citenamefont {Xu}, \citenamefont {Du},\ and\ \citenamefont {Huang}}]{xuImprovingCoherentPopulation2019}%
  \BibitemOpen
  \bibfield  {author} {\bibinfo {author} {\bibfnamefont {J.}~\bibnamefont {Xu}}, \bibinfo {author} {\bibfnamefont {Y.-X.}\ \bibnamefont {Du}},\ and\ \bibinfo {author} {\bibfnamefont {W.}~\bibnamefont {Huang}},\ }\bibfield  {title} {\bibinfo {title} {Improving coherent population transfer via a stricter adiabatic condition},\ }\href {https://doi.org/10.1103/PhysRevA.100.023848} {\bibfield  {journal} {\bibinfo  {journal} {Physical Review A}\ }\textbf {\bibinfo {volume} {100}},\ \bibinfo {pages} {023848} (\bibinfo {year} {2019})}\BibitemShut {NoStop}%
\bibitem [{\citenamefont {Fehse}\ \emph {et~al.}(2023)\citenamefont {Fehse}, \citenamefont {David}, \citenamefont {{Pioro-Ladri{\`e}re}},\ and\ \citenamefont {Coish}}]{fehseGeneralizedFastQuasiadiabatic2023}%
  \BibitemOpen
  \bibfield  {author} {\bibinfo {author} {\bibfnamefont {F.}~\bibnamefont {Fehse}}, \bibinfo {author} {\bibfnamefont {M.}~\bibnamefont {David}}, \bibinfo {author} {\bibfnamefont {M.}~\bibnamefont {{Pioro-Ladri{\`e}re}}},\ and\ \bibinfo {author} {\bibfnamefont {W.~A.}\ \bibnamefont {Coish}},\ }\bibfield  {title} {\bibinfo {title} {Generalized fast quasiadiabatic population transfer for improved qubit readout, shuttling, and noise mitigation},\ }\href {https://doi.org/10.1103/PhysRevB.107.245303} {\bibfield  {journal} {\bibinfo  {journal} {Physical Review B}\ }\textbf {\bibinfo {volume} {107}},\ \bibinfo {pages} {245303} (\bibinfo {year} {2023})}\BibitemShut {NoStop}%
\bibitem [{\citenamefont {Lima}\ and\ \citenamefont {Burkard}(2024)}]{limaSuperadiabaticLandauZenerTransitions2024}%
  \BibitemOpen
  \bibfield  {author} {\bibinfo {author} {\bibfnamefont {J.~R.~F.}\ \bibnamefont {Lima}}\ and\ \bibinfo {author} {\bibfnamefont {G.}~\bibnamefont {Burkard}},\ }\href@noop {} {\bibinfo {title} {Superadiabatic {{Landau-Zener}} transitions}} (\bibinfo {year} {2024}),\ \Eprint {https://arxiv.org/abs/2408.03173} {2408.03173} \BibitemShut {NoStop}%
\bibitem [{\citenamefont {Richerme}(2013)}]{richermeExperimentalPerformanceQuantum2013}%
  \BibitemOpen
  \bibfield  {author} {\bibinfo {author} {\bibfnamefont {P.}~\bibnamefont {Richerme}},\ }\bibfield  {title} {\bibinfo {title} {Experimental performance of a quantum simulator: {{Optimizing}} adiabatic evolution and identifying many-body ground states},\ }\bibfield  {journal} {\bibinfo  {journal} {Physical Review A}\ }\textbf {\bibinfo {volume} {88}},\ \href {https://doi.org/10.1103/PhysRevA.88.012334} {10.1103/PhysRevA.88.012334} (\bibinfo {year} {2013})\BibitemShut {NoStop}%
\bibitem [{\citenamefont {Roland}(2002)}]{rolandQuantumSearchLocal2002}%
  \BibitemOpen
  \bibfield  {author} {\bibinfo {author} {\bibfnamefont {J.}~\bibnamefont {Roland}},\ }\bibfield  {title} {\bibinfo {title} {Quantum search by local adiabatic evolution},\ }\bibfield  {journal} {\bibinfo  {journal} {Physical Review A}\ }\textbf {\bibinfo {volume} {65}},\ \href {https://doi.org/10.1103/PhysRevA.65.042308} {10.1103/PhysRevA.65.042308} (\bibinfo {year} {2002})\BibitemShut {NoStop}%
\bibitem [{\citenamefont {{Mart{\'i}nez-Garaot}}\ \emph {et~al.}(2015)\citenamefont {{Mart{\'i}nez-Garaot}}, \citenamefont {Ruschhaupt}, \citenamefont {Gillet}, \citenamefont {Busch},\ and\ \citenamefont {Muga}}]{martinez-garaotFastQuasiadiabaticDynamics2015}%
  \BibitemOpen
  \bibfield  {author} {\bibinfo {author} {\bibfnamefont {S.}~\bibnamefont {{Mart{\'i}nez-Garaot}}}, \bibinfo {author} {\bibfnamefont {A.}~\bibnamefont {Ruschhaupt}}, \bibinfo {author} {\bibfnamefont {J.}~\bibnamefont {Gillet}}, \bibinfo {author} {\bibfnamefont {{\relax Th}.}~\bibnamefont {Busch}},\ and\ \bibinfo {author} {\bibfnamefont {J.~G.}\ \bibnamefont {Muga}},\ }\bibfield  {title} {\bibinfo {title} {Fast quasiadiabatic dynamics},\ }\href {https://doi.org/10.1103/PhysRevA.92.043406} {\bibfield  {journal} {\bibinfo  {journal} {Physical Review A}\ }\textbf {\bibinfo {volume} {92}},\ \bibinfo {pages} {043406} (\bibinfo {year} {2015})}\BibitemShut {NoStop}%
\bibitem [{\citenamefont {Chen}(2022)}]{chenSpeedingQuantumAdiabatic2022}%
  \BibitemOpen
  \bibfield  {author} {\bibinfo {author} {\bibfnamefont {J.-F.}\ \bibnamefont {Chen}},\ }\bibfield  {title} {\bibinfo {title} {Speeding up quantum adiabatic processes with a dynamical quantum geometric tensor},\ }\href {https://doi.org/10.1103/PhysRevResearch.4.023252} {\bibfield  {journal} {\bibinfo  {journal} {Physical Review Research}\ }\textbf {\bibinfo {volume} {4}},\ \bibinfo {pages} {023252} (\bibinfo {year} {2022})}\BibitemShut {NoStop}%
\bibitem [{\citenamefont {{Fern{\'a}ndez-Fern{\'a}ndez}}\ \emph {et~al.}(2022)\citenamefont {{Fern{\'a}ndez-Fern{\'a}ndez}}, \citenamefont {Ban},\ and\ \citenamefont {Platero}}]{fernandez-fernandezQuantumControlHole2022}%
  \BibitemOpen
  \bibfield  {author} {\bibinfo {author} {\bibfnamefont {D.}~\bibnamefont {{Fern{\'a}ndez-Fern{\'a}ndez}}}, \bibinfo {author} {\bibfnamefont {Y.}~\bibnamefont {Ban}},\ and\ \bibinfo {author} {\bibfnamefont {G.}~\bibnamefont {Platero}},\ }\bibfield  {title} {\bibinfo {title} {Quantum {{Control}} of {{Hole Spin Qubits}} in {{Double Quantum Dots}}},\ }\href {https://doi.org/10.1103/PhysRevApplied.18.054090} {\bibfield  {journal} {\bibinfo  {journal} {Physical Review Applied}\ }\textbf {\bibinfo {volume} {18}},\ \bibinfo {pages} {054090} (\bibinfo {year} {2022})}\BibitemShut {NoStop}%
\bibitem [{\citenamefont {Meinersen}\ \emph {et~al.}(2024)\citenamefont {Meinersen}, \citenamefont {Bosco},\ and\ \citenamefont {{Rimbach-Russ}}}]{meinersenQuantumGeometricProtocols2024}%
  \BibitemOpen
  \bibfield  {author} {\bibinfo {author} {\bibfnamefont {C.~V.}\ \bibnamefont {Meinersen}}, \bibinfo {author} {\bibfnamefont {S.}~\bibnamefont {Bosco}},\ and\ \bibinfo {author} {\bibfnamefont {M.}~\bibnamefont {{Rimbach-Russ}}},\ }\href {https://doi.org/10.48550/arXiv.2409.03084} {\bibinfo {title} {Quantum geometric protocols for fast high-fidelity adiabatic state transfer}} (\bibinfo {year} {2024}),\ \Eprint {https://arxiv.org/abs/2409.03084} {2409.03084} \BibitemShut {NoStop}%
\bibitem [{\citenamefont {Kolodrubetz}\ \emph {et~al.}(2017)\citenamefont {Kolodrubetz}, \citenamefont {Sels}, \citenamefont {Mehta},\ and\ \citenamefont {Polkovnikov}}]{kolodrubetzGeometryNonadiabaticResponse2017}%
  \BibitemOpen
  \bibfield  {author} {\bibinfo {author} {\bibfnamefont {M.}~\bibnamefont {Kolodrubetz}}, \bibinfo {author} {\bibfnamefont {D.}~\bibnamefont {Sels}}, \bibinfo {author} {\bibfnamefont {P.}~\bibnamefont {Mehta}},\ and\ \bibinfo {author} {\bibfnamefont {A.}~\bibnamefont {Polkovnikov}},\ }\bibfield  {title} {\bibinfo {title} {Geometry and non-adiabatic response in quantum and classical systems},\ }\href {https://doi.org/10.1016/j.physrep.2017.07.001} {\bibfield  {journal} {\bibinfo  {journal} {Physics Reports}\ }\bibinfo {series} {Geometry and Non-Adiabatic Response in Quantum and Classical Systems},\ \textbf {\bibinfo {volume} {697}},\ \bibinfo {pages} {1} (\bibinfo {year} {2017})}\BibitemShut {NoStop}%
\bibitem [{\citenamefont {Bukov}\ \emph {et~al.}(2019)\citenamefont {Bukov}, \citenamefont {Sels},\ and\ \citenamefont {Polkovnikov}}]{bukovGeometricSpeedLimit2019}%
  \BibitemOpen
  \bibfield  {author} {\bibinfo {author} {\bibfnamefont {M.}~\bibnamefont {Bukov}}, \bibinfo {author} {\bibfnamefont {D.}~\bibnamefont {Sels}},\ and\ \bibinfo {author} {\bibfnamefont {A.}~\bibnamefont {Polkovnikov}},\ }\bibfield  {title} {\bibinfo {title} {Geometric {{Speed Limit}} of {{Accessible Many-Body State Preparation}}},\ }\href {https://doi.org/10.1103/PhysRevX.9.011034} {\bibfield  {journal} {\bibinfo  {journal} {Physical Review X}\ }\textbf {\bibinfo {volume} {9}},\ \bibinfo {pages} {011034} (\bibinfo {year} {2019})}\BibitemShut {NoStop}%
\bibitem [{\citenamefont {Liska}\ and\ \citenamefont {Gritsev}(2021)}]{liskaHiddenSymmetriesBianchi2021}%
  \BibitemOpen
  \bibfield  {author} {\bibinfo {author} {\bibfnamefont {D.}~\bibnamefont {Liska}}\ and\ \bibinfo {author} {\bibfnamefont {V.}~\bibnamefont {Gritsev}},\ }\bibfield  {title} {\bibinfo {title} {Hidden symmetries, the {{Bianchi}} classification and geodesics of the quantum geometric ground-state manifolds},\ }\href {https://doi.org/10.21468/SciPostPhys.10.1.020} {\bibfield  {journal} {\bibinfo  {journal} {SciPost Physics}\ }\textbf {\bibinfo {volume} {10}},\ \bibinfo {pages} {020} (\bibinfo {year} {2021})}\BibitemShut {NoStop}%
\bibitem [{\citenamefont {De~Smet}\ \emph {et~al.}(2024)\citenamefont {De~Smet}, \citenamefont {Matsumoto}, \citenamefont {Zwerver}, \citenamefont {Tryputen}, \citenamefont {{de Snoo}}, \citenamefont {Amitonov}, \citenamefont {Sammak}, \citenamefont {Samkharadze}, \citenamefont {G{\"u}l}, \citenamefont {Wasserman}, \citenamefont {{Rimbach-Russ}}, \citenamefont {Scappucci},\ and\ \citenamefont {Vandersypen}}]{desmetHighfidelitySinglespinShuttling2024}%
  \BibitemOpen
  \bibfield  {author} {\bibinfo {author} {\bibfnamefont {M.}~\bibnamefont {De~Smet}}, \bibinfo {author} {\bibfnamefont {Y.}~\bibnamefont {Matsumoto}}, \bibinfo {author} {\bibfnamefont {A.-M.~J.}\ \bibnamefont {Zwerver}}, \bibinfo {author} {\bibfnamefont {L.}~\bibnamefont {Tryputen}}, \bibinfo {author} {\bibfnamefont {S.~L.}\ \bibnamefont {{de Snoo}}}, \bibinfo {author} {\bibfnamefont {S.~V.}\ \bibnamefont {Amitonov}}, \bibinfo {author} {\bibfnamefont {A.}~\bibnamefont {Sammak}}, \bibinfo {author} {\bibfnamefont {N.}~\bibnamefont {Samkharadze}}, \bibinfo {author} {\bibfnamefont {{\"O}.}~\bibnamefont {G{\"u}l}}, \bibinfo {author} {\bibfnamefont {R.~N.~M.}\ \bibnamefont {Wasserman}}, \bibinfo {author} {\bibfnamefont {M.}~\bibnamefont {{Rimbach-Russ}}}, \bibinfo {author} {\bibfnamefont {G.}~\bibnamefont {Scappucci}},\ and\ \bibinfo {author} {\bibfnamefont {L.~M.~K.}\ \bibnamefont {Vandersypen}},\ }\href@noop {} {\bibinfo {title} {High-fidelity single-spin shuttling in silicon}} (\bibinfo {year} {2024}),\ \Eprint
  {https://arxiv.org/abs/2406.07267} {2406.07267} \BibitemShut {NoStop}%
\bibitem [{\citenamefont {{van Riggelen-Doelman}}\ \emph {et~al.}(2024)\citenamefont {{van Riggelen-Doelman}}, \citenamefont {Wang}, \citenamefont {{de Snoo}}, \citenamefont {Lawrie}, \citenamefont {Hendrickx}, \citenamefont {{Rimbach-Russ}}, \citenamefont {Sammak}, \citenamefont {Scappucci}, \citenamefont {D{\'e}prez},\ and\ \citenamefont {Veldhorst}}]{vanriggelen-doelmanCoherentSpinQubit2024}%
  \BibitemOpen
  \bibfield  {author} {\bibinfo {author} {\bibfnamefont {F.}~\bibnamefont {{van Riggelen-Doelman}}}, \bibinfo {author} {\bibfnamefont {C.-A.}\ \bibnamefont {Wang}}, \bibinfo {author} {\bibfnamefont {S.~L.}\ \bibnamefont {{de Snoo}}}, \bibinfo {author} {\bibfnamefont {W.~I.~L.}\ \bibnamefont {Lawrie}}, \bibinfo {author} {\bibfnamefont {N.~W.}\ \bibnamefont {Hendrickx}}, \bibinfo {author} {\bibfnamefont {M.}~\bibnamefont {{Rimbach-Russ}}}, \bibinfo {author} {\bibfnamefont {A.}~\bibnamefont {Sammak}}, \bibinfo {author} {\bibfnamefont {G.}~\bibnamefont {Scappucci}}, \bibinfo {author} {\bibfnamefont {C.}~\bibnamefont {D{\'e}prez}},\ and\ \bibinfo {author} {\bibfnamefont {M.}~\bibnamefont {Veldhorst}},\ }\bibfield  {title} {\bibinfo {title} {Coherent spin qubit shuttling through germanium quantum dots},\ }\href {https://doi.org/10.1038/s41467-024-49358-y} {\bibfield  {journal} {\bibinfo  {journal} {Nature Communications}\ }\textbf {\bibinfo {volume} {15}},\ \bibinfo {pages} {1} (\bibinfo {year} {2024})}\BibitemShut
  {NoStop}%
\bibitem [{\citenamefont {{Fern{\'a}ndez-Fern{\'a}ndez}}\ \emph {et~al.}(2024)\citenamefont {{Fern{\'a}ndez-Fern{\'a}ndez}}, \citenamefont {Ban},\ and\ \citenamefont {Platero}}]{fernandez-fernandezFlyingSpinQubits2024}%
  \BibitemOpen
  \bibfield  {author} {\bibinfo {author} {\bibfnamefont {D.}~\bibnamefont {{Fern{\'a}ndez-Fern{\'a}ndez}}}, \bibinfo {author} {\bibfnamefont {Y.}~\bibnamefont {Ban}},\ and\ \bibinfo {author} {\bibfnamefont {G.}~\bibnamefont {Platero}},\ }\bibfield  {title} {\bibinfo {title} {Flying {{Spin Qubits}} in {{Quantum Dot Arrays Driven}} by {{Spin-Orbit Interaction}}},\ }\href {https://doi.org/10.22331/q-2024-11-21-1533} {\bibfield  {journal} {\bibinfo  {journal} {Quantum}\ }\textbf {\bibinfo {volume} {8}},\ \bibinfo {pages} {1533} (\bibinfo {year} {2024})},\ \Eprint {https://arxiv.org/abs/2312.04631} {2312.04631} \BibitemShut {NoStop}%
\bibitem [{\citenamefont {Ban}\ \emph {et~al.}(2012)\citenamefont {Ban}, \citenamefont {Chen}, \citenamefont {Sherman},\ and\ \citenamefont {Muga}}]{banFastRobustSpin2012}%
  \BibitemOpen
  \bibfield  {author} {\bibinfo {author} {\bibfnamefont {Y.}~\bibnamefont {Ban}}, \bibinfo {author} {\bibfnamefont {X.}~\bibnamefont {Chen}}, \bibinfo {author} {\bibfnamefont {E.~Y.}\ \bibnamefont {Sherman}},\ and\ \bibinfo {author} {\bibfnamefont {J.~G.}\ \bibnamefont {Muga}},\ }\bibfield  {title} {\bibinfo {title} {Fast and {{Robust Spin Manipulation}} in a {{Quantum Dot}} by {{Electric Fields}}},\ }\href {https://doi.org/10.1103/PhysRevLett.109.206602} {\bibfield  {journal} {\bibinfo  {journal} {Physical Review Letters}\ }\textbf {\bibinfo {volume} {109}},\ \bibinfo {pages} {206602} (\bibinfo {year} {2012})}\BibitemShut {NoStop}%
\bibitem [{\citenamefont {Nielsen}\ \emph {et~al.}(2006)\citenamefont {Nielsen}, \citenamefont {Dowling}, \citenamefont {Gu},\ and\ \citenamefont {Doherty}}]{nielsenOptimalControlGeometry2006}%
  \BibitemOpen
  \bibfield  {author} {\bibinfo {author} {\bibfnamefont {M.~A.}\ \bibnamefont {Nielsen}}, \bibinfo {author} {\bibfnamefont {M.~R.}\ \bibnamefont {Dowling}}, \bibinfo {author} {\bibfnamefont {M.}~\bibnamefont {Gu}},\ and\ \bibinfo {author} {\bibfnamefont {A.~C.}\ \bibnamefont {Doherty}},\ }\bibfield  {title} {\bibinfo {title} {Optimal control, geometry, and quantum computing},\ }\href {https://doi.org/10.1103/PhysRevA.73.062323} {\bibfield  {journal} {\bibinfo  {journal} {Physical Review A}\ }\textbf {\bibinfo {volume} {73}},\ \bibinfo {pages} {062323} (\bibinfo {year} {2006})}\BibitemShut {NoStop}%
\bibitem [{\citenamefont {Neupert}\ \emph {et~al.}(2013)\citenamefont {Neupert}, \citenamefont {Chamon},\ and\ \citenamefont {Mudry}}]{neupertMeasuringQuantumGeometry2013}%
  \BibitemOpen
  \bibfield  {author} {\bibinfo {author} {\bibfnamefont {T.}~\bibnamefont {Neupert}}, \bibinfo {author} {\bibfnamefont {C.}~\bibnamefont {Chamon}},\ and\ \bibinfo {author} {\bibfnamefont {C.}~\bibnamefont {Mudry}},\ }\bibfield  {title} {\bibinfo {title} {Measuring the quantum geometry of {{Bloch}} bands with current noise},\ }\href {https://doi.org/10.1103/PhysRevB.87.245103} {\bibfield  {journal} {\bibinfo  {journal} {Physical Review B}\ }\textbf {\bibinfo {volume} {87}},\ \bibinfo {pages} {245103} (\bibinfo {year} {2013})}\BibitemShut {NoStop}%
\bibitem [{\citenamefont {Kolodrubetz}\ \emph {et~al.}(2013)\citenamefont {Kolodrubetz}, \citenamefont {Gritsev},\ and\ \citenamefont {Polkovnikov}}]{kolodrubetzClassifyingMeasuringGeometry2013}%
  \BibitemOpen
  \bibfield  {author} {\bibinfo {author} {\bibfnamefont {M.}~\bibnamefont {Kolodrubetz}}, \bibinfo {author} {\bibfnamefont {V.}~\bibnamefont {Gritsev}},\ and\ \bibinfo {author} {\bibfnamefont {A.}~\bibnamefont {Polkovnikov}},\ }\bibfield  {title} {\bibinfo {title} {Classifying and measuring geometry of a quantum ground state manifold},\ }\href {https://doi.org/10.1103/PhysRevB.88.064304} {\bibfield  {journal} {\bibinfo  {journal} {Physical Review B}\ }\textbf {\bibinfo {volume} {88}},\ \bibinfo {pages} {064304} (\bibinfo {year} {2013})}\BibitemShut {NoStop}%
\bibitem [{\citenamefont {Lambert}\ and\ \citenamefont {S{\o}rensen}(2023)}]{lambertClassicalQuantumInformation2023}%
  \BibitemOpen
  \bibfield  {author} {\bibinfo {author} {\bibfnamefont {J.}~\bibnamefont {Lambert}}\ and\ \bibinfo {author} {\bibfnamefont {E.~S.}\ \bibnamefont {S{\o}rensen}},\ }\bibfield  {title} {\bibinfo {title} {From classical to quantum information geometry: A guide for physicists},\ }\href {https://doi.org/10.1088/1367-2630/aceb14} {\bibfield  {journal} {\bibinfo  {journal} {New Journal of Physics}\ }\textbf {\bibinfo {volume} {25}},\ \bibinfo {pages} {081201} (\bibinfo {year} {2023})}\BibitemShut {NoStop}%
\bibitem [{\citenamefont {Petiziol}\ \emph {et~al.}(2024)\citenamefont {Petiziol}, \citenamefont {Mintert},\ and\ \citenamefont {Wimberger}}]{petiziolQuantumControlEffective2024}%
  \BibitemOpen
  \bibfield  {author} {\bibinfo {author} {\bibfnamefont {F.}~\bibnamefont {Petiziol}}, \bibinfo {author} {\bibfnamefont {F.}~\bibnamefont {Mintert}},\ and\ \bibinfo {author} {\bibfnamefont {S.}~\bibnamefont {Wimberger}},\ }\bibfield  {title} {\bibinfo {title} {Quantum control by effective counterdiabatic driving},\ }\href {https://doi.org/10.1209/0295-5075/ad19e3} {\bibfield  {journal} {\bibinfo  {journal} {Europhysics Letters}\ }\textbf {\bibinfo {volume} {145}},\ \bibinfo {pages} {15001} (\bibinfo {year} {2024})},\ \Eprint {https://arxiv.org/abs/2402.04936} {2402.04936} \BibitemShut {NoStop}%
\bibitem [{\citenamefont {Sun}\ and\ \citenamefont {Zheng}(2019)}]{sunDistinctBoundQuantum2019}%
  \BibitemOpen
  \bibfield  {author} {\bibinfo {author} {\bibfnamefont {S.}~\bibnamefont {Sun}}\ and\ \bibinfo {author} {\bibfnamefont {Y.}~\bibnamefont {Zheng}},\ }\bibfield  {title} {\bibinfo {title} {Distinct {{Bound}} of the {{Quantum Speed Limit}} via the {{Gauge Invariant Distance}}},\ }\href {https://doi.org/10.1103/PhysRevLett.123.180403} {\bibfield  {journal} {\bibinfo  {journal} {Physical Review Letters}\ }\textbf {\bibinfo {volume} {123}},\ \bibinfo {pages} {180403} (\bibinfo {year} {2019})}\BibitemShut {NoStop}%
\bibitem [{\citenamefont {Sharipov}\ \emph {et~al.}(2024)\citenamefont {Sharipov}, \citenamefont {Tiutiakina}, \citenamefont {Gorsky}, \citenamefont {Gritsev},\ and\ \citenamefont {Polkovnikov}}]{sharipovHilbertSpaceGeometry2024}%
  \BibitemOpen
  \bibfield  {author} {\bibinfo {author} {\bibfnamefont {R.}~\bibnamefont {Sharipov}}, \bibinfo {author} {\bibfnamefont {A.}~\bibnamefont {Tiutiakina}}, \bibinfo {author} {\bibfnamefont {A.}~\bibnamefont {Gorsky}}, \bibinfo {author} {\bibfnamefont {V.}~\bibnamefont {Gritsev}},\ and\ \bibinfo {author} {\bibfnamefont {A.}~\bibnamefont {Polkovnikov}},\ }\href {https://doi.org/10.48550/arXiv.2411.11968} {\bibinfo {title} {Hilbert space geometry and quantum chaos}} (\bibinfo {year} {2024}),\ \Eprint {https://arxiv.org/abs/2411.11968} {2411.11968} \BibitemShut {NoStop}%
\bibitem [{\citenamefont {Cerfontaine}\ \emph {et~al.}(2021)\citenamefont {Cerfontaine}, \citenamefont {Hangleiter},\ and\ \citenamefont {Bluhm}}]{cerfontaineFilterFunctionsQuantum2021}%
  \BibitemOpen
  \bibfield  {author} {\bibinfo {author} {\bibfnamefont {P.}~\bibnamefont {Cerfontaine}}, \bibinfo {author} {\bibfnamefont {T.}~\bibnamefont {Hangleiter}},\ and\ \bibinfo {author} {\bibfnamefont {H.}~\bibnamefont {Bluhm}},\ }\bibfield  {title} {\bibinfo {title} {Filter {{Functions}} for {{Quantum Processes}} under {{Correlated Noise}}},\ }\href {https://doi.org/10.1103/PhysRevLett.127.170403} {\bibfield  {journal} {\bibinfo  {journal} {Physical Review Letters}\ }\textbf {\bibinfo {volume} {127}},\ \bibinfo {pages} {170403} (\bibinfo {year} {2021})}\BibitemShut {NoStop}%
\bibitem [{\citenamefont {Burkard}\ \emph {et~al.}(2023)\citenamefont {Burkard}, \citenamefont {Ladd}, \citenamefont {Pan}, \citenamefont {Nichol},\ and\ \citenamefont {Petta}}]{burkardSemiconductorSpinQubits2023}%
  \BibitemOpen
  \bibfield  {author} {\bibinfo {author} {\bibfnamefont {G.}~\bibnamefont {Burkard}}, \bibinfo {author} {\bibfnamefont {T.~D.}\ \bibnamefont {Ladd}}, \bibinfo {author} {\bibfnamefont {A.}~\bibnamefont {Pan}}, \bibinfo {author} {\bibfnamefont {J.~M.}\ \bibnamefont {Nichol}},\ and\ \bibinfo {author} {\bibfnamefont {J.~R.}\ \bibnamefont {Petta}},\ }\bibfield  {title} {\bibinfo {title} {Semiconductor spin qubits},\ }\href {https://doi.org/10.1103/RevModPhys.95.025003} {\bibfield  {journal} {\bibinfo  {journal} {Reviews of Modern Physics}\ }\textbf {\bibinfo {volume} {95}},\ \bibinfo {pages} {025003} (\bibinfo {year} {2023})}\BibitemShut {NoStop}%
\bibitem [{\citenamefont {Green}\ \emph {et~al.}(2013)\citenamefont {Green}, \citenamefont {Sastrawan}, \citenamefont {Uys},\ and\ \citenamefont {Biercuk}}]{greenArbitraryQuantumControl2013a}%
  \BibitemOpen
  \bibfield  {author} {\bibinfo {author} {\bibfnamefont {T.~J.}\ \bibnamefont {Green}}, \bibinfo {author} {\bibfnamefont {J.}~\bibnamefont {Sastrawan}}, \bibinfo {author} {\bibfnamefont {H.}~\bibnamefont {Uys}},\ and\ \bibinfo {author} {\bibfnamefont {M.~J.}\ \bibnamefont {Biercuk}},\ }\bibfield  {title} {\bibinfo {title} {Arbitrary quantum control of qubits in the presence of universal noise},\ }\href {https://doi.org/10.1088/1367-2630/15/9/095004} {\bibfield  {journal} {\bibinfo  {journal} {New Journal of Physics}\ }\textbf {\bibinfo {volume} {15}},\ \bibinfo {pages} {095004} (\bibinfo {year} {2013})}\BibitemShut {NoStop}%
\bibitem [{\citenamefont {Hansen}\ \emph {et~al.}(2023)\citenamefont {Hansen}, \citenamefont {Seedhouse}, \citenamefont {Saraiva}, \citenamefont {Dzurak},\ and\ \citenamefont {Yang}}]{hansenAccessingFullCapabilities2023}%
  \BibitemOpen
  \bibfield  {author} {\bibinfo {author} {\bibfnamefont {I.}~\bibnamefont {Hansen}}, \bibinfo {author} {\bibfnamefont {A.~E.}\ \bibnamefont {Seedhouse}}, \bibinfo {author} {\bibfnamefont {A.}~\bibnamefont {Saraiva}}, \bibinfo {author} {\bibfnamefont {A.~S.}\ \bibnamefont {Dzurak}},\ and\ \bibinfo {author} {\bibfnamefont {C.~H.}\ \bibnamefont {Yang}},\ }\bibfield  {title} {\bibinfo {title} {Accessing the full capabilities of filter functions: {{Tool}} for detailed noise and quantum control susceptibility analysis},\ }\href {https://doi.org/10.1103/PhysRevA.108.012426} {\bibfield  {journal} {\bibinfo  {journal} {Physical Review A}\ }\textbf {\bibinfo {volume} {108}},\ \bibinfo {pages} {012426} (\bibinfo {year} {2023})}\BibitemShut {NoStop}%
\bibitem [{\citenamefont {Barnes}\ \emph {et~al.}(2022)\citenamefont {Barnes}, \citenamefont {{Calderon-Vargas}}, \citenamefont {Dong}, \citenamefont {Li}, \citenamefont {Zeng},\ and\ \citenamefont {Zhuang}}]{barnesDynamicallyCorrectedGates2022}%
  \BibitemOpen
  \bibfield  {author} {\bibinfo {author} {\bibfnamefont {E.}~\bibnamefont {Barnes}}, \bibinfo {author} {\bibfnamefont {F.~A.}\ \bibnamefont {{Calderon-Vargas}}}, \bibinfo {author} {\bibfnamefont {W.}~\bibnamefont {Dong}}, \bibinfo {author} {\bibfnamefont {B.}~\bibnamefont {Li}}, \bibinfo {author} {\bibfnamefont {J.}~\bibnamefont {Zeng}},\ and\ \bibinfo {author} {\bibfnamefont {F.}~\bibnamefont {Zhuang}},\ }\bibfield  {title} {\bibinfo {title} {Dynamically corrected gates from geometric space curves},\ }\href {https://doi.org/10.1088/2058-9565/ac4421} {\bibfield  {journal} {\bibinfo  {journal} {Quantum Science and Technology}\ }\textbf {\bibinfo {volume} {7}},\ \bibinfo {pages} {023001} (\bibinfo {year} {2022})}\BibitemShut {NoStop}%
\bibitem [{\citenamefont {Rach}\ \emph {et~al.}(2015)\citenamefont {Rach}, \citenamefont {M{\"u}ller}, \citenamefont {Calarco},\ and\ \citenamefont {Montangero}}]{rachDressingChoppedrandombasisOptimization2015}%
  \BibitemOpen
  \bibfield  {author} {\bibinfo {author} {\bibfnamefont {N.}~\bibnamefont {Rach}}, \bibinfo {author} {\bibfnamefont {M.~M.}\ \bibnamefont {M{\"u}ller}}, \bibinfo {author} {\bibfnamefont {T.}~\bibnamefont {Calarco}},\ and\ \bibinfo {author} {\bibfnamefont {S.}~\bibnamefont {Montangero}},\ }\bibfield  {title} {\bibinfo {title} {Dressing the chopped-random-basis optimization: {{A}} bandwidth-limited access to the trap-free landscape},\ }\href {https://doi.org/10.1103/PhysRevA.92.062343} {\bibfield  {journal} {\bibinfo  {journal} {Physical Review A}\ }\textbf {\bibinfo {volume} {92}},\ \bibinfo {pages} {062343} (\bibinfo {year} {2015})}\BibitemShut {NoStop}%
\bibitem [{\citenamefont {Zanardi}\ and\ \citenamefont {Paunkovi{\'c}}(2006)}]{zanardiGroundStateOverlap2006}%
  \BibitemOpen
  \bibfield  {author} {\bibinfo {author} {\bibfnamefont {P.}~\bibnamefont {Zanardi}}\ and\ \bibinfo {author} {\bibfnamefont {N.}~\bibnamefont {Paunkovi{\'c}}},\ }\bibfield  {title} {\bibinfo {title} {Ground state overlap and quantum phase transitions},\ }\href {https://doi.org/10.1103/PhysRevE.74.031123} {\bibfield  {journal} {\bibinfo  {journal} {Physical Review E}\ }\textbf {\bibinfo {volume} {74}},\ \bibinfo {pages} {031123} (\bibinfo {year} {2006})}\BibitemShut {NoStop}%
\bibitem [{\citenamefont {Zanardi}\ \emph {et~al.}(2007)\citenamefont {Zanardi}, \citenamefont {Giorda},\ and\ \citenamefont {Cozzini}}]{zanardiInformationTheoreticDifferentialGeometry2007}%
  \BibitemOpen
  \bibfield  {author} {\bibinfo {author} {\bibfnamefont {P.}~\bibnamefont {Zanardi}}, \bibinfo {author} {\bibfnamefont {P.}~\bibnamefont {Giorda}},\ and\ \bibinfo {author} {\bibfnamefont {M.}~\bibnamefont {Cozzini}},\ }\bibfield  {title} {\bibinfo {title} {Information-{{Theoretic Differential Geometry}} of {{Quantum Phase Transitions}}},\ }\href {https://doi.org/10.1103/PhysRevLett.99.100603} {\bibfield  {journal} {\bibinfo  {journal} {Physical Review Letters}\ }\textbf {\bibinfo {volume} {99}},\ \bibinfo {pages} {100603} (\bibinfo {year} {2007})}\BibitemShut {NoStop}%
\bibitem [{\citenamefont {{Alvarez-Jimenez}}\ \emph {et~al.}(2017)\citenamefont {{Alvarez-Jimenez}}, \citenamefont {Dector},\ and\ \citenamefont {Vergara}}]{alvarez-jimenezQuantumInformationMetric2017}%
  \BibitemOpen
  \bibfield  {author} {\bibinfo {author} {\bibfnamefont {J.}~\bibnamefont {{Alvarez-Jimenez}}}, \bibinfo {author} {\bibfnamefont {A.}~\bibnamefont {Dector}},\ and\ \bibinfo {author} {\bibfnamefont {J.~D.}\ \bibnamefont {Vergara}},\ }\bibfield  {title} {\bibinfo {title} {Quantum information metric and {{Berry}} curvature from a {{Lagrangian}} approach},\ }\href {https://doi.org/10.1007/JHEP03(2017)044} {\bibfield  {journal} {\bibinfo  {journal} {Journal of High Energy Physics}\ }\textbf {\bibinfo {volume} {2017}},\ \bibinfo {pages} {44} (\bibinfo {year} {2017})}\BibitemShut {NoStop}%
\bibitem [{\citenamefont {Michielis}\ \emph {et~al.}(2023)\citenamefont {Michielis}, \citenamefont {Ferraro}, \citenamefont {Prati}, \citenamefont {Hutin}, \citenamefont {Bertrand}, \citenamefont {Charbon}, \citenamefont {Ibberson},\ and\ \citenamefont {{Gonzalez-Zalba}}}]{michielisSiliconSpinQubits2023}%
  \BibitemOpen
  \bibfield  {author} {\bibinfo {author} {\bibfnamefont {M.~D.}\ \bibnamefont {Michielis}}, \bibinfo {author} {\bibfnamefont {E.}~\bibnamefont {Ferraro}}, \bibinfo {author} {\bibfnamefont {E.}~\bibnamefont {Prati}}, \bibinfo {author} {\bibfnamefont {L.}~\bibnamefont {Hutin}}, \bibinfo {author} {\bibfnamefont {B.}~\bibnamefont {Bertrand}}, \bibinfo {author} {\bibfnamefont {E.}~\bibnamefont {Charbon}}, \bibinfo {author} {\bibfnamefont {D.~J.}\ \bibnamefont {Ibberson}},\ and\ \bibinfo {author} {\bibfnamefont {M.~F.}\ \bibnamefont {{Gonzalez-Zalba}}},\ }\bibfield  {title} {\bibinfo {title} {Silicon spin qubits from laboratory to industry},\ }\href {https://doi.org/10.1088/1361-6463/acd8c7} {\bibfield  {journal} {\bibinfo  {journal} {Journal of Physics D: Applied Physics}\ }\textbf {\bibinfo {volume} {56}},\ \bibinfo {pages} {363001} (\bibinfo {year} {2023})}\BibitemShut {NoStop}%
\bibitem [{\citenamefont {Wang}\ \emph {et~al.}(2023)\citenamefont {Wang}, \citenamefont {Rooney},\ and\ \citenamefont {Jiang}}]{wangAutomatedCharacterizationDouble2023}%
  \BibitemOpen
  \bibfield  {author} {\bibinfo {author} {\bibfnamefont {W.}~\bibnamefont {Wang}}, \bibinfo {author} {\bibfnamefont {J.~D.}\ \bibnamefont {Rooney}},\ and\ \bibinfo {author} {\bibfnamefont {H.}~\bibnamefont {Jiang}},\ }\href@noop {} {\bibinfo {title} {Automated {{Characterization}} of a {{Double Quantum Dot}} using the {{Hubbard Model}}}} (\bibinfo {year} {2023}),\ \Eprint {https://arxiv.org/abs/2309.03400} {2309.03400} \BibitemShut {NoStop}%
\bibitem [{\citenamefont {Xue}\ \emph {et~al.}(2022)\citenamefont {Xue}, \citenamefont {Russ}, \citenamefont {Samkharadze}, \citenamefont {Undseth}, \citenamefont {Sammak}, \citenamefont {Scappucci},\ and\ \citenamefont {Vandersypen}}]{xueQuantumLogicSpin2022}%
  \BibitemOpen
  \bibfield  {author} {\bibinfo {author} {\bibfnamefont {X.}~\bibnamefont {Xue}}, \bibinfo {author} {\bibfnamefont {M.}~\bibnamefont {Russ}}, \bibinfo {author} {\bibfnamefont {N.}~\bibnamefont {Samkharadze}}, \bibinfo {author} {\bibfnamefont {B.}~\bibnamefont {Undseth}}, \bibinfo {author} {\bibfnamefont {A.}~\bibnamefont {Sammak}}, \bibinfo {author} {\bibfnamefont {G.}~\bibnamefont {Scappucci}},\ and\ \bibinfo {author} {\bibfnamefont {L.~M.~K.}\ \bibnamefont {Vandersypen}},\ }\bibfield  {title} {\bibinfo {title} {Quantum logic with spin qubits crossing the surface code threshold},\ }\href {https://doi.org/10.1038/s41586-021-04273-w} {\bibfield  {journal} {\bibinfo  {journal} {Nature}\ }\textbf {\bibinfo {volume} {601}},\ \bibinfo {pages} {343} (\bibinfo {year} {2022})}\BibitemShut {NoStop}%
\bibitem [{\citenamefont {{von Horstig}}\ \emph {et~al.}(2024)\citenamefont {{von Horstig}}, \citenamefont {Peri}, \citenamefont {Barraud}, \citenamefont {Robinson}, \citenamefont {Benito}, \citenamefont {Martins},\ and\ \citenamefont {{Gonzalez-Zalba}}}]{vonhorstigElectricalReadoutSpins2024}%
  \BibitemOpen
  \bibfield  {author} {\bibinfo {author} {\bibfnamefont {F.-E.}\ \bibnamefont {{von Horstig}}}, \bibinfo {author} {\bibfnamefont {L.}~\bibnamefont {Peri}}, \bibinfo {author} {\bibfnamefont {S.}~\bibnamefont {Barraud}}, \bibinfo {author} {\bibfnamefont {J.~A.~W.}\ \bibnamefont {Robinson}}, \bibinfo {author} {\bibfnamefont {M.}~\bibnamefont {Benito}}, \bibinfo {author} {\bibfnamefont {F.}~\bibnamefont {Martins}},\ and\ \bibinfo {author} {\bibfnamefont {M.~F.}\ \bibnamefont {{Gonzalez-Zalba}}},\ }\href {https://doi.org/10.48550/arXiv.2403.12888} {\bibinfo {title} {Electrical readout of spins in the absence of spin blockade}} (\bibinfo {year} {2024}),\ \Eprint {https://arxiv.org/abs/2403.12888} {2403.12888} \BibitemShut {NoStop}%
\bibitem [{\citenamefont {Polizzi}(2020)}]{polizziFEASTEigenvalueSolver2020}%
  \BibitemOpen
  \bibfield  {author} {\bibinfo {author} {\bibfnamefont {E.}~\bibnamefont {Polizzi}},\ }\href {https://doi.org/10.48550/arXiv.2002.04807} {\bibinfo {title} {{{FEAST Eigenvalue Solver}} v4.0 {{User Guide}}}} (\bibinfo {year} {2020}),\ \Eprint {https://arxiv.org/abs/2002.04807} {2002.04807} \BibitemShut {NoStop}%
\bibitem [{\citenamefont {Bartel}\ and\ \citenamefont {Korch}(2023)}]{bartelGenerationLogicDesigns2023}%
  \BibitemOpen
  \bibfield  {author} {\bibinfo {author} {\bibfnamefont {S.}~\bibnamefont {Bartel}}\ and\ \bibinfo {author} {\bibfnamefont {M.}~\bibnamefont {Korch}},\ }\bibfield  {title} {\bibinfo {title} {Generation of logic designs for efficiently solving ordinary differential equations on field programmable gate arrays},\ }\href {https://doi.org/10.1002/spe.3043} {\bibfield  {journal} {\bibinfo  {journal} {Software: Practice and Experience}\ }\textbf {\bibinfo {volume} {53}},\ \bibinfo {pages} {27} (\bibinfo {year} {2023})}\BibitemShut {NoStop}%
\bibitem [{\citenamefont {Scappucci}\ \emph {et~al.}(2021)\citenamefont {Scappucci}, \citenamefont {Kloeffel}, \citenamefont {Zwanenburg}, \citenamefont {Loss}, \citenamefont {Myronov}, \citenamefont {Zhang}, \citenamefont {De~Franceschi}, \citenamefont {Katsaros},\ and\ \citenamefont {Veldhorst}}]{scappucciGermaniumQuantumInformation2021}%
  \BibitemOpen
  \bibfield  {author} {\bibinfo {author} {\bibfnamefont {G.}~\bibnamefont {Scappucci}}, \bibinfo {author} {\bibfnamefont {C.}~\bibnamefont {Kloeffel}}, \bibinfo {author} {\bibfnamefont {F.~A.}\ \bibnamefont {Zwanenburg}}, \bibinfo {author} {\bibfnamefont {D.}~\bibnamefont {Loss}}, \bibinfo {author} {\bibfnamefont {M.}~\bibnamefont {Myronov}}, \bibinfo {author} {\bibfnamefont {J.-J.}\ \bibnamefont {Zhang}}, \bibinfo {author} {\bibfnamefont {S.}~\bibnamefont {De~Franceschi}}, \bibinfo {author} {\bibfnamefont {G.}~\bibnamefont {Katsaros}},\ and\ \bibinfo {author} {\bibfnamefont {M.}~\bibnamefont {Veldhorst}},\ }\bibfield  {title} {\bibinfo {title} {The germanium quantum information route},\ }\href {https://doi.org/10.1038/s41578-020-00262-z} {\bibfield  {journal} {\bibinfo  {journal} {Nature Reviews Materials}\ }\textbf {\bibinfo {volume} {6}},\ \bibinfo {pages} {926} (\bibinfo {year} {2021})}\BibitemShut {NoStop}%
\bibitem [{\citenamefont {Stano}\ and\ \citenamefont {Loss}(2022)}]{stanoReviewPerformanceMetrics2022}%
  \BibitemOpen
  \bibfield  {author} {\bibinfo {author} {\bibfnamefont {P.}~\bibnamefont {Stano}}\ and\ \bibinfo {author} {\bibfnamefont {D.}~\bibnamefont {Loss}},\ }\bibfield  {title} {\bibinfo {title} {Review of performance metrics of spin qubits in gated semiconducting nanostructures},\ }\href {https://doi.org/10.1038/s42254-022-00484-w} {\bibfield  {journal} {\bibinfo  {journal} {Nature Reviews Physics}\ }\textbf {\bibinfo {volume} {4}},\ \bibinfo {pages} {672} (\bibinfo {year} {2022})}\BibitemShut {NoStop}%
\bibitem [{\citenamefont {Undseth}\ \emph {et~al.}(2023{\natexlab{a}})\citenamefont {Undseth}, \citenamefont {{Pietx-Casas}}, \citenamefont {Raymenants}, \citenamefont {Mehmandoost}, \citenamefont {Madzik}, \citenamefont {Philips}, \citenamefont {{de Snoo}}, \citenamefont {Michalak}, \citenamefont {Amitonov}, \citenamefont {Tryputen}, \citenamefont {Wuetz}, \citenamefont {Fezzi}, \citenamefont {Esposti}, \citenamefont {Sammak}, \citenamefont {Scappucci},\ and\ \citenamefont {Vandersypen}}]{undsethHotterEasierUnexpected2023}%
  \BibitemOpen
  \bibfield  {author} {\bibinfo {author} {\bibfnamefont {B.}~\bibnamefont {Undseth}}, \bibinfo {author} {\bibfnamefont {O.}~\bibnamefont {{Pietx-Casas}}}, \bibinfo {author} {\bibfnamefont {E.}~\bibnamefont {Raymenants}}, \bibinfo {author} {\bibfnamefont {M.}~\bibnamefont {Mehmandoost}}, \bibinfo {author} {\bibfnamefont {M.~T.}\ \bibnamefont {Madzik}}, \bibinfo {author} {\bibfnamefont {S.~G.~J.}\ \bibnamefont {Philips}}, \bibinfo {author} {\bibfnamefont {S.~L.}\ \bibnamefont {{de Snoo}}}, \bibinfo {author} {\bibfnamefont {D.~J.}\ \bibnamefont {Michalak}}, \bibinfo {author} {\bibfnamefont {S.~V.}\ \bibnamefont {Amitonov}}, \bibinfo {author} {\bibfnamefont {L.}~\bibnamefont {Tryputen}}, \bibinfo {author} {\bibfnamefont {B.~P.}\ \bibnamefont {Wuetz}}, \bibinfo {author} {\bibfnamefont {V.}~\bibnamefont {Fezzi}}, \bibinfo {author} {\bibfnamefont {D.~D.}\ \bibnamefont {Esposti}}, \bibinfo {author} {\bibfnamefont {A.}~\bibnamefont {Sammak}}, \bibinfo {author} {\bibfnamefont {G.}~\bibnamefont {Scappucci}},\ and\
  \bibinfo {author} {\bibfnamefont {L.~M.~K.}\ \bibnamefont {Vandersypen}},\ }\bibfield  {title} {\bibinfo {title} {Hotter is {{Easier}}: {{Unexpected Temperature Dependence}} of {{Spin Qubit Frequencies}}},\ }\href {https://doi.org/10.1103/PhysRevX.13.041015} {\bibfield  {journal} {\bibinfo  {journal} {Physical Review X}\ }\textbf {\bibinfo {volume} {13}},\ \bibinfo {pages} {041015} (\bibinfo {year} {2023}{\natexlab{a}})}\BibitemShut {NoStop}%
\bibitem [{\citenamefont {Huang}\ \emph {et~al.}(2024)\citenamefont {Huang}, \citenamefont {Su}, \citenamefont {Lim}, \citenamefont {Feng}, \citenamefont {{van Straaten}}, \citenamefont {Severin}, \citenamefont {Gilbert}, \citenamefont {Dumoulin~Stuyck}, \citenamefont {Tanttu}, \citenamefont {Serrano}, \citenamefont {Cifuentes}, \citenamefont {Hansen}, \citenamefont {Seedhouse}, \citenamefont {Vahapoglu}, \citenamefont {Leon}, \citenamefont {Abrosimov}, \citenamefont {Pohl}, \citenamefont {Thewalt}, \citenamefont {Hudson}, \citenamefont {Escott}, \citenamefont {Ares}, \citenamefont {Bartlett}, \citenamefont {Morello}, \citenamefont {Saraiva}, \citenamefont {Laucht}, \citenamefont {Dzurak},\ and\ \citenamefont {Yang}}]{huangHighfidelitySpinQubit2024}%
  \BibitemOpen
  \bibfield  {author} {\bibinfo {author} {\bibfnamefont {J.~Y.}\ \bibnamefont {Huang}}, \bibinfo {author} {\bibfnamefont {R.~Y.}\ \bibnamefont {Su}}, \bibinfo {author} {\bibfnamefont {W.~H.}\ \bibnamefont {Lim}}, \bibinfo {author} {\bibfnamefont {M.}~\bibnamefont {Feng}}, \bibinfo {author} {\bibfnamefont {B.}~\bibnamefont {{van Straaten}}}, \bibinfo {author} {\bibfnamefont {B.}~\bibnamefont {Severin}}, \bibinfo {author} {\bibfnamefont {W.}~\bibnamefont {Gilbert}}, \bibinfo {author} {\bibfnamefont {N.}~\bibnamefont {Dumoulin~Stuyck}}, \bibinfo {author} {\bibfnamefont {T.}~\bibnamefont {Tanttu}}, \bibinfo {author} {\bibfnamefont {S.}~\bibnamefont {Serrano}}, \bibinfo {author} {\bibfnamefont {J.~D.}\ \bibnamefont {Cifuentes}}, \bibinfo {author} {\bibfnamefont {I.}~\bibnamefont {Hansen}}, \bibinfo {author} {\bibfnamefont {A.~E.}\ \bibnamefont {Seedhouse}}, \bibinfo {author} {\bibfnamefont {E.}~\bibnamefont {Vahapoglu}}, \bibinfo {author} {\bibfnamefont {R.~C.~C.}\ \bibnamefont {Leon}}, \bibinfo {author}
  {\bibfnamefont {N.~V.}\ \bibnamefont {Abrosimov}}, \bibinfo {author} {\bibfnamefont {H.-J.}\ \bibnamefont {Pohl}}, \bibinfo {author} {\bibfnamefont {M.~L.~W.}\ \bibnamefont {Thewalt}}, \bibinfo {author} {\bibfnamefont {F.~E.}\ \bibnamefont {Hudson}}, \bibinfo {author} {\bibfnamefont {C.~C.}\ \bibnamefont {Escott}}, \bibinfo {author} {\bibfnamefont {N.}~\bibnamefont {Ares}}, \bibinfo {author} {\bibfnamefont {S.~D.}\ \bibnamefont {Bartlett}}, \bibinfo {author} {\bibfnamefont {A.}~\bibnamefont {Morello}}, \bibinfo {author} {\bibfnamefont {A.}~\bibnamefont {Saraiva}}, \bibinfo {author} {\bibfnamefont {A.}~\bibnamefont {Laucht}}, \bibinfo {author} {\bibfnamefont {A.~S.}\ \bibnamefont {Dzurak}},\ and\ \bibinfo {author} {\bibfnamefont {C.~H.}\ \bibnamefont {Yang}},\ }\bibfield  {title} {\bibinfo {title} {High-fidelity spin qubit operation and algorithmic initialization above 1 {{K}}},\ }\href {https://doi.org/10.1038/s41586-024-07160-2} {\bibfield  {journal} {\bibinfo  {journal} {Nature}\ }\textbf {\bibinfo
  {volume} {627}},\ \bibinfo {pages} {772} (\bibinfo {year} {2024})}\BibitemShut {NoStop}%
\bibitem [{\citenamefont {Russ}\ and\ \citenamefont {Burkard}(2017)}]{russThreeelectronSpinQubits2017}%
  \BibitemOpen
  \bibfield  {author} {\bibinfo {author} {\bibfnamefont {M.}~\bibnamefont {Russ}}\ and\ \bibinfo {author} {\bibfnamefont {G.}~\bibnamefont {Burkard}},\ }\bibfield  {title} {\bibinfo {title} {Three-electron spin qubits},\ }\href {https://doi.org/10.1088/1361-648X/aa761f} {\bibfield  {journal} {\bibinfo  {journal} {Journal of Physics: Condensed Matter}\ }\textbf {\bibinfo {volume} {29}},\ \bibinfo {pages} {393001} (\bibinfo {year} {2017})}\BibitemShut {NoStop}%
\bibitem [{\citenamefont {Bosco}\ and\ \citenamefont {{Rimbach-Russ}}(2024)}]{boscoExchangeOnlySpinOrbitQubits2024}%
  \BibitemOpen
  \bibfield  {author} {\bibinfo {author} {\bibfnamefont {S.}~\bibnamefont {Bosco}}\ and\ \bibinfo {author} {\bibfnamefont {M.}~\bibnamefont {{Rimbach-Russ}}},\ }\href@noop {} {\bibinfo {title} {Exchange-{{Only Spin-Orbit Qubits}} in {{Silicon}} and {{Germanium}}}} (\bibinfo {year} {2024}),\ \Eprint {https://arxiv.org/abs/2410.05461} {2410.05461} \BibitemShut {NoStop}%
\bibitem [{\citenamefont {{Rimbach-Russ}}\ \emph {et~al.}(2024)\citenamefont {{Rimbach-Russ}}, \citenamefont {John}, \citenamefont {van Straaten},\ and\ \citenamefont {Bosco}}]{rimbach-russSpinlessSpinQubit2024}%
  \BibitemOpen
  \bibfield  {author} {\bibinfo {author} {\bibfnamefont {M.}~\bibnamefont {{Rimbach-Russ}}}, \bibinfo {author} {\bibfnamefont {V.}~\bibnamefont {John}}, \bibinfo {author} {\bibfnamefont {B.}~\bibnamefont {van Straaten}},\ and\ \bibinfo {author} {\bibfnamefont {S.}~\bibnamefont {Bosco}},\ }\href {https://doi.org/10.48550/arXiv.2412.13658} {\bibinfo {title} {A spinless spin qubit}} (\bibinfo {year} {2024}),\ \Eprint {https://arxiv.org/abs/2412.13658} {2412.13658} \BibitemShut {NoStop}%
\bibitem [{\citenamefont {Foulk}\ \emph {et~al.}(2025)\citenamefont {Foulk}, \citenamefont {Hoffman}, \citenamefont {Laubscher},\ and\ \citenamefont {Sarma}}]{foulkSingletonlyAlwaysonGapless2025}%
  \BibitemOpen
  \bibfield  {author} {\bibinfo {author} {\bibfnamefont {N.~L.}\ \bibnamefont {Foulk}}, \bibinfo {author} {\bibfnamefont {S.}~\bibnamefont {Hoffman}}, \bibinfo {author} {\bibfnamefont {K.}~\bibnamefont {Laubscher}},\ and\ \bibinfo {author} {\bibfnamefont {S.~D.}\ \bibnamefont {Sarma}},\ }\href {https://doi.org/10.48550/arXiv.2501.18589} {\bibinfo {title} {Singlet-only {{Always-on Gapless Exchange Qubits}} with {{Baseband Control}}}} (\bibinfo {year} {2025}),\ \Eprint {https://arxiv.org/abs/2501.18589} {2501.18589} \BibitemShut {NoStop}%
\bibitem [{\citenamefont {Nguyen}\ \emph {et~al.}(2025)\citenamefont {Nguyen}, \citenamefont {{Rimbach-Russ}}, \citenamefont {Vandersypen},\ and\ \citenamefont {Bosco}}]{nguyenSinglestepHighfidelityThreequbit2025}%
  \BibitemOpen
  \bibfield  {author} {\bibinfo {author} {\bibfnamefont {M.~T.~P.}\ \bibnamefont {Nguyen}}, \bibinfo {author} {\bibfnamefont {M.}~\bibnamefont {{Rimbach-Russ}}}, \bibinfo {author} {\bibfnamefont {L.~M.~K.}\ \bibnamefont {Vandersypen}},\ and\ \bibinfo {author} {\bibfnamefont {S.}~\bibnamefont {Bosco}},\ }\href {https://doi.org/10.48550/arXiv.2503.12182} {\bibinfo {title} {Single-step high-fidelity three-qubit gates by anisotropic chiral interactions}} (\bibinfo {year} {2025}),\ \Eprint {https://arxiv.org/abs/2503.12182} {2503.12182} \BibitemShut {NoStop}%
\bibitem [{\citenamefont {Lanza}\ \emph {et~al.}(2020)\citenamefont {Lanza}, \citenamefont {Smets}, \citenamefont {Huyghebaert},\ and\ \citenamefont {Li}}]{lanzaYieldVariabilityReliability2020}%
  \BibitemOpen
  \bibfield  {author} {\bibinfo {author} {\bibfnamefont {M.}~\bibnamefont {Lanza}}, \bibinfo {author} {\bibfnamefont {Q.}~\bibnamefont {Smets}}, \bibinfo {author} {\bibfnamefont {C.}~\bibnamefont {Huyghebaert}},\ and\ \bibinfo {author} {\bibfnamefont {L.-J.}\ \bibnamefont {Li}},\ }\bibfield  {title} {\bibinfo {title} {Yield, variability, reliability, and stability of two-dimensional materials based solid-state electronic devices},\ }\href {https://doi.org/10.1038/s41467-020-19053-9} {\bibfield  {journal} {\bibinfo  {journal} {Nature Communications}\ }\textbf {\bibinfo {volume} {11}},\ \bibinfo {pages} {5689} (\bibinfo {year} {2020})}\BibitemShut {NoStop}%
\bibitem [{\citenamefont {Zwerver}\ \emph {et~al.}(2022)\citenamefont {Zwerver}, \citenamefont {Kr{\"a}henmann}, \citenamefont {Watson}, \citenamefont {Lampert}, \citenamefont {George}, \citenamefont {Pillarisetty}, \citenamefont {Bojarski}, \citenamefont {Amin}, \citenamefont {Amitonov}, \citenamefont {Boter}, \citenamefont {Caudillo}, \citenamefont {{Correas-Serrano}}, \citenamefont {Dehollain}, \citenamefont {Droulers}, \citenamefont {Henry}, \citenamefont {Kotlyar}, \citenamefont {Lodari}, \citenamefont {L{\"u}thi}, \citenamefont {Michalak}, \citenamefont {Mueller}, \citenamefont {Neyens}, \citenamefont {Roberts}, \citenamefont {Samkharadze}, \citenamefont {Zheng}, \citenamefont {Zietz}, \citenamefont {Scappucci}, \citenamefont {Veldhorst}, \citenamefont {Vandersypen},\ and\ \citenamefont {Clarke}}]{zwerverQubitsMadeAdvanced2022}%
  \BibitemOpen
  \bibfield  {author} {\bibinfo {author} {\bibfnamefont {A.~M.~J.}\ \bibnamefont {Zwerver}}, \bibinfo {author} {\bibfnamefont {T.}~\bibnamefont {Kr{\"a}henmann}}, \bibinfo {author} {\bibfnamefont {T.~F.}\ \bibnamefont {Watson}}, \bibinfo {author} {\bibfnamefont {L.}~\bibnamefont {Lampert}}, \bibinfo {author} {\bibfnamefont {H.~C.}\ \bibnamefont {George}}, \bibinfo {author} {\bibfnamefont {R.}~\bibnamefont {Pillarisetty}}, \bibinfo {author} {\bibfnamefont {S.~A.}\ \bibnamefont {Bojarski}}, \bibinfo {author} {\bibfnamefont {P.}~\bibnamefont {Amin}}, \bibinfo {author} {\bibfnamefont {S.~V.}\ \bibnamefont {Amitonov}}, \bibinfo {author} {\bibfnamefont {J.~M.}\ \bibnamefont {Boter}}, \bibinfo {author} {\bibfnamefont {R.}~\bibnamefont {Caudillo}}, \bibinfo {author} {\bibfnamefont {D.}~\bibnamefont {{Correas-Serrano}}}, \bibinfo {author} {\bibfnamefont {J.~P.}\ \bibnamefont {Dehollain}}, \bibinfo {author} {\bibfnamefont {G.}~\bibnamefont {Droulers}}, \bibinfo {author} {\bibfnamefont {E.~M.}\ \bibnamefont {Henry}},
  \bibinfo {author} {\bibfnamefont {R.}~\bibnamefont {Kotlyar}}, \bibinfo {author} {\bibfnamefont {M.}~\bibnamefont {Lodari}}, \bibinfo {author} {\bibfnamefont {F.}~\bibnamefont {L{\"u}thi}}, \bibinfo {author} {\bibfnamefont {D.~J.}\ \bibnamefont {Michalak}}, \bibinfo {author} {\bibfnamefont {B.~K.}\ \bibnamefont {Mueller}}, \bibinfo {author} {\bibfnamefont {S.}~\bibnamefont {Neyens}}, \bibinfo {author} {\bibfnamefont {J.}~\bibnamefont {Roberts}}, \bibinfo {author} {\bibfnamefont {N.}~\bibnamefont {Samkharadze}}, \bibinfo {author} {\bibfnamefont {G.}~\bibnamefont {Zheng}}, \bibinfo {author} {\bibfnamefont {O.~K.}\ \bibnamefont {Zietz}}, \bibinfo {author} {\bibfnamefont {G.}~\bibnamefont {Scappucci}}, \bibinfo {author} {\bibfnamefont {M.}~\bibnamefont {Veldhorst}}, \bibinfo {author} {\bibfnamefont {L.~M.~K.}\ \bibnamefont {Vandersypen}},\ and\ \bibinfo {author} {\bibfnamefont {J.~S.}\ \bibnamefont {Clarke}},\ }\bibfield  {title} {\bibinfo {title} {Qubits made by advanced semiconductor manufacturing},\ }\href
  {https://doi.org/10.1038/s41928-022-00727-9} {\bibfield  {journal} {\bibinfo  {journal} {Nature Electronics}\ }\textbf {\bibinfo {volume} {5}},\ \bibinfo {pages} {184} (\bibinfo {year} {2022})}\BibitemShut {NoStop}%
\bibitem [{\citenamefont {Cifuentes}\ \emph {et~al.}(2024)\citenamefont {Cifuentes}, \citenamefont {Tanttu}, \citenamefont {Gilbert}, \citenamefont {Huang}, \citenamefont {Vahapoglu}, \citenamefont {Leon}, \citenamefont {Serrano}, \citenamefont {Otter}, \citenamefont {Dunmore}, \citenamefont {Mai}, \citenamefont {Schlattner}, \citenamefont {Feng}, \citenamefont {Itoh}, \citenamefont {Abrosimov}, \citenamefont {Pohl}, \citenamefont {Thewalt}, \citenamefont {Laucht}, \citenamefont {Yang}, \citenamefont {Escott}, \citenamefont {Lim}, \citenamefont {Hudson}, \citenamefont {Rahman}, \citenamefont {Dzurak},\ and\ \citenamefont {Saraiva}}]{cifuentesBoundsElectronSpin2024}%
  \BibitemOpen
  \bibfield  {author} {\bibinfo {author} {\bibfnamefont {J.~D.}\ \bibnamefont {Cifuentes}}, \bibinfo {author} {\bibfnamefont {T.}~\bibnamefont {Tanttu}}, \bibinfo {author} {\bibfnamefont {W.}~\bibnamefont {Gilbert}}, \bibinfo {author} {\bibfnamefont {J.~Y.}\ \bibnamefont {Huang}}, \bibinfo {author} {\bibfnamefont {E.}~\bibnamefont {Vahapoglu}}, \bibinfo {author} {\bibfnamefont {R.~C.~C.}\ \bibnamefont {Leon}}, \bibinfo {author} {\bibfnamefont {S.}~\bibnamefont {Serrano}}, \bibinfo {author} {\bibfnamefont {D.}~\bibnamefont {Otter}}, \bibinfo {author} {\bibfnamefont {D.}~\bibnamefont {Dunmore}}, \bibinfo {author} {\bibfnamefont {P.~Y.}\ \bibnamefont {Mai}}, \bibinfo {author} {\bibfnamefont {F.}~\bibnamefont {Schlattner}}, \bibinfo {author} {\bibfnamefont {M.}~\bibnamefont {Feng}}, \bibinfo {author} {\bibfnamefont {K.}~\bibnamefont {Itoh}}, \bibinfo {author} {\bibfnamefont {N.}~\bibnamefont {Abrosimov}}, \bibinfo {author} {\bibfnamefont {H.-J.}\ \bibnamefont {Pohl}}, \bibinfo {author} {\bibfnamefont
  {M.}~\bibnamefont {Thewalt}}, \bibinfo {author} {\bibfnamefont {A.}~\bibnamefont {Laucht}}, \bibinfo {author} {\bibfnamefont {C.~H.}\ \bibnamefont {Yang}}, \bibinfo {author} {\bibfnamefont {C.~C.}\ \bibnamefont {Escott}}, \bibinfo {author} {\bibfnamefont {W.~H.}\ \bibnamefont {Lim}}, \bibinfo {author} {\bibfnamefont {F.~E.}\ \bibnamefont {Hudson}}, \bibinfo {author} {\bibfnamefont {R.}~\bibnamefont {Rahman}}, \bibinfo {author} {\bibfnamefont {A.~S.}\ \bibnamefont {Dzurak}},\ and\ \bibinfo {author} {\bibfnamefont {A.}~\bibnamefont {Saraiva}},\ }\bibfield  {title} {\bibinfo {title} {Bounds to electron spin qubit variability for scalable {{CMOS}} architectures},\ }\href {https://doi.org/10.1038/s41467-024-48557-x} {\bibfield  {journal} {\bibinfo  {journal} {Nature Communications}\ }\textbf {\bibinfo {volume} {15}},\ \bibinfo {pages} {4299} (\bibinfo {year} {2024})}\BibitemShut {NoStop}%
\bibitem [{\citenamefont {Siegel}\ \emph {et~al.}(2024)\citenamefont {Siegel}, \citenamefont {Strikis},\ and\ \citenamefont {Fogarty}}]{siegelEarlyFaultTolerance2024a}%
  \BibitemOpen
  \bibfield  {author} {\bibinfo {author} {\bibfnamefont {A.}~\bibnamefont {Siegel}}, \bibinfo {author} {\bibfnamefont {A.}~\bibnamefont {Strikis}},\ and\ \bibinfo {author} {\bibfnamefont {M.}~\bibnamefont {Fogarty}},\ }\bibfield  {title} {\bibinfo {title} {Towards {{Early Fault Tolerance}} on a 2 {\texttimes} {{N Array}} of {{Qubits Equipped}} with {{Shuttling}}},\ }\href {https://doi.org/10.1103/PRXQuantum.5.040328} {\bibfield  {journal} {\bibinfo  {journal} {PRX Quantum}\ }\textbf {\bibinfo {volume} {5}},\ \bibinfo {pages} {040328} (\bibinfo {year} {2024})}\BibitemShut {NoStop}%
\bibitem [{\citenamefont {Moses}\ \emph {et~al.}(2023)\citenamefont {Moses}, \citenamefont {Baldwin}, \citenamefont {Allman}, \citenamefont {Ancona}, \citenamefont {Ascarrunz}, \citenamefont {Barnes}, \citenamefont {Bartolotta}, \citenamefont {Bjork}, \citenamefont {Blanchard}, \citenamefont {Bohn}, \citenamefont {Bohnet}, \citenamefont {Brown}, \citenamefont {Burdick}, \citenamefont {Burton}, \citenamefont {Campbell}, \citenamefont {J.~P.~Campora}, \citenamefont {Carron}, \citenamefont {Chambers}, \citenamefont {Chan}, \citenamefont {Chen}, \citenamefont {Chernoguzov}, \citenamefont {Chertkov}, \citenamefont {Colina}, \citenamefont {Curtis}, \citenamefont {Daniel}, \citenamefont {DeCross}, \citenamefont {Deen}, \citenamefont {Delaney}, \citenamefont {Dreiling}, \citenamefont {Ertsgaard}, \citenamefont {Esposito}, \citenamefont {Estey}, \citenamefont {Fabrikant}, \citenamefont {Figgatt}, \citenamefont {Foltz}, \citenamefont {{Foss-Feig}}, \citenamefont {Francois}, \citenamefont {Gaebler}, \citenamefont
  {Gatterman}, \citenamefont {Gilbreth}, \citenamefont {Giles}, \citenamefont {Glynn}, \citenamefont {Hall}, \citenamefont {Hankin}, \citenamefont {Hansen}, \citenamefont {Hayes}, \citenamefont {Higashi}, \citenamefont {Hoffman}, \citenamefont {Horning}, \citenamefont {Hout}, \citenamefont {Jacobs}, \citenamefont {Johansen}, \citenamefont {Jones}, \citenamefont {Karcz}, \citenamefont {Klein}, \citenamefont {Lauria}, \citenamefont {Lee}, \citenamefont {Liefer}, \citenamefont {Lu}, \citenamefont {Lucchetti}, \citenamefont {Lytle}, \citenamefont {Malm}, \citenamefont {Matheny}, \citenamefont {Mathewson}, \citenamefont {Mayer}, \citenamefont {Miller}, \citenamefont {Mills}, \citenamefont {Neyenhuis}, \citenamefont {Nugent}, \citenamefont {Olson}, \citenamefont {Parks}, \citenamefont {Price}, \citenamefont {Price}, \citenamefont {Pugh}, \citenamefont {Ransford}, \citenamefont {Reed}, \citenamefont {Roman}, \citenamefont {Rowe}, \citenamefont {{Ryan-Anderson}}, \citenamefont {Sanders}, \citenamefont {Sedlacek},
  \citenamefont {Shevchuk}, \citenamefont {Siegfried}, \citenamefont {Skripka}, \citenamefont {Spaun}, \citenamefont {Sprenkle}, \citenamefont {Stutz}, \citenamefont {Swallows}, \citenamefont {Tobey}, \citenamefont {Tran}, \citenamefont {Tran}, \citenamefont {Vogt}, \citenamefont {Volin}, \citenamefont {Walker}, \citenamefont {Zolot},\ and\ \citenamefont {Pino}}]{mosesRaceTrackTrappedIonQuantum2023}%
  \BibitemOpen
  \bibfield  {author} {\bibinfo {author} {\bibfnamefont {S.~A.}\ \bibnamefont {Moses}}, \bibinfo {author} {\bibfnamefont {C.~H.}\ \bibnamefont {Baldwin}}, \bibinfo {author} {\bibfnamefont {M.~S.}\ \bibnamefont {Allman}}, \bibinfo {author} {\bibfnamefont {R.}~\bibnamefont {Ancona}}, \bibinfo {author} {\bibfnamefont {L.}~\bibnamefont {Ascarrunz}}, \bibinfo {author} {\bibfnamefont {C.}~\bibnamefont {Barnes}}, \bibinfo {author} {\bibfnamefont {J.}~\bibnamefont {Bartolotta}}, \bibinfo {author} {\bibfnamefont {B.}~\bibnamefont {Bjork}}, \bibinfo {author} {\bibfnamefont {P.}~\bibnamefont {Blanchard}}, \bibinfo {author} {\bibfnamefont {M.}~\bibnamefont {Bohn}}, \bibinfo {author} {\bibfnamefont {J.~G.}\ \bibnamefont {Bohnet}}, \bibinfo {author} {\bibfnamefont {N.~C.}\ \bibnamefont {Brown}}, \bibinfo {author} {\bibfnamefont {N.~Q.}\ \bibnamefont {Burdick}}, \bibinfo {author} {\bibfnamefont {W.~C.}\ \bibnamefont {Burton}}, \bibinfo {author} {\bibfnamefont {S.~L.}\ \bibnamefont {Campbell}}, \bibinfo {author}
  {\bibfnamefont {I.~I.~I.}\ \bibnamefont {J.~P.~Campora}}, \bibinfo {author} {\bibfnamefont {C.}~\bibnamefont {Carron}}, \bibinfo {author} {\bibfnamefont {J.}~\bibnamefont {Chambers}}, \bibinfo {author} {\bibfnamefont {J.~W.}\ \bibnamefont {Chan}}, \bibinfo {author} {\bibfnamefont {Y.~H.}\ \bibnamefont {Chen}}, \bibinfo {author} {\bibfnamefont {A.}~\bibnamefont {Chernoguzov}}, \bibinfo {author} {\bibfnamefont {E.}~\bibnamefont {Chertkov}}, \bibinfo {author} {\bibfnamefont {J.}~\bibnamefont {Colina}}, \bibinfo {author} {\bibfnamefont {J.~P.}\ \bibnamefont {Curtis}}, \bibinfo {author} {\bibfnamefont {R.}~\bibnamefont {Daniel}}, \bibinfo {author} {\bibfnamefont {M.}~\bibnamefont {DeCross}}, \bibinfo {author} {\bibfnamefont {D.}~\bibnamefont {Deen}}, \bibinfo {author} {\bibfnamefont {C.}~\bibnamefont {Delaney}}, \bibinfo {author} {\bibfnamefont {J.~M.}\ \bibnamefont {Dreiling}}, \bibinfo {author} {\bibfnamefont {C.~T.}\ \bibnamefont {Ertsgaard}}, \bibinfo {author} {\bibfnamefont {J.}~\bibnamefont {Esposito}},
  \bibinfo {author} {\bibfnamefont {B.}~\bibnamefont {Estey}}, \bibinfo {author} {\bibfnamefont {M.}~\bibnamefont {Fabrikant}}, \bibinfo {author} {\bibfnamefont {C.}~\bibnamefont {Figgatt}}, \bibinfo {author} {\bibfnamefont {C.}~\bibnamefont {Foltz}}, \bibinfo {author} {\bibfnamefont {M.}~\bibnamefont {{Foss-Feig}}}, \bibinfo {author} {\bibfnamefont {D.}~\bibnamefont {Francois}}, \bibinfo {author} {\bibfnamefont {J.~P.}\ \bibnamefont {Gaebler}}, \bibinfo {author} {\bibfnamefont {T.~M.}\ \bibnamefont {Gatterman}}, \bibinfo {author} {\bibfnamefont {C.~N.}\ \bibnamefont {Gilbreth}}, \bibinfo {author} {\bibfnamefont {J.}~\bibnamefont {Giles}}, \bibinfo {author} {\bibfnamefont {E.}~\bibnamefont {Glynn}}, \bibinfo {author} {\bibfnamefont {A.}~\bibnamefont {Hall}}, \bibinfo {author} {\bibfnamefont {A.~M.}\ \bibnamefont {Hankin}}, \bibinfo {author} {\bibfnamefont {A.}~\bibnamefont {Hansen}}, \bibinfo {author} {\bibfnamefont {D.}~\bibnamefont {Hayes}}, \bibinfo {author} {\bibfnamefont {B.}~\bibnamefont {Higashi}},
  \bibinfo {author} {\bibfnamefont {I.~M.}\ \bibnamefont {Hoffman}}, \bibinfo {author} {\bibfnamefont {B.}~\bibnamefont {Horning}}, \bibinfo {author} {\bibfnamefont {J.~J.}\ \bibnamefont {Hout}}, \bibinfo {author} {\bibfnamefont {R.}~\bibnamefont {Jacobs}}, \bibinfo {author} {\bibfnamefont {J.}~\bibnamefont {Johansen}}, \bibinfo {author} {\bibfnamefont {L.}~\bibnamefont {Jones}}, \bibinfo {author} {\bibfnamefont {J.}~\bibnamefont {Karcz}}, \bibinfo {author} {\bibfnamefont {T.}~\bibnamefont {Klein}}, \bibinfo {author} {\bibfnamefont {P.}~\bibnamefont {Lauria}}, \bibinfo {author} {\bibfnamefont {P.}~\bibnamefont {Lee}}, \bibinfo {author} {\bibfnamefont {D.}~\bibnamefont {Liefer}}, \bibinfo {author} {\bibfnamefont {S.~T.}\ \bibnamefont {Lu}}, \bibinfo {author} {\bibfnamefont {D.}~\bibnamefont {Lucchetti}}, \bibinfo {author} {\bibfnamefont {C.}~\bibnamefont {Lytle}}, \bibinfo {author} {\bibfnamefont {A.}~\bibnamefont {Malm}}, \bibinfo {author} {\bibfnamefont {M.}~\bibnamefont {Matheny}}, \bibinfo {author}
  {\bibfnamefont {B.}~\bibnamefont {Mathewson}}, \bibinfo {author} {\bibfnamefont {K.}~\bibnamefont {Mayer}}, \bibinfo {author} {\bibfnamefont {D.~B.}\ \bibnamefont {Miller}}, \bibinfo {author} {\bibfnamefont {M.}~\bibnamefont {Mills}}, \bibinfo {author} {\bibfnamefont {B.}~\bibnamefont {Neyenhuis}}, \bibinfo {author} {\bibfnamefont {L.}~\bibnamefont {Nugent}}, \bibinfo {author} {\bibfnamefont {S.}~\bibnamefont {Olson}}, \bibinfo {author} {\bibfnamefont {J.}~\bibnamefont {Parks}}, \bibinfo {author} {\bibfnamefont {G.~N.}\ \bibnamefont {Price}}, \bibinfo {author} {\bibfnamefont {Z.}~\bibnamefont {Price}}, \bibinfo {author} {\bibfnamefont {M.}~\bibnamefont {Pugh}}, \bibinfo {author} {\bibfnamefont {A.}~\bibnamefont {Ransford}}, \bibinfo {author} {\bibfnamefont {A.~P.}\ \bibnamefont {Reed}}, \bibinfo {author} {\bibfnamefont {C.}~\bibnamefont {Roman}}, \bibinfo {author} {\bibfnamefont {M.}~\bibnamefont {Rowe}}, \bibinfo {author} {\bibfnamefont {C.}~\bibnamefont {{Ryan-Anderson}}}, \bibinfo {author} {\bibfnamefont
  {S.}~\bibnamefont {Sanders}}, \bibinfo {author} {\bibfnamefont {J.}~\bibnamefont {Sedlacek}}, \bibinfo {author} {\bibfnamefont {P.}~\bibnamefont {Shevchuk}}, \bibinfo {author} {\bibfnamefont {P.}~\bibnamefont {Siegfried}}, \bibinfo {author} {\bibfnamefont {T.}~\bibnamefont {Skripka}}, \bibinfo {author} {\bibfnamefont {B.}~\bibnamefont {Spaun}}, \bibinfo {author} {\bibfnamefont {R.~T.}\ \bibnamefont {Sprenkle}}, \bibinfo {author} {\bibfnamefont {R.~P.}\ \bibnamefont {Stutz}}, \bibinfo {author} {\bibfnamefont {M.}~\bibnamefont {Swallows}}, \bibinfo {author} {\bibfnamefont {R.~I.}\ \bibnamefont {Tobey}}, \bibinfo {author} {\bibfnamefont {A.}~\bibnamefont {Tran}}, \bibinfo {author} {\bibfnamefont {T.}~\bibnamefont {Tran}}, \bibinfo {author} {\bibfnamefont {E.}~\bibnamefont {Vogt}}, \bibinfo {author} {\bibfnamefont {C.}~\bibnamefont {Volin}}, \bibinfo {author} {\bibfnamefont {J.}~\bibnamefont {Walker}}, \bibinfo {author} {\bibfnamefont {A.~M.}\ \bibnamefont {Zolot}},\ and\ \bibinfo {author} {\bibfnamefont
  {J.~M.}\ \bibnamefont {Pino}},\ }\bibfield  {title} {\bibinfo {title} {A {{Race-Track Trapped-Ion Quantum Processor}}},\ }\href {https://doi.org/10.1103/PhysRevX.13.041052} {\bibfield  {journal} {\bibinfo  {journal} {Physical Review X}\ }\textbf {\bibinfo {volume} {13}},\ \bibinfo {pages} {041052} (\bibinfo {year} {2023})}\BibitemShut {NoStop}%
\bibitem [{\citenamefont {Krzywda}\ and\ \citenamefont {Cywi{\'n}ski}(2024)}]{krzywdaDecoherenceElectronSpin2024}%
  \BibitemOpen
  \bibfield  {author} {\bibinfo {author} {\bibfnamefont {J.~A.}\ \bibnamefont {Krzywda}}\ and\ \bibinfo {author} {\bibfnamefont {{\L}.}~\bibnamefont {Cywi{\'n}ski}},\ }\href {https://doi.org/10.48550/arXiv.2405.12185} {\bibinfo {title} {Decoherence of electron spin qubit during transfer between two semiconductor quantum dots at low magnetic fields}} (\bibinfo {year} {2024}),\ \Eprint {https://arxiv.org/abs/2405.12185} {2405.12185} \BibitemShut {NoStop}%
\bibitem [{\citenamefont {Undseth}\ \emph {et~al.}(2023{\natexlab{b}})\citenamefont {Undseth}, \citenamefont {Xue}, \citenamefont {Mehmandoost}, \citenamefont {{Rimbach-Russ}}, \citenamefont {Eendebak}, \citenamefont {Samkharadze}, \citenamefont {Sammak}, \citenamefont {Dobrovitski}, \citenamefont {Scappucci},\ and\ \citenamefont {Vandersypen}}]{undsethNonlinearResponseCrosstalk2023}%
  \BibitemOpen
  \bibfield  {author} {\bibinfo {author} {\bibfnamefont {B.}~\bibnamefont {Undseth}}, \bibinfo {author} {\bibfnamefont {X.}~\bibnamefont {Xue}}, \bibinfo {author} {\bibfnamefont {M.}~\bibnamefont {Mehmandoost}}, \bibinfo {author} {\bibfnamefont {M.}~\bibnamefont {{Rimbach-Russ}}}, \bibinfo {author} {\bibfnamefont {P.~T.}\ \bibnamefont {Eendebak}}, \bibinfo {author} {\bibfnamefont {N.}~\bibnamefont {Samkharadze}}, \bibinfo {author} {\bibfnamefont {A.}~\bibnamefont {Sammak}}, \bibinfo {author} {\bibfnamefont {V.~V.}\ \bibnamefont {Dobrovitski}}, \bibinfo {author} {\bibfnamefont {G.}~\bibnamefont {Scappucci}},\ and\ \bibinfo {author} {\bibfnamefont {L.~M.}\ \bibnamefont {Vandersypen}},\ }\bibfield  {title} {\bibinfo {title} {Nonlinear {{Response}} and {{Crosstalk}} of {{Electrically Driven Silicon Spin Qubits}}},\ }\href {https://doi.org/10.1103/PhysRevApplied.19.044078} {\bibfield  {journal} {\bibinfo  {journal} {Physical Review Applied}\ }\textbf {\bibinfo {volume} {19}},\ \bibinfo {pages} {044078} (\bibinfo
  {year} {2023}{\natexlab{b}})}\BibitemShut {NoStop}%
\bibitem [{\citenamefont {Losert}(2024)}]{losertStrategiesEnhancingSpinShuttling2024}%
  \BibitemOpen
  \bibfield  {author} {\bibinfo {author} {\bibfnamefont {M.~P.}\ \bibnamefont {Losert}},\ }\bibfield  {title} {\bibinfo {title} {Strategies for {{Enhancing Spin-Shuttling Fidelities}} in {{Si}}/{{SiGe Quantum Wells}} with {{Random-Alloy Disorder}}},\ }\bibfield  {journal} {\bibinfo  {journal} {PRX Quantum}\ }\textbf {\bibinfo {volume} {5}},\ \href {https://doi.org/10.1103/PRXQuantum.5.040322} {10.1103/PRXQuantum.5.040322} (\bibinfo {year} {2024})\BibitemShut {NoStop}%
\bibitem [{\citenamefont {N{\'e}meth}\ \emph {et~al.}(2024)\citenamefont {N{\'e}meth}, \citenamefont {Bandaru}, \citenamefont {Alves}, \citenamefont {Losert}, \citenamefont {Brann}, \citenamefont {Eskandari}, \citenamefont {Soomro}, \citenamefont {Vivrekar}, \citenamefont {Eriksson},\ and\ \citenamefont {Friesen}}]{nemethOmnidirectionalShuttlingAvoid2024}%
  \BibitemOpen
  \bibfield  {author} {\bibinfo {author} {\bibfnamefont {R.}~\bibnamefont {N{\'e}meth}}, \bibinfo {author} {\bibfnamefont {V.~K.}\ \bibnamefont {Bandaru}}, \bibinfo {author} {\bibfnamefont {P.}~\bibnamefont {Alves}}, \bibinfo {author} {\bibfnamefont {M.~P.}\ \bibnamefont {Losert}}, \bibinfo {author} {\bibfnamefont {E.}~\bibnamefont {Brann}}, \bibinfo {author} {\bibfnamefont {O.~M.}\ \bibnamefont {Eskandari}}, \bibinfo {author} {\bibfnamefont {H.}~\bibnamefont {Soomro}}, \bibinfo {author} {\bibfnamefont {A.}~\bibnamefont {Vivrekar}}, \bibinfo {author} {\bibfnamefont {M.~A.}\ \bibnamefont {Eriksson}},\ and\ \bibinfo {author} {\bibfnamefont {M.}~\bibnamefont {Friesen}},\ }\href {https://doi.org/10.48550/arXiv.2412.09574} {\bibinfo {title} {Omnidirectional shuttling to avoid valley excitations in {{Si}}/{{SiGe}} quantum wells}} (\bibinfo {year} {2024}),\ \Eprint {https://arxiv.org/abs/2412.09574} {2412.09574} \BibitemShut {NoStop}%
\bibitem [{\citenamefont {Thayil}\ \emph {et~al.}(2024)\citenamefont {Thayil}, \citenamefont {Ermoneit},\ and\ \citenamefont {Kantner}}]{thayilTheoryValleySplitting2024}%
  \BibitemOpen
  \bibfield  {author} {\bibinfo {author} {\bibfnamefont {A.}~\bibnamefont {Thayil}}, \bibinfo {author} {\bibfnamefont {L.}~\bibnamefont {Ermoneit}},\ and\ \bibinfo {author} {\bibfnamefont {M.}~\bibnamefont {Kantner}},\ }\href {https://doi.org/10.48550/arXiv.2412.20618} {\bibinfo {title} {Theory of {{Valley Splitting}} in {{Si}}/{{SiGe Spin-Qubits}}: {{Interplay}} of {{Strain}}, {{Resonances}} and {{Random Alloy Disorder}}}} (\bibinfo {year} {2024}),\ \Eprint {https://arxiv.org/abs/2412.20618} {2412.20618} \BibitemShut {NoStop}%
\bibitem [{\citenamefont {Santos}\ \emph {et~al.}(2021)\citenamefont {Santos}, \citenamefont {{Villas-Boas}},\ and\ \citenamefont {Bachelard}}]{santosQuantumAdiabaticBrachistochrone2021}%
  \BibitemOpen
  \bibfield  {author} {\bibinfo {author} {\bibfnamefont {A.~C.}\ \bibnamefont {Santos}}, \bibinfo {author} {\bibfnamefont {C.~J.}\ \bibnamefont {{Villas-Boas}}},\ and\ \bibinfo {author} {\bibfnamefont {R.}~\bibnamefont {Bachelard}},\ }\bibfield  {title} {\bibinfo {title} {Quantum adiabatic brachistochrone for open systems},\ }\href {https://doi.org/10.1103/PhysRevA.103.012206} {\bibfield  {journal} {\bibinfo  {journal} {Physical Review A}\ }\textbf {\bibinfo {volume} {103}},\ \bibinfo {pages} {012206} (\bibinfo {year} {2021})}\BibitemShut {NoStop}%
\bibitem [{\citenamefont {Figueiredo~Roque}\ \emph {et~al.}(2021)\citenamefont {Figueiredo~Roque}, \citenamefont {Clerk},\ and\ \citenamefont {Ribeiro}}]{figueiredoroqueEngineeringFastHighfidelity2021}%
  \BibitemOpen
  \bibfield  {author} {\bibinfo {author} {\bibfnamefont {T.}~\bibnamefont {Figueiredo~Roque}}, \bibinfo {author} {\bibfnamefont {A.~A.}\ \bibnamefont {Clerk}},\ and\ \bibinfo {author} {\bibfnamefont {H.}~\bibnamefont {Ribeiro}},\ }\bibfield  {title} {\bibinfo {title} {Engineering fast high-fidelity quantum operations with constrained interactions},\ }\href {https://doi.org/10.1038/s41534-020-00349-z} {\bibfield  {journal} {\bibinfo  {journal} {npj Quantum Information}\ }\textbf {\bibinfo {volume} {7}},\ \bibinfo {pages} {1} (\bibinfo {year} {2021})}\BibitemShut {NoStop}%
\bibitem [{\citenamefont {Pandey}\ \emph {et~al.}(2020)\citenamefont {Pandey}, \citenamefont {Claeys}, \citenamefont {Campbell}, \citenamefont {Polkovnikov},\ and\ \citenamefont {Sels}}]{pandeyAdiabaticEigenstateDeformations2020a}%
  \BibitemOpen
  \bibfield  {author} {\bibinfo {author} {\bibfnamefont {M.}~\bibnamefont {Pandey}}, \bibinfo {author} {\bibfnamefont {P.~W.}\ \bibnamefont {Claeys}}, \bibinfo {author} {\bibfnamefont {D.~K.}\ \bibnamefont {Campbell}}, \bibinfo {author} {\bibfnamefont {A.}~\bibnamefont {Polkovnikov}},\ and\ \bibinfo {author} {\bibfnamefont {D.}~\bibnamefont {Sels}},\ }\bibfield  {title} {\bibinfo {title} {Adiabatic {{Eigenstate Deformations}} as a {{Sensitive Probe}} for {{Quantum Chaos}}},\ }\href {https://doi.org/10.1103/PhysRevX.10.041017} {\bibfield  {journal} {\bibinfo  {journal} {Physical Review X}\ }\textbf {\bibinfo {volume} {10}},\ \bibinfo {pages} {041017} (\bibinfo {year} {2020})}\BibitemShut {NoStop}%
\bibitem [{\citenamefont {Vicentini}\ \emph {et~al.}(2022)\citenamefont {Vicentini}, \citenamefont {Hofmann}, \citenamefont {Szab{\'o}}, \citenamefont {Wu}, \citenamefont {Roth}, \citenamefont {Giuliani}, \citenamefont {Pescia}, \citenamefont {Nys}, \citenamefont {{Vargas-Calder{\'o}n}}, \citenamefont {Astrakhantsev},\ and\ \citenamefont {Carleo}}]{vicentiniNetKetMachineLearning2022}%
  \BibitemOpen
  \bibfield  {author} {\bibinfo {author} {\bibfnamefont {F.}~\bibnamefont {Vicentini}}, \bibinfo {author} {\bibfnamefont {D.}~\bibnamefont {Hofmann}}, \bibinfo {author} {\bibfnamefont {A.}~\bibnamefont {Szab{\'o}}}, \bibinfo {author} {\bibfnamefont {D.}~\bibnamefont {Wu}}, \bibinfo {author} {\bibfnamefont {C.}~\bibnamefont {Roth}}, \bibinfo {author} {\bibfnamefont {C.}~\bibnamefont {Giuliani}}, \bibinfo {author} {\bibfnamefont {G.}~\bibnamefont {Pescia}}, \bibinfo {author} {\bibfnamefont {J.}~\bibnamefont {Nys}}, \bibinfo {author} {\bibfnamefont {V.}~\bibnamefont {{Vargas-Calder{\'o}n}}}, \bibinfo {author} {\bibfnamefont {N.}~\bibnamefont {Astrakhantsev}},\ and\ \bibinfo {author} {\bibfnamefont {G.}~\bibnamefont {Carleo}},\ }\bibfield  {title} {\bibinfo {title} {{{NetKet}} 3: {{Machine Learning Toolbox}} for {{Many-Body Quantum Systems}}},\ }\href {https://doi.org/10.21468/SciPostPhysCodeb.7} {\bibfield  {journal} {\bibinfo  {journal} {SciPost Physics Codebases}\ ,\ \bibinfo {pages} {7}} (\bibinfo {year}
  {2022})}\BibitemShut {NoStop}%
\bibitem [{\citenamefont {Dawid}\ \emph {et~al.}(2023)\citenamefont {Dawid}, \citenamefont {Arnold}, \citenamefont {Requena}, \citenamefont {Gresch}, \citenamefont {P{\l}odzie{\'n}}, \citenamefont {Donatella}, \citenamefont {Nicoli}, \citenamefont {Stornati}, \citenamefont {Koch}, \citenamefont {B{\"u}ttner}, \citenamefont {Oku{\l}a}, \citenamefont {{Mu{\~n}oz-Gil}}, \citenamefont {{Vargas-Hern{\'a}ndez}}, \citenamefont {{Cervera-Lierta}}, \citenamefont {Carrasquilla}, \citenamefont {Dunjko}, \citenamefont {Gabri{\'e}}, \citenamefont {Huembeli}, \citenamefont {van Nieuwenburg}, \citenamefont {Vicentini}, \citenamefont {Wang}, \citenamefont {Wetzel}, \citenamefont {Carleo}, \citenamefont {Greplov{\'a}}, \citenamefont {Krems}, \citenamefont {Marquardt}, \citenamefont {Tomza}, \citenamefont {Lewenstein},\ and\ \citenamefont {Dauphin}}]{dawidModernApplicationsMachine2023}%
  \BibitemOpen
  \bibfield  {author} {\bibinfo {author} {\bibfnamefont {A.}~\bibnamefont {Dawid}}, \bibinfo {author} {\bibfnamefont {J.}~\bibnamefont {Arnold}}, \bibinfo {author} {\bibfnamefont {B.}~\bibnamefont {Requena}}, \bibinfo {author} {\bibfnamefont {A.}~\bibnamefont {Gresch}}, \bibinfo {author} {\bibfnamefont {M.}~\bibnamefont {P{\l}odzie{\'n}}}, \bibinfo {author} {\bibfnamefont {K.}~\bibnamefont {Donatella}}, \bibinfo {author} {\bibfnamefont {K.~A.}\ \bibnamefont {Nicoli}}, \bibinfo {author} {\bibfnamefont {P.}~\bibnamefont {Stornati}}, \bibinfo {author} {\bibfnamefont {R.}~\bibnamefont {Koch}}, \bibinfo {author} {\bibfnamefont {M.}~\bibnamefont {B{\"u}ttner}}, \bibinfo {author} {\bibfnamefont {R.}~\bibnamefont {Oku{\l}a}}, \bibinfo {author} {\bibfnamefont {G.}~\bibnamefont {{Mu{\~n}oz-Gil}}}, \bibinfo {author} {\bibfnamefont {R.~A.}\ \bibnamefont {{Vargas-Hern{\'a}ndez}}}, \bibinfo {author} {\bibfnamefont {A.}~\bibnamefont {{Cervera-Lierta}}}, \bibinfo {author} {\bibfnamefont {J.}~\bibnamefont {Carrasquilla}},
  \bibinfo {author} {\bibfnamefont {V.}~\bibnamefont {Dunjko}}, \bibinfo {author} {\bibfnamefont {M.}~\bibnamefont {Gabri{\'e}}}, \bibinfo {author} {\bibfnamefont {P.}~\bibnamefont {Huembeli}}, \bibinfo {author} {\bibfnamefont {E.}~\bibnamefont {van Nieuwenburg}}, \bibinfo {author} {\bibfnamefont {F.}~\bibnamefont {Vicentini}}, \bibinfo {author} {\bibfnamefont {L.}~\bibnamefont {Wang}}, \bibinfo {author} {\bibfnamefont {S.~J.}\ \bibnamefont {Wetzel}}, \bibinfo {author} {\bibfnamefont {G.}~\bibnamefont {Carleo}}, \bibinfo {author} {\bibfnamefont {E.}~\bibnamefont {Greplov{\'a}}}, \bibinfo {author} {\bibfnamefont {R.}~\bibnamefont {Krems}}, \bibinfo {author} {\bibfnamefont {F.}~\bibnamefont {Marquardt}}, \bibinfo {author} {\bibfnamefont {M.}~\bibnamefont {Tomza}}, \bibinfo {author} {\bibfnamefont {M.}~\bibnamefont {Lewenstein}},\ and\ \bibinfo {author} {\bibfnamefont {A.}~\bibnamefont {Dauphin}},\ }\href {https://doi.org/10.48550/arXiv.2204.04198} {\bibinfo {title} {Modern applications of machine learning in
  quantum sciences}} (\bibinfo {year} {2023}),\ \Eprint {https://arxiv.org/abs/2204.04198} {2204.04198} \BibitemShut {NoStop}%
\bibitem [{\citenamefont {Ozawa}\ and\ \citenamefont {Mera}(2021)}]{ozawaRelationsTopologyQuantum2021}%
  \BibitemOpen
  \bibfield  {author} {\bibinfo {author} {\bibfnamefont {T.}~\bibnamefont {Ozawa}}\ and\ \bibinfo {author} {\bibfnamefont {B.}~\bibnamefont {Mera}},\ }\bibfield  {title} {\bibinfo {title} {Relations between topology and the quantum metric for {{Chern}} insulators},\ }\href {https://doi.org/10.1103/PhysRevB.104.045103} {\bibfield  {journal} {\bibinfo  {journal} {Physical Review B}\ }\textbf {\bibinfo {volume} {104}},\ \bibinfo {pages} {045103} (\bibinfo {year} {2021})},\ \Eprint {https://arxiv.org/abs/2103.11582} {2103.11582} \BibitemShut {NoStop}%
\bibitem [{\citenamefont {Orel}\ and\ \citenamefont {Perne}(2012)}]{orelComputationsHalfrangeChebyshev2012}%
  \BibitemOpen
  \bibfield  {author} {\bibinfo {author} {\bibfnamefont {B.}~\bibnamefont {Orel}}\ and\ \bibinfo {author} {\bibfnamefont {A.}~\bibnamefont {Perne}},\ }\bibfield  {title} {\bibinfo {title} {Computations with half-range {{Chebyshev}} polynomials},\ }\href {https://doi.org/10.1016/j.cam.2011.10.006} {\bibfield  {journal} {\bibinfo  {journal} {Journal of Computational and Applied Mathematics}\ }\textbf {\bibinfo {volume} {236}},\ \bibinfo {pages} {1753} (\bibinfo {year} {2012})}\BibitemShut {NoStop}%
\end{thebibliography}%

\end{document}